\documentclass{article}
\usepackage[utf8]{inputenc} 
\usepackage[T1]{fontenc}    
\usepackage{hyperref}       
\usepackage{url}            
\usepackage{booktabs}       
\usepackage{amsfonts}       
\usepackage{nicefrac}       
\usepackage{microtype}      
\usepackage{titlesec}
\usepackage[sorting=none,maxbibnames=5,giveninits]{biblatex}
\usepackage{mathtools}
\usepackage{graphicx}
\usepackage{xcolor}
\usepackage{float}
\usepackage{multicol}
\usepackage{fullpage}
\usepackage{subfigure}
\usepackage{colortbl}
\usepackage{amsmath,amsfonts,amssymb,amsthm}
\usepackage{outlines}
\usepackage[misc]{ifsym}
\usepackage{csvsimple}
\usepackage{makecell}
\usepackage{wasysym}
\usepackage{array}
\usepackage{authblk}

\hypersetup{
    colorlinks,
    citecolor=black,
    filecolor=black,
    linkcolor=black,
    urlcolor=black
}

\title{A Multi-Modal AI System for Screening Mammography: Integrating 2D and 3D Imaging to Improve Breast Cancer Detection in a Prospective Clinical Study}

\author[1]{Jungkyu Park}
\author[2]{Jan Witowski}
\author[3]{Yanqi Xu}
\author[4]{Hari Trivedi}
\author[4]{Judy Gichoya}
\author[4]{Beatrice Brown-Mulry}
\author[5]{Malte Westerhoff}
\author[2]{Linda Moy}
\author[2]{Laura Heacock}
\author[2]{Alana Lewin}
\author[1,2,3]{Krzysztof J. Geras\thanks{k.j.geras@nyu.edu}}

\affil[1]{Vilcek Institute of Graduate Biomedical Sciences, NYU Grossman School of Medicine, New York, NY, USA}
\affil[2]{Department of Radiology, NYU Grossman School of Medicine, New York, NY, USA}
\affil[3]{Center for Data Science, New York University, New York, NY, USA}
\affil[4]{HITI Lab, Emory University, Atlanta, GA, USA}
\affil[5]{Visage Imaging, Inc., San Diego, CA, USA}

\date{}
\begin{document}

\maketitle

\newrefsegment

\begin{abstract} 
Although digital breast tomosynthesis (DBT) improves diagnostic performance over full-field digital mammography (FFDM), false-positive recalls remain a concern in breast cancer screening.
We developed a multi-modal artificial intelligence system integrating FFDM, synthetic mammography, and DBT to provide breast-level predictions and bounding-box localizations of suspicious findings.
Our AI system, trained on approximately 500,000 mammography exams, achieved 0.945 AUROC on an internal test set.
It demonstrated capacity to reduce recalls by 31.7\% and radiologist workload by 43.8\% while maintaining 100\% sensitivity, underscoring its potential to improve clinical workflows. 
External validation confirmed strong generalizability, reducing the gap to a perfect AUROC by 35.31\%-69.14\% relative to strong baselines.
In prospective deployment across 18 sites, the system reduced recall rates for low-risk cases.
An improved version, trained on over 750,000 exams with additional labels, further reduced the gap by 18.86\%-56.62\% across large external datasets. 
Overall, these results underscore the importance of utilizing all available imaging modalities, demonstrate the potential for clinical impact, and indicate feasibility of further reduction of the test error with increased training set when using large-capacity neural networks.
\end{abstract}

\maketitle

\section{Introduction}\label{sec1}

Breast cancer is the leading cause of cancer-related deaths among women worldwide~\cite{globocan2020}, with a lifetime risk of approximately 13\%~\cite{howlader2020seer}. 
Screening mammography aims to detect cancer at its earliest stage. It has been shown to reduce the mortality rate for breast cancer~\cite{kopans2002beyond, duffy2002impact}, although dense tissue can mask cancer~\cite{mandelson2000breast}.
In the US, approximately 40 million mammography exams are performed annually~\cite{mqsa2021national}.
The optimal recall rate for screening mammography is between 5\% and 12\%~\cite{schell2007evidence, lehman2017national}. 
Recalled patients undergo additional imaging, and 1-2\% undergo a breast biopsy~\cite{kopans2015open} which amounts to over 1.5 million women~\cite{elmore2015diagnostic,silverstein2009s,silverstein2009special}.
However, only 20-40\% of biopsies yield a diagnosis of cancer~\cite{kopans2015open}.
False-positive mammograms and biopsies cost \$2.8 billion and \$2.18 billion annually in the US, respectively~\cite{vlahiotis2018analysis}.
Additionally, biopsies are associated with pain and emotional distress~\cite{hemmer2008stereotactic,maxwell2000imaging}, which decreases short-term quality of life and adherence to future screening recommendations ~\cite{humphrey2014percutaneous,miglioretti2024association}.

Despite increased cancer detection and decreased recall rate due to the introduction of digital breast tomosynthesis (DBT)~\cite{kopans2014digital, mcdonald2016effectiveness, rafferty2016breast, conant2019association, conant2020five, bahl2020breast}, substantial variability in sensitivity and specificity between radiologists persists~\cite{elmore2009variability, lehman2015diagnostic}. 
For example, 1 in 8 breast cancers are missed during interpretation in community practices in the US~\cite{lehman2017national}, and interpretation errors in mammography contribute to up to 25\% of missed detectable breast cancers~\cite{bird1992analysis, majid2003missed, weber2016characteristics, broeders2003use}. 
Additionally, DBT interpretation time is almost doubled compared to that of Full-Field Digital Mammography (FFDM, or ``2D mammography''), due to the increased number of images~\cite{aase2019randomized}. 
This reduces workflow efficiency and contributes to fatigue in radiologists.

Artificial Intelligence (AI)~\cite{geras2019artificial} could assist radiologists by highlighting suspicious lesions~\cite{gao2019new,park2021lessons,park2023efficient,konz2023competition,buda2021data}, reducing false-positive recalls and filtering exams with a low likelihood of cancer~\cite{dembrower2020effect,kyono2020improving,raya2021ai}, and increasing diagnostic accuracy when integrated with radiologists' assessments~\cite{wu2019deep, shen2021interpretable,shen2021ultrasound}.
Many AI systems have been developed to support mammography interpretation~\cite{mammo, wu2019deep, mckinney2020international, shen2019globally, shen2021interpretable, wu2020improving, park2021lessons,park2023efficient}. 
In a recent randomized controlled trial in Sweden (MASAI), AI-supported mammography screening resulted in cancer detection rate similar to standard double reading without increasing the recall rate~\cite{laang2023artificial}. 
A different prospective clinical trial in Sweden (ScreenTrustCAD) demonstrated that double reading by one radiologist with AI results in 4\% increase in screening-detected cancers compared to double reading by two radiologists~\cite{dembrower2023artificial}. 
Despite these promising results, most studies on AI-assisted mammography interpretation have been conducted in European settings with double reading as standard of care. 
In contrast, the impact of AI in a single-reader workflow representative of the US screening environment, remains unexplored.
Our study addresses this gap by evaluating AI-assisted single-reader screening within the US system.

Despite recent advances, the practical deployment of AI systems for mammography remains challenging due to limitations in achieving sufficient accuracy, generalizability, and trust in users. 
Much of the effort in AI for mammography has been dedicated to developing systems that use FFDM~\cite{wu2019deep, shen2021interpretable, mckinney2020international, shen2019globally, wu2020improving} or synthetic mammography~\cite{matthews2020multi}, without fully leveraging the depth information provided by DBT.
Additionally, much of the effort in developing DBT-based models has been dedicated to approaches that compress the full 3D input into one or a few 2D images using techniques such as maximum-intensity projection~\cite{singh2020adaptation} or dynamic feature image~\cite{liang2019joint}, or trainable summarization algorithms~\cite{tardy2021trainable}.
While computationally efficient, these compression-based approaches can conceal subtle or obscured lesions that are best seen in specific slices.
Furthermore, many studies on AI models were conducted with relatively small datasets~\cite{samala2016mass, lai2020dbt, fotin2016detection, mendel2019transfer}, which limits the applicability of their conclusions.

\begin{figure}
    \centering
    \includegraphics[width=0.98\linewidth]{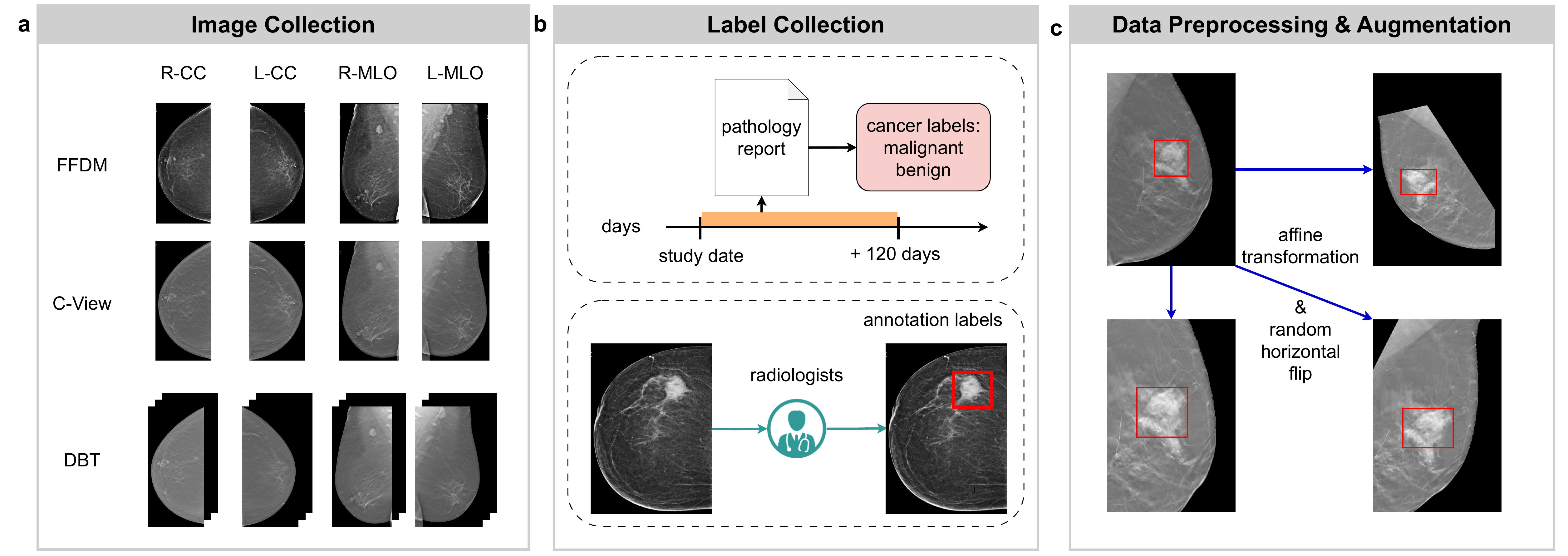}
    \includegraphics[width=0.98\linewidth]{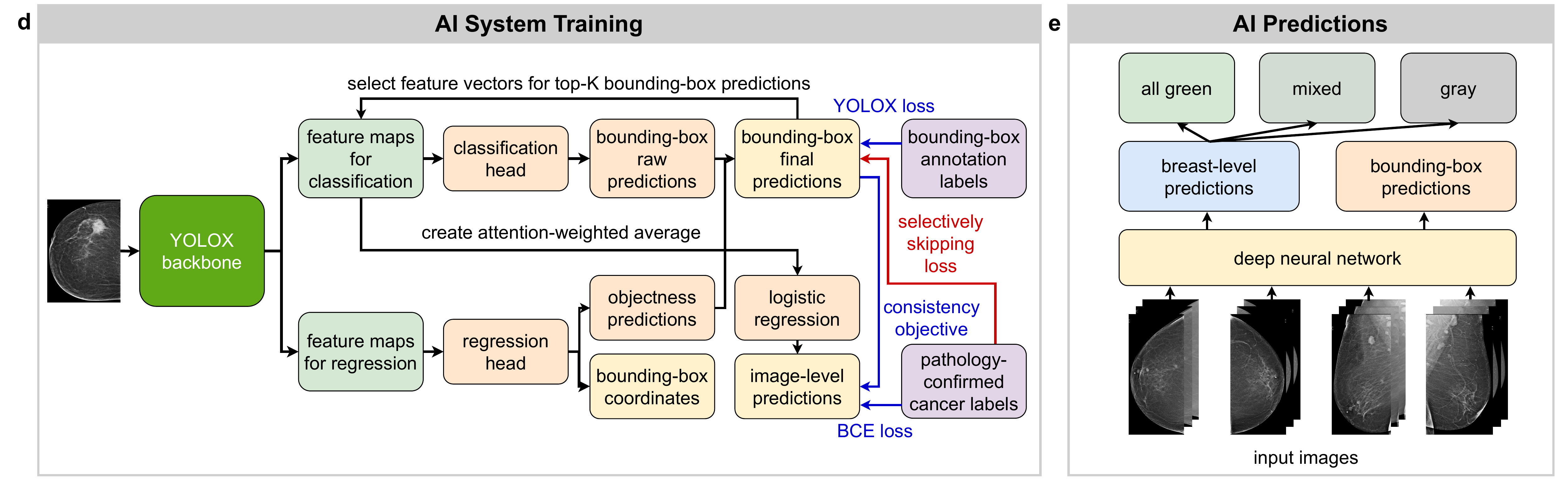}\\
    \hspace{0pt} 
    \includegraphics[width=0.98\linewidth]{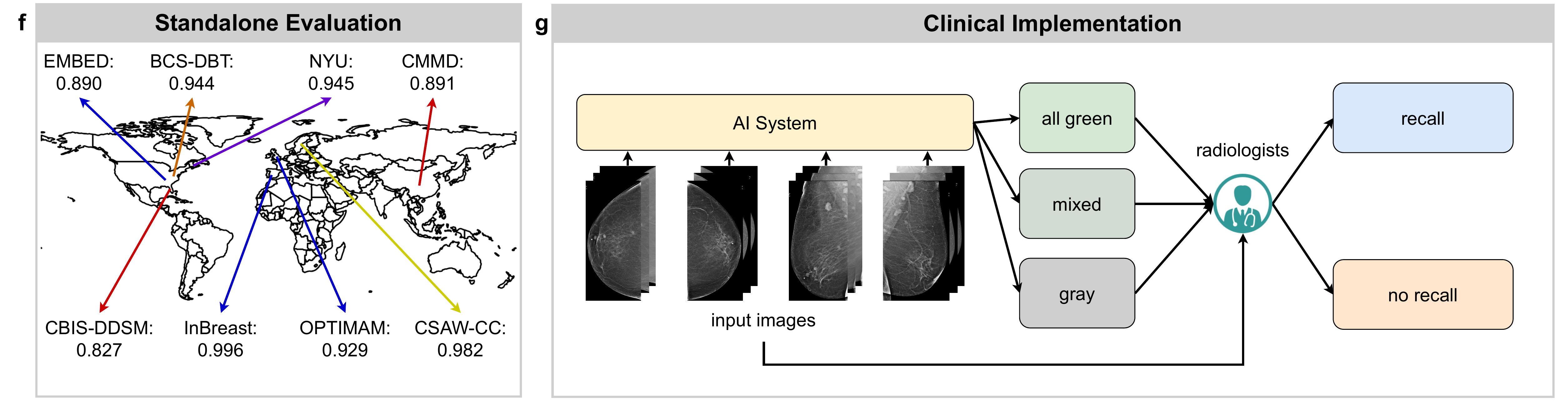}
    \caption{An overview of the AI system. \textbf{a} To build the AI system, we collected screening and diagnostic mammography images that contain FFDM, C-View and DBT images. \textbf{b} For each breast, we determined a cancer label based on the pathology reports for the patient within a timeframe of 0 to 120 days from the study date. \textbf{c} To enhance the diversity of the training data, images underwent data augmentation, including affine transformations and random horizontal flips. \textbf{d} Each neural network model in the proposed AI system is trained to create not only bounding-box predictions but also image-level prediction by aggregating information from the top bounding box predictions. \textbf{e} The AI system analyzes all images for each exam and generates probability predictions at the breast level. An exam is classified as ``all green'' if both breasts receive predictions indicating a low likelihood of cancer. It is labeled ``mixed'' when only one breast has a low prediction, while the exam is categorized as ``gray'' if neither breast is deemed unlikely to have cancer. \textbf{f} We evaluated the system on an internal test set (AUROC: 0.945, 95\% CI: 0.930, 0.960, N=38,368 breasts) as well as seven external datasets collected across three continents. \textbf{g} In clinical implementation, radiologists review the input images along with the model's predictions (``all green,'' ``mixed,'' or ``gray'') to make informed decisions about whether to recall the patient for additional imaging or follow-up.}
    \label{fig:extended-abstract}
\end{figure}

In this study, we present an AI system (Fig.~\ref{fig:extended-abstract}) designed to detect malignant and benign lesions and to compute image-level and breast-level probabilities for the presence of malignant and benign lesions. 
To train and evaluate our AI system, we collected a dataset, referred to as the ``NYU Comprehensive Mammography Dataset,'' that includes both screening and diagnostic exams acquired with three imaging modalities: FFDM, DBT, and 2D images synthesized from DBT using Hologic's C-View\texttrademark \phantom{ } algorithm. 
While all DBT exams performed on Hologic systems, the FFDM exams were acquired using a variety of mammography systems from Hologic and Siemens.
Our initial model was trained on 496,832 screening and diagnostic mammography exams completed at NYU Langone Health, leveraging pathology-confirmed labels that indicate the presence of cancer, as well as bounding-box annotations that indicate its location.

Our AI system provides breast-level predictions suitable for triage, and highlights the location of suspicious findings on 2D and 3D imaging with bounding-box predictions.
To enhance interpretability for radiologists, we encourage a strong correlation between these predictions through the training objective.
Furthermore, the system fully leverages the complementary information provided by all three mammography modalities by ensembling modality-specific models.
The system is flexible and can handle exams with missing modalities. 
Importantly, our DBT models utilize all DBT slices to preserve depth-specific information.

The model's performance was rigorously evaluated using task-specific metrics for binary classification and detection.  
Moreover, to assess the efficacy of our AI system in the clinical setting, we applied it to prospective screening mammography interpretation at a tertiary care center. 
We evaluated the effect of interpretation with AI support on the recall rate.
Additionally, we measured how this compares to national benchmarks. 
In doing so, we found a subset of examinations in which the recall rate could be significantly decreased with AI assistance during interpretation.

This initial model was designed as a preliminary system, built with the data and resources readily available at the time. 
Based on its promising performance in retrospective and prospective studies, with institutional support, we expanded our dataset and our collection of annotation labels, enabling the development of an improved version of the model.
The improved model was trained on an expanded dataset of 768,493 exams, incorporating not only additional exams but also more comprehensive annotation labels.
Our initial model achieved AUROCs ranging from 0.827 to 0.996 across seven external datasets (Table~\ref{tab:auc_table}).
This improved model outperforms the first version of the system and reduces the gap to a perfect AUROC by 18.86\%-56.62\% on large external datasets (Supplementary Table~\ref{tab:test_v1_v2}), paving the way for broader clinical adoption.
The performance of the improved model underscores the value of iterative refinement and expanded datasets in advancing AI systems for medical imaging.

\section{Results}\label{sec2}

\subsection{Datasets}

To perform this study, we collected the ``NYU Comprehensive Mammography Dataset'', comprising FFDM, DBT, and synthetic 2D mammography (C-View) images. 
Older screening exams contain only FFDM whereas newer ones contain FFDM, C-View and DBT.
All screening exams contain CC and MLO views for both breasts, while diagnostic exams may contain additional views (e.g., LM, ML, XCCL, XCC, TAN, XCCM, AT, RL, and RM).
Diagnostic exams may have varying numbers of views and are not required to include all three modalities.
The statistics of the mammography views is shown in Supplementary Table~\ref{tab:stats_views}.
Our dataset consists of images acquired using various mammography systems: Siemens Mammomat Novation DR, Siemens Mammomat Inspiration, HOLOGIC Lorad Selenia, and HOLOGIC Selenia Dimensions. All exams involving DBT were collected using the Hologic scanners, while the FFDM exams were collected with Hologic and Siemens scanners.

We developed two versions of models using V1 and V2 of the NYU Comprehensive Mammography Dataset. 
The V1 dataset comprises 519,757 exams from 235,288 patients imaged between 2010 and 2020 at NYU Langone Health and includes fewer exams with limited bounding-box annotations compared to V2.
The initial model trained on the V1 dataset, referred to as the V1 model, has already been applied in a clinical setting.
The V2 dataset consists of 833,997 exams from 319,621 patients imaged between 2010 and 2022 at NYU Langone Health and was designed to evaluate the effect of increasing training data; it features a larger number of exams along with substantially more bounding-box annotations. 
However, the V2 model trained on the V2 dataset has not yet been evaluated prospectively.
Both V1 and V2 datasets are divided into three subsets: a training-validation set used during the training of individual models, an ensemble-selection set for creating ensembles of models, and a test set for evaluation. 
The detailed description of the different subsets can be found in Section~\ref{sec:dataset}.

At the initial interpretation, each exam was assigned a BI-RADS label indicating the assessment of findings on mammography. 
Additionally, each breast received two binary labels derived from associated pathology reports based on biopsies, created within 120 days of imaging: the malignant label (positive if biopsy-confirmed cancer is present) and the benign label (positive if benign findings are reported).
Furthermore, we collected bounding-box labels from 52 radiologists from NYU Langone Health.

Finally, we evaluated bounding-box predictions across modalities using specialized ``multi-modal detection test subsets'' as described below.
Ensembling predictions across different modalities is feasible only when the images share identical breast placement; therefore, we identified FFDM-C-View-DBT triplets with confirmed pixel-level alignment.
For accurate breast-level evaluation, we retained only breasts that had both CC and MLO views available within the identified triplets to form ``multi-modal detection test subsets''.
This process yielded 79,579 aligned triplets (39,358 breasts from 19,680 exams) in the V1 test set and 183,578 aligned triplets (90,769 breasts from 45,389 exams) in the V2 test set. 
More details can be found in the Supplementary Section~\ref{sec:matching_modalities}.

To assess its generalizability, we evaluated the AI system on multiple external datasets presented in Table~\ref{tab:external_dataset}: the Chinese Mammography Database (CMMD)~\cite{Cui2021}, OPTIMAM~\cite{halling2020optimam}, and CSAW-CC~\cite{dembrower2020multi}, EMBED~\cite{jeong2023emory}, CBIS-DDSM~\cite{lee2017curated, shen2019deep, Sawyer-Lee2016}, INbreast~\cite{moreira2012inbreast}, and BCS-DBT~\cite{buda2021data, buda2020data}
The BCS-DBT dataset contains DBT images, and the other external datasets contain FFDM images.
Further information on the external datasets is in Section~\ref{sec:external}.

\begin{table}
    \centering
\caption{Breakdown of studies and labels in external data sets used to evaluate the model performance.
The numbers for CBIS-DDSM and INbreast datasets represent examples in the test set.
The study counts for OPTIMAM reflect a subset of the full dataset to which we were granted access. 
The numbers for the EMBED represent a subset of the full dataset as described in section~\ref{sec:external}.
}
\label{tab:external_dataset}
    \begin{tabular}{p{3.3cm}  |p{1.5cm}p{2cm} p{2.5cm} p{3cm} }
         \hline
         Dataset&  Imaging modality&Total studies & Studies with cancer findings &Studies with bounding-box annotations for cancer \\ \hline
         OPTIMAM&  FFDM&11,633&  4,004& N/A \\
         CMMD&   FFDM&1,774&  1,310& N/A \\
         CSAW-CC&   FFDM&23,395&  524& N/A \\
         EMBED&   FFDM&9,998&  121& 27 \\
         CBIS-DDSM &   FFDM&188&  92& N/A \\
         INbreast &   FFDM&31&  15& N/A \\
         BCS-DBT &  DBT&5,610 & 89 & 89  \\ 
         \hline
    \end{tabular}

\end{table}

\subsection{AI system overview}

Our AI system (Fig.~\ref{fig:extended-abstract}) is designed to enhance radiologists' performance by providing accurate breast-level cancer predictions and precise lesion localization across multiple modalities. For example, radiologists can utilize the breast-level predictions to quickly assess the likelihood of cancer and the bounding-box predictions to focus on specific areas. The system can act as a support tool that complements the radiologist's expertise, aiming to improve diagnostic accuracy and/or reduce reading time.

Our AI system uses a modified YOLOX~\cite{ge2021yolox} architecture, adapted specifically to mammography analysis. The vanilla YOLOX generates only bounding-box predictions. We extended its functionality by aggregating hidden representations from the top bounding-box predictions to produce overall image-level predictions for cancer presence.
In addition, to fully exploit the depth information in DBT images, our approach processes all DBT slices.
Specifically, while the YOLOX models generate predictions on individual 2D images, our system aggregates these outputs by removing duplicate bounding-box predictions across slices (Supplementary Section~\ref{sec:mss}).
The image-level predictions for DBT are created by using these top-K bounding-box predictions for the whole DBT image.
The detailed description of the AI system architecture can be found in Section~\ref{sec:yolox}.

Additionally, the system analyzes all mammography images in screening exams including FFDM, C-View and DBT, thereby harnessing the unique strengths of each imaging modality. 
Specifically, each modality-specific model computes a breast-level prediction by averaging all image-level outputs from its available views; these predictions are then combined via a weighted average to form the final multi-modal ensemble.
Likewise, bounding-box predictions are also ensembled across modalities to achieve accurate lesion localization with minimal false-positive predictions. 

Our model is trained using both pathology-confirmed cancer labels and bounding-box annotations provided by radiologists. 
This dual-training strategy enables the system to learn from breast-level outcomes including findings that are invisible to human eyes while utilizing the richer and more precise training signal from the bounding-box annotations for clearly visible cancers.
Moreover, by training on a diverse set of cases that include both screening and diagnostic exams, the AI system is equipped to generalize across different patient populations and clinical scenarios as evidenced by its strong performance on external datasets.

\subsection{AI system stand-alone performance}

The V1 model achieved 0.945 AUROC [95\% confidence interval (CI): 0.930 to 0.960] and 0.275 AUPRC [CI: 0.193 to 0.370] when classifying breasts in the V1 test set. 
The V2 model achieved 0.953 AUROC [CI: 0.938 to 0.967] and 0.333 AUPRC [CI: 0.271 to 0.395] when classifying breasts in the V2 test set. 
The classification performance of our AI system, compared to prior work, is shown in Tables~\ref{tab:auc_table}~and~\ref{tab:auc_table_duke}.

\begin{figure}[ht]
    \centering
    \includegraphics[width=\linewidth]{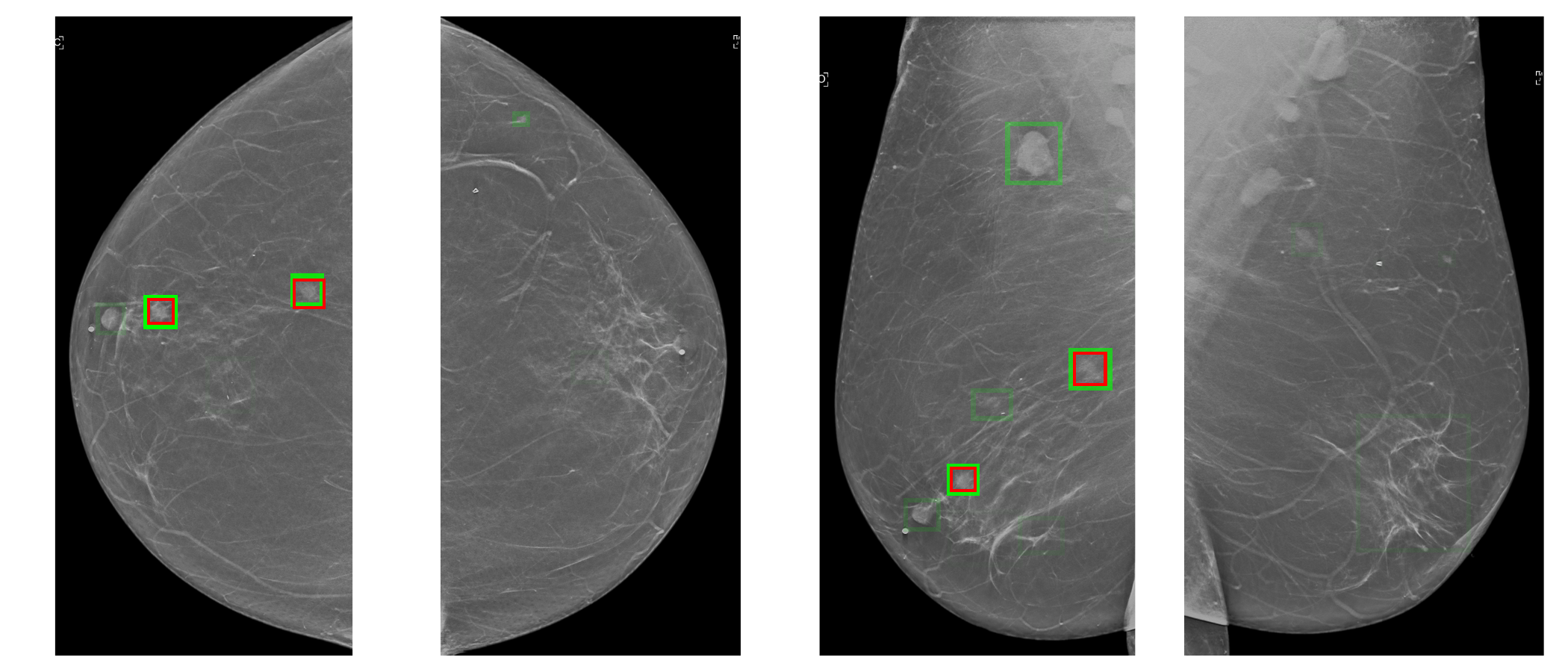} \\
    \caption{Visualization of the bounding-box predictions of the AI system (V2 model). From left to right, R-CC, L-CC, R-MLO, and L-MLO views are displayed. The bounding-box predictions generated from FFDM, C-View, DBT images are ensembled together and displayed on C-View images. The boxes with the highest malignancy prediction (displayed in the brightest green color) closely match the ground-truth lesions (shown as red boxes). 
    This example contains two spiculated masses at right 9:00, anterior depth and right 9:00, posterior depth which underwent ultrasound core biopsies yielding invasive mammary carcinoma with lobular and ductal features. 
    Additionally, the AI system detected a benign lymph node in the right axilla in the top portion of the R-MLO view.
    }
    \label{fig:visualization}
\end{figure}

For bounding-box predictions, we report AUFROC\_1 (Section~\ref{sec:stats}), which is the area under the free-response receiver operating characteristic curve for the interval on the x-axis between 0 and 1 false-positive predictions per image on the respective test sets.
Additionally, we report the sensitivity at three different thresholds per model that leads to 0.5, 1, and 2 false-positive predictions per image on the respective test sets.
More details of the FROC analysis can be found in Section~\ref{sec:stats}.
The model performance in bounding-box prediction is displayed in Table~\ref{tab:detection}, using the multi-modal detection test subsets for the internal test sets.
The V1 model achieved 0.848 AUFROC\_1 [CI: 0.793 to 0.903] when detecting breasts with malignant lesions in the V1 test set. 
The V2 model achieved 0.945 AUFROC\_1  [CI: 0.923 to 0.962] when detecting breasts with malignant lesions in the V2 test set. 
An exam with visualizations of bounding-box predictions is shown in Figure~\ref{fig:visualization}.

Supplementary Tables~\ref{tab:test_gmic_v1}~and~\ref{tab:test_v1_v2} show the p-values of permutation tests, AUC differences, and error reductions (Sec.~\ref{sec:stats}) between models.
The V1 model significantly outperforms the Globally-Aware Multiple Instance Classifier (GMIC)~\cite{shen2019globally,shen2021interpretable} in CMMD, OPTIMAM, CSAW-CC, EMBED, and CBIS-DDSM for AUROC and AUPRC.
Similarly, the V1 model significantly outperforms the 3D Globally-Aware Multiple Instance Classifier (3D-GMIC)~\cite{park2023efficient} in BCS-DBT for AUROC and AUPRC.
Additionally, the V2 model significantly outperforms the V1 model in OPTIMAM, CSAW-CC, EMBED, and BCS-DBT for AUROC and AUPRC.
Lastly, the differences between the AUFROC\_1 of the V1 model and the V2 model are statistically significant in EMBED and BCS-DBT.

Table~\ref{tab:ablation_modality} presents the model performance using different subsets of imaging modalities, highlighting the contribution of each modality.
Additionally, Table~\ref{tab:ablation_modality_2} shows the top-1 model performance for each modality, highlighting the effectiveness of each modality in capturing information relevant to diagnosis.

\subsection{Retrospective application to finding low-risk patients} 

Our AI system generates predictions as continuous values between 0 and 1. 
To make these predictions actionable, we apply a threshold that determines whether the prediction suggests a high enough likelihood of abnormality to warrant further examination by a radiologist. 
Using exam-level predictions (maximum of two breast-level predictions) from the V1 test set, we retrospectively analyzed the impact of different thresholds by hypothetically excluding exams below each threshold from being reviewed by radiologists.
Each threshold presents a trade-off: while higher thresholds reduce the number of potential recalls, they also increase the risk of missing cancers.

\begin{figure}[ht]
    \centering
    \begin{picture}(0,0) 
    \put(-3,120){\textbf{a}} 
    \end{picture}
    \includegraphics[width=0.33\linewidth]{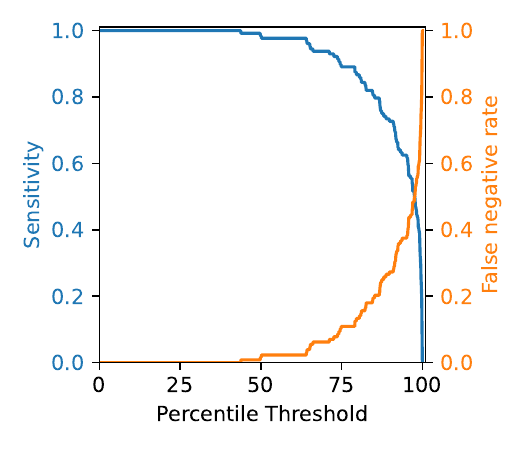}
    \begin{picture}(0,0)
    \put(-3,120){\textbf{b}}
    \end{picture}
    \includegraphics[width=0.33\linewidth]{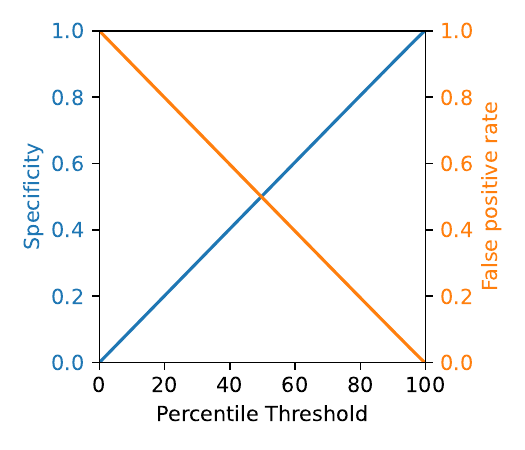}
    \begin{picture}(0,0)
    \put(-3,120){\textbf{c}}
    \end{picture}
    \raisebox{-4pt}{\includegraphics[width=0.3\linewidth]{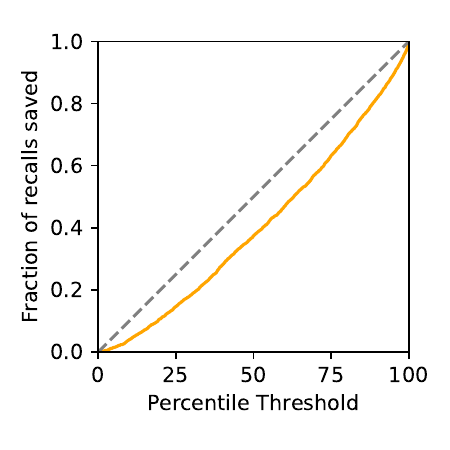}}

    \caption{Retrospective performance analysis of the V1 model on the V1 test set. 
    Exam-level predictions were generated by taking the maximum of the two breast-level predictions for each exam. 
    Thresholds were applied at various AI score percentiles to evaluate sensitivity, specificity, false-negative rate, false positive rate, and recall savings, assuming that exams below the threshold would not be reviewed by a radiologist.
    \textbf{a} Sensitivity (blue, left axis) and false-negative rate (orange, right axis) as a function of the AI score percentile threshold. 
    Sensitivity decreases and the false-negative rate increases as the threshold rises.
    \textbf{b} Specificity (blue, left axis) and false positive rate (orange, right axis) plotted against the AI score percentile threshold. 
    They appear as almost flat lines as the number of exams with cancer is very small compared to the entire test set, leading to minimal variation in these metrics across thresholds.
    \textbf{c} Fraction of recalls saved (orange) as a function of the AI score percentile threshold. 
    At a 43.8th percentile threshold for AI as a standalone reader, 31.7\% of recalls originally made by the radiologist could hypothetically be avoided without missing any breast cancers, illustrating the potential of AI to reduce the recall rates in this retrospective analysis. 
    }
    \label{fig:percentile_result}
\end{figure}

We evaluated the performance across a range of threshold values, which correspond to the 0th-100th percentile of the patient population in the test set (Fig.~\ref{fig:percentile_result}).
Among these, a threshold rejecting the bottom 43.8\% of exams, corresponding to the lowest prediction among positive cases, is noteworthy, as it enables the system to reject the maximum number of exams while ensuring no breast cancers are missed in this retrospective analysis.
This threshold prevents 31.7\% of unnecessary recalls and potentially reduces radiologist workload by 43.8\%.
This analysis is retrospective and has not yet been clinically validated.

\begin{table}
    \centering
\caption{Breast-level classification performance for identifying breasts with cancer across different data sets with 95\% confidence intervals. The performance on internal test sets is from the full models using all imaging modalities (FFDM, C-View, DBT). The external datasets in this table only contain FFDM images, and thus the performance on these datasets are from the subset of the models that utilize FFDM images. The results present breast-level estimates for the receiver operating characteristic curve (AUROC) and the areas under the precision-recall curve (AUPRC). AUPRC is highly sensitive to the class imbalance within each dataset. $^\dagger$ denotes cases in which the V1 model significantly outperforms GMIC, and $^\ddagger$ denotes cases in which the V2 model significantly outperforms the V1 model. Statistical significance was determined using the permutation test. The significance level is 0.05. The p-values are shown in Supplementary Table~\ref{tab:test_gmic_v1} and Supplementary Table~\ref{tab:test_v1_v2}.}
\label{tab:auc_table}
    \begin{tabular}{cc|ccc}
    \hline
         Data set&   AUC
&  GMIC (top-5 ensemble)
& V1 model&V2 model\\ \hline
         NYU V1 test set&  ROC&  N/A& 0.945 (0.930 - 0.960)\hphantom{$^\dagger$}&N/A
\\
         NYU V1 test set&  PR&  N/A& 0.275 (0.193 - 0.370)\hphantom{$^\dagger$}&N/A\\
         NYU V2 test set& ROC
& N/A
& N/A
& 0.953 (0.938 - 0.967)\hphantom{$^\ddagger$}\\
         NYU V2 test set& PR& N/A& N/A& 0.333 (0.271 - 0.395)\hphantom{$^\ddagger$}\\
          CMMD
&  ROC
&   0.831 (0.815 - 0.846)
&  0.891 (0.879 - 0.903)$^\dagger$&0.892 (0.879 - 0.904)\hphantom{$^\ddagger$}\\
         CMMD
&  PR&  0.859 (0.842 - 0.875)
& 0.908 (0.896 - 0.920)$^\dagger$&0.909 (0.896 - 0.921)\hphantom{$^\ddagger$}\\
         OPTIMAM
&  ROC
&  0.832 (0.825 - 0.840)
& 0.929 (0.925 - 0.933)$^\dagger$&0.942 (0.939 - 0.946)$^\ddagger$
\\
         OPTIMAM
&  PR&  0.633 (0.619 - 0.648)
& 0.799 (0.789 - 0.809)$^\dagger$&0.828 (0.819 - 0.838)$^\ddagger$\\
         CSAW-CC
&  ROC
&  0.943 (0.931 - 0.954)
& 0.982 (0.976 - 0.988)$^\dagger$&0.988 (0.983 - 0.993)$^\ddagger$\\
         CSAW-CC
&  PR&  0.495 (0.447 - 0.543)
& 0.763 (0.727 - 0.796)$^\dagger$&0.797 (0.762 - 0.829)$^\ddagger$\\
         EMBED
&  ROC
&  0.782 (0.736 - 0.824)& 0.890 (0.859 - 0.918)$^\dagger$&0.922 (0.897 - 0.946)$^\ddagger$\\
 EMBED
& PR& 0.064 (0.039 - 0.111)&0.213 (0.146 - 0.292)$^\dagger$&0.289 (0.208 - 0.383)$^\ddagger$\\
 InBreast& ROC
& 0.980 (0.940 - 1.000)&0.996 (0.984 - 1.000)\hphantom{$^\dagger$}&1.000 (1.000 - 1.000)\hphantom{$^\ddagger$}\\
 InBreast& 
PR& 0.957 (0.856 - 1.000)&0.991 (0.962 - 1.000)\hphantom{$^\dagger$}&1.000 (1.000 - 1.000)\hphantom{$^\ddagger$}\\
 CBIS-DDSM& ROC& 0.610 (0.529 - 0.690)&0.827 (0.768 - 0.881)$^\dagger$&0.791 (0.727 - 0.854)\hphantom{$^\ddagger$}\\
 CBIS-DDSM& PR& 0.569 (0.470 - 0.670)&0.842 (0.778 - 0.895)$^\dagger$&0.813 (0.741 - 0.875)\hphantom{$^\ddagger$}\\ \hline
    \end{tabular}

\end{table}

\begin{table}
    \centering
\caption{Image-level classification performance for identifying breasts with cancer along with 95\% confidence intervals. The BCS-DBT dataset only contains DBT images. In this table, the performances of the subset of the models that utilize DBT images are displayed. The results present image-level estimates for the receiver operating characteristic curve (AUROC) and the areas under the precision-recall curve (AUPRC). $^\dagger$ denotes cases in which the V1 model significantly outperforms GMIC, and $^\ddagger$ denotes cases in which the V2 model significantly outperforms the V1 model. Statistical significance was determined using the permutation test. The significance level is 0.05. The p-values are shown in Supplementary Table~\ref{tab:test_gmic_v1} and Supplementary Table~\ref{tab:test_v1_v2}.}
\label{tab:auc_table_duke}
    \begin{tabular}{cc|ccc}
    \hline
         Data set&   AUC
&  3D-GMIC (top-5 ensemble)
& V1 model&V2 model\\ \hline
         NYU V1 test set&  ROC&  N/A& 0.902 (0.880 - 0.921)\hphantom{$^\dagger$}&N/A
\\
         NYU V1 test set&  PR& N/A&0.246 (0.190 - 0.309)\hphantom{$^\ddagger$}&N/A
\\
 NYU V2 test set& ROC& N/A& N/A&0.921 (0.907 - 0.934)\hphantom{$^\ddagger$}\\
 NYU V2 test set& PR& N/A& N/A&0.186 (0.151 - 0.229)\hphantom{$^\ddagger$}\\ 
         BCS-DBT&  ROC&  0.857 (0.827 - 0.888)& 0.944 (0.927 - 0.961)$^\dagger$&0.976 (0.964 - 0.985)$^\ddagger$
\\
         BCS-DBT&  PR& 0.165 (0.115 - 0.226)&0.405 (0.324 - 0.490)$^\dagger$&0.490 (0.411 - 0.574)$^\ddagger$\\ \hline
    \end{tabular}
    
\end{table}

\begin{table}
    \centering
\caption{Performance of the AI system in making bounding-box predictions for malignant lesions with
95\% confidence intervals. 
For both NYU V1 and V2 test sets, multi-modal detection test subsets of the respective test sets are used in this table.
For EMBED, exams with missing annotations (positive pathology-confirmed cancer label but no bounding-box annotations) are excluded when computing the evaluations in this table. 
The lesion-level sensitivity measures the percentage of malignant lesions that are correctly detected.
The breast-level sensitivity measures the percentage of breasts with at least one malignant lesion that are correctly identified, requiring at least one true-positive bounding-box prediction across any lesion on any view. 
$^\ddagger$ in the AUFROC\_1 column for the external datasets denotes cases in which V2 model significantly outperforms the V1 model. Statistical significance was determined using the permutation test. The significance level is 0.05. The p-values are shown in Table~\ref{tab:test_v1_v2}.
Abbreviations: FP, false positive.
}
\label{tab:detection}
\resizebox{\textwidth}{!}{%
    \begin{tabular}{lp{2.1cm} p{1.7cm}  p{1.5cm} |p{2.2cm} p{2.6cm} p{2.2cm} p{2.4cm}} 
    \hline
            Model&Dataset&  Imaging modalities used&   Sensitivity level& 
  AUFROC\_1 &Sensitivity at 0.5 FP per image&
  Sensitivity at 1 FP per image&
 Sensitivity at 2 FP per image
\\ \hline
            V1&V1 test set&  FFDM&  Breast&  0.751 \newline (0.681 - 0.821) &0.796 \newline (0.716 - 0.874)
&  0.857 \newline (0.792 - 0.925)& 0.898 \newline (0.837 - 0.955)
\\
            V2&V2 test set&  FFDM&  
Breast&  
0.908 \newline (0.879 - 0.935) &0.938 \newline (0.907 - 0.966)
&  
0.955 \newline (0.926 - 0.979)& 
0.971 \newline (0.946 - 0.988)
\\ \hline
            V1&V1 test set&  C-View&  Breast&  0.776 \newline (0.709 - 0.844) &0.813 \newline (0.733 - 0.890)
&  0.875 \newline (0.808 - 0.941)& 0.927 \newline (0.870 - 0.977)
\\
            V2&V2 test set&  C-View&  Breast&  0.884 \newline (0.853 - 0.915) &0.915 \newline (0.879 - 0.949)
&  0.955 \newline (0.929 - 0.979)& 0.972 \newline (0.949 - 0.991)
\\ \hline
            V1&V1 test set&  DBT&  Breast&  0.847 \newline (0.790 - 0.904)&0.880 \newline (0.812 - 0.947)&  0.924 \newline (0.872 - 0.976)& 0.978 \newline (0.945 - 1.000)\\
            V2&V2 test set&  DBT&  Breast&  0.915 \newline (0.887 - 0.939)&0.943 \newline (0.912 - 0.969)&  0.963 \newline (0.939 - 0.985)& 0.976 \newline (0.954 - 0.992)\\ \hline
            V1&V1 test set&  all three \newline modalities& Breast& 0.848 \newline (0.793 - 0.903) &0.885 \newline (0.822 - 0.949)
& 0.948 \newline (0.901 - 0.989)& 0.979 \newline (0.947 - 1.000)
\\
            V2&V2 test set&  all three \newline modalities& 
 Breast&  0.945 \newline (0.923 - 0.962) &0.976 \newline (0.954 - 0.992)
&  0.984 \newline (0.966 - 0.996)& 0.992 \newline (0.979 - 1.000)
\\
    \hline
           V1&EMBED \newline subset&  FFDM& Breast&  0.907 \newline (0.852 - 0.962) &0.963 \newline (0.893 - 1.000)
&  1.000 \newline (1.000 - 1.000)& 1.000 \newline(1.000 - 1.000)
\\
           V2&EMBED \newline subset&  FFDM& 
Breast&  
0.942$^\ddagger$ \newline (0.907 - 0.973) &1.000 \newline (1.000 - 1.000)
&  
1.000 \newline (1.000 - 1.000)& 
1.000 \newline (1.000 - 1.000)\\ \hline
            V1&BCS-DBT&  DBT&  Lesion&  0.824 \newline (0.775 - 0.867)
 &0.865 \newline (0.814 - 0.914)
&  0.905 \newline (0.859 - 0.948)& 0.926 \newline (0.885 - 0.963)
\\
            V2&BCS-DBT&  DBT&  Lesion&  0.884$^\ddagger$ \newline (0.845 - 0.924) &0.900 \newline (0.858 - 0.943)&  0.932 \newline (0.893 - 0.966)& 0.942 \newline (0.909 - 0.972)\\ \hline
    \end{tabular}%
    }

\end{table}

\begin{table}[ht]
\centering
\caption{Breast-level classification performance of the different subsets of the V1 and the V2 models evaluated on their respective test sets with 95\% confidence intervals. Rows labeled by imaging modality (e.g. FFDM) represent the performance of model subsets that exclusively utilize that imaging modality for predictions. This comparison highlights the impact of incorporating additional imaging modalities as inputs to our models.}
\begin{tabular}{ll|rr}
\hline
 {modality}                  &  AUC&{ V1 model on V1 test set } & { V2 model on V2 test set} \\ \hline \hline
 {FFDM}            &  ROC&0.910 (0.883 - 0.936)& 0.934 (0.919 - 0.949)
\\
 {FFDM}            &  PR&0.235 (0.158 - 0.325)& 0.308 (0.245 - 0.371)
\\
 {C-View}                    &  ROC&0.915 (0.894 - 0.936)& 0.912 (0.891 - 0.932)
\\
 {C-View}                    &  PR&0.214 (0.138 - 0.304)& 0.248 (0.190 - 0.308)
\\
 {DBT}                       &  ROC&0.933 (0.911 - 0.953)& 0.946 (0.929 - 0.962)
\\ 
 {DBT}                       &  PR&0.298 (0.218 - 0.404)& 0.261 (0.207 - 0.324)
\\ \hline
 {FFDM + C-View}              &  ROC&0.923 (0.901 - 0.944)& 0.940 (0.924 - 0.953)
\\
 {FFDM + C-View}              &  PR&0.250 (0.171 - 0.339)& 0.311 (0.249 - 0.374)
\\
 {FFDM + DBT}                &  ROC&0.945 (0.931 - 0.960)& 0.954 (0.938 - 0.968)
\\
 {FFDM + DBT}                &  PR&0.273 (0.192 - 0.369)& 0.330 (0.270 - 0.392)
\\
 {C-View + DBT}                &  ROC&0.938 (0.918 - 0.955)& 0.947 (0.930 - 0.963)
\\ 
 {C-View + DBT}                &  PR&0.266 (0.185 - 0.363)& 0.304 (0.246 - 0.366)
\\ \hline
 {all 3 modalities (1 model per modality)}        &  ROC&0.940 (0.921 - 0.958)& 0.949 (0.933 - 0.964)
\\
 {all 3 modalities (1 model per modality)}        &  PR&0.285 (0.205 - 0.387)& 0.329 (0.265 - 0.391)
\\
 {all 3 modalities (all models)}        &  ROC&0.945 (0.930 - 0.960)& 0.953 (0.938 - 0.967)
\\ 
 {all 3 modalities (all models)}        &  PR&0.275 (0.193 - 0.370)& 0.333 (0.271 - 0.395)\\ \hline
\end{tabular}
\label{tab:ablation_modality}
\end{table}

\begin{table}[ht]
\centering
\caption{Breast-level classification performance of top-1 models for each imaging modality with 95\% confidence intervals. Among all individual models trained on the V1 and V2 datasets, individual models with the highest performance on the respective ensemble-selection datasets were chosen for this ablation study. These models are not necessarily included in the original ensembles. This comparison highlights each modality's effectiveness in capturing diagnostic information.}
\begin{tabular}{ll|rr}
\hline
 {modality}                  &  AUC&{V1 model on V1 test set} & {V2 model on V2 test set} \\ \hline \hline
 {FFDM}            &  ROC&0.900 (0.873 - 0.927)&0.925 (0.906 - 0.942)
\\
 {FFDM}            &  PR&0.174 (0.111 - 0.260)& 0.291 (0.231 - 0.355)
\\
 {C-View}                    &  ROC&0.906 (0.882 - 0.930)& 0.913 (0.893 - 0.931)
\\
 {C-View}                    &  PR&0.190 (0.120 - 0.272)& 0.222 (0.170 - 0.279)
\\
 {DBT}                       &  ROC&0.920 (0.893 - 0.946)& 0.940 (0.922 - 0.956)
\\ 
 {DBT}                       &  PR&0.277 (0.197 - 0.372)& 0.253 (0.198 - 0.313)\\  \hline
\end{tabular}
\label{tab:ablation_modality_2}
\end{table}

\subsection{Clinical Implementation}

The V1 version of our AI system was evaluated in a prospective clinical study. It was applied to 2D/3D screening mammography exams performed at 18 sites from August 2022 to April 2023. Screening mammography exams that contained a CC and MLO view in FFDM, DBT, and C-View were included in the study. 
Exclusions included exams with breast implants, male patients, prior mastectomy, or single-breast imaging, as the AI system was not validated for these groups.
Additionally, we excluded exams that included same-day screening ultrasound to prevent potential bias from sonographically detectable findings.
All mammograms were interpreted using a high-resolution viewing workstation with a 10-megapixel monitor that met Mammography Quality Standards Act (MQSA) standards, using Visage Client 7.1.18. Before the study, all radiologists underwent AI training by a fellowship-trained breast imager who had assisted with the development and deployment of the AI model.

\begin{figure}
    \centering
    \includegraphics[width=0.8\linewidth]{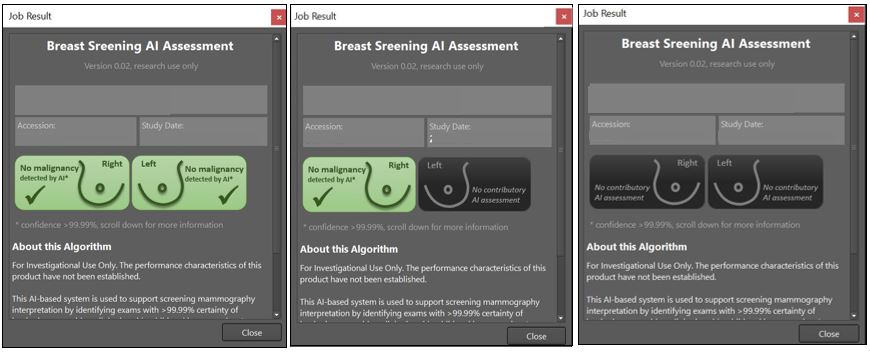}
    \caption{The AI assessment box viewed by radiologists during the interpretation of a screening mammogram. 
    A result is assigned to each breast, which can be either green or gray.
    A green result indicates that the AI model's output is below the operating point threshold, while a gray result signifies that the AI findings are ``noncontributory''.}
    \label{fig:green-gray-mixed-box}
\end{figure}

The AI model classified each breast as ``green'' (below the operating point threshold) or as ``gray'' (above the operating point threshold, characterized as ``no contributory AI assessment can be made'') (Figure~\ref{fig:green-gray-mixed-box}). 
While our model can generate regions of interest, there is uncertainty about the optimal deployment strategy for AI in mammography. 
Studies have suggested region-of-interest predictions may confer automation bias compared to exam-level or breast-level predictions~\cite{dratsch2023automation,vered2023effects,gaube2023non}. 
In this initial clinical deployment, we chose to display only breast-level predictions as follows. 
An operating point was set using the V1 test set to maximize green cases while maintaining 100\% sensitivity for cancers among gray cases.
This was determined by the lowest model prediction among the breasts with cancer, corresponding to the 34.86th percentile of the V1 test set, and incorporating a safety margin of 1 percentile, yielding a final threshold at the 33.86th percentile.
This threshold leads to 100\% sensitivity and 34\% specificity.
During the study, 33.2\% of exams had both breasts classified as green (``all green''), 31.5\% had one breast classified green and one breast was classified gray (``mixed''), and 35.2\% had both breasts classified ``gray'' (``all gray'').

The AI model assisted in interpreting 40,603 screening mammography exams by 20 fellowship-trained breast imagers (3-38 years of experience); each interpreting 645-7471 exams with AI assistance. Abnormal interpretation rate (AIR), or recall rate, was compared to the performance of the same radiologists before AI implementation on 40,415 mammographic studies read from December 2021 to July 2022, where each reader interpreted 429-7131 exams.

\begin{figure}
    \centering
    \includegraphics[width=0.8\linewidth]{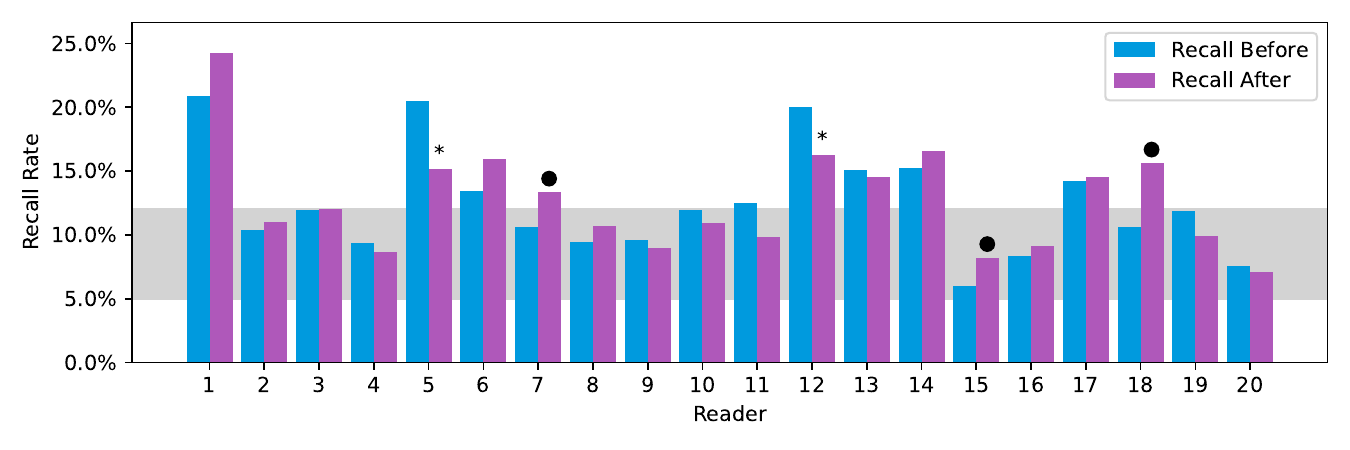}
    \caption{Comparison of AIR by Reader for all cases.
    8 of the 20 interpreting radiologists (3-28 years of experience) demonstrated recall rates >12\% in the 8 month study period prior to clinical implementation of the AI model. 
    After the introduction of the AI model, 2 of these 8 radiologists (6 and 21 years of experience; interpreted 429 and 1491 exams prior to AI implementation and 645 and 1212 exams after AI implementation, respectively) demonstrated a significant reduction in recall rate (p = 0.02406, z = -2.256, Cohen's h = -0.139 and p = 0.01115, z = -2.538, Cohen's h = -0.099 respectively). 3 of 20 radiologists demonstrated a significant increase (p = 0.01750, z = 2.376, Cohen's h = 0.088; p < 0.00001, z = 5.016, Cohen's h = 0.083; and p < 0.00001, z = 7.631, Cohen's h = 0.148 respectively) in recall rate (3, 8, 28 years of experience; interpreted 1574, 7131, and 5990 prior to AI implementation and 1363, 7471, and 4641 exams after AI implementation, respectively). The remaining 15 of 20 radiologists demonstrated no significant change to their recall rates after the AI clinical implementation (Supplementary Table~\ref{tab:my_table}). 
    * signifies statistically significant decrease in AIR after AI implementation. \CIRCLE \phantom{ } signifies statistically significant increase in AIR after AI implementation. Gray shading signifies the ideal AIR 5-12\% according to national benchmarks.}
    \label{fig:air-all-cases-by-reader}
\end{figure}

\begin{figure}
    \centering
    \includegraphics[width=0.8\linewidth]{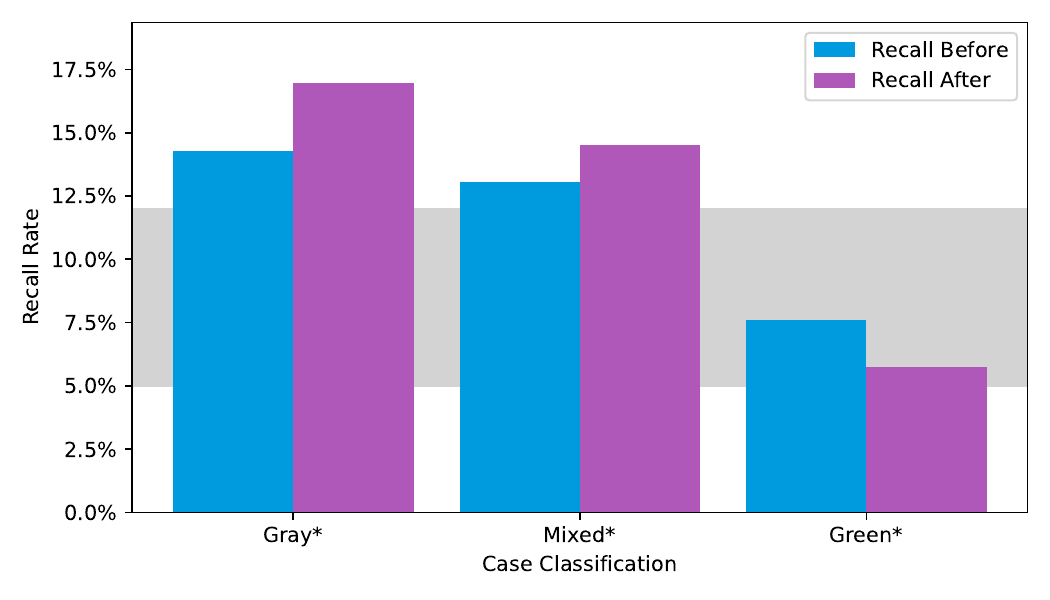}
    \caption{Comparison of AIR before and after AI implementation by case classification. 
    The recall rates increased after AI implementation for gray and mixed cases but decreased after AI implementation for green cases.
    Gray shading signifies the ideal AIR 5-12\% according to national benchmarks. }
    \label{fig:air-all-cases}
\end{figure}

\begin{figure}
    \centering
    \includegraphics[width=0.8\linewidth]{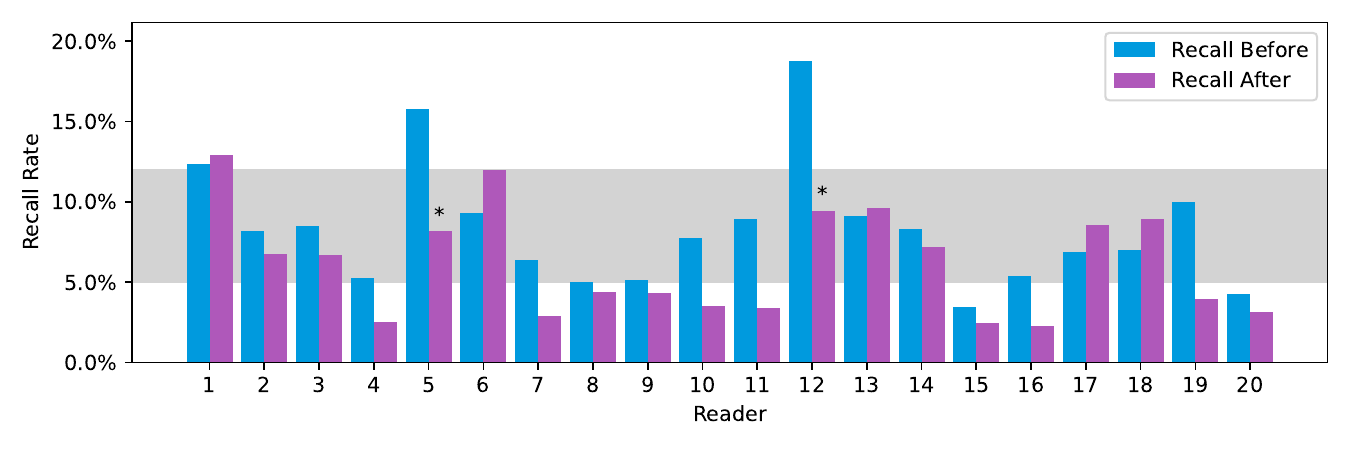}
    \caption{Comparison of AIR by reader for cases with both breasts designated by AI as ``Green''. 
    After the introduction of the AI model, 2 of 3 radiologists whose recall rates were above the national benchmark of 5-12\% demonstrated a significant reduction in the recall rates: 15.8\% (CI: 10.3\% - 21.2\%) to 8.2\% (CI: 4.9\% - 11.8\%) (p = 0.01434, z = -2.449, Cohen's h = -0.237; interpreted 184 green exams prior to AI implementation and 245 green exams after AI implementation, 6 years of experience) and 18.8\% (CI: 15.6\% - 22.1\%) to 9.4\% (CI: 7.0\% - 12.1\%) (p = 0.00003, z = -4.169, Cohen's h = -0.272; interpreted 570 green exams prior to AI implementation and 445 green exams after AI implementation, 21 years of experience).
    * signifies statistically significant decrease in AIR after AI implementation among the radiologists whose original recall rates prior to AI implementation were above the national benchmark. Gray shading signifies the ideal AIR 5-12\% according to national benchmarks.}
    \label{fig:air-by-reader-all-green}
\end{figure}

\begin{figure}
    \centering
    \includegraphics[width=0.8\linewidth]{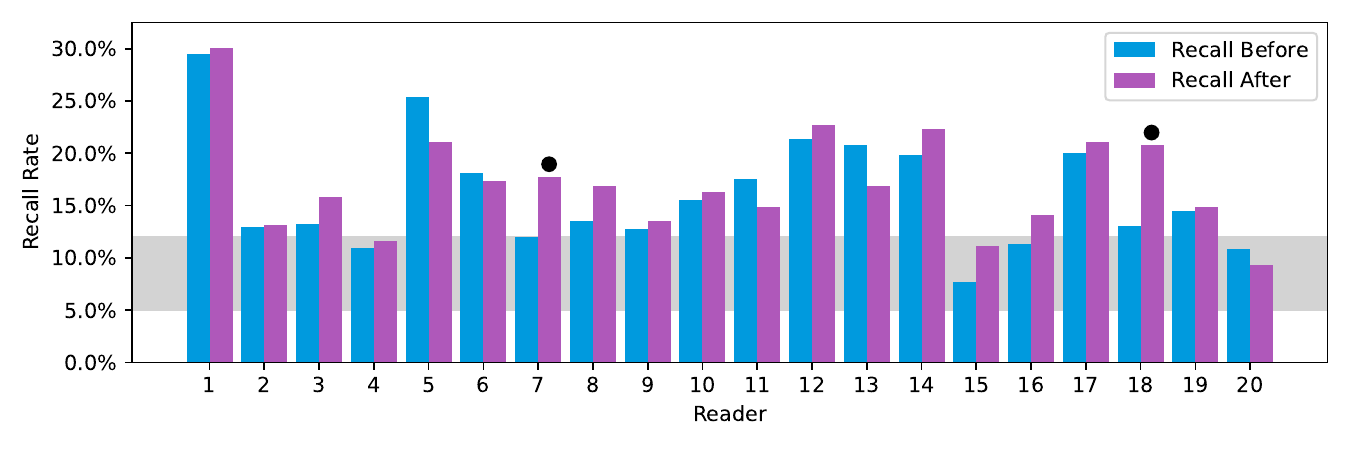}
    \caption{Comparison of AIR by reader for cases with both breasts designated by AI as ``Gray''. 
    15 of 20 radiologists (3-35 years of experience; interpreted 690-5953 gray exams) demonstrated increase in recall rates after AI clinical implementation, of which 2 of 15 radiologists showed significant increase in recall rates: 12.0\% (CI: 10.8\% - 13.2\%) to 17.7\% (CI: 16.4\% - 19.2\%) (p<0.00001, z = 6.153, Cohen's h = 0.161; interpreted 2828 gray exams prior to AI implementation and 3125 gray exams after AI implementation, 8 years of experience) and 13.1\% (CI: 11.6\% - 14.6\%) to 20.8\% (CI: 18.6\% - 22.8\%) (p<0.00001, z = 5.911 Cohen's h = 0.206; interpreted 1737 gray exams prior to AI implementation and 1566 gray exams after AI implementation, 28 years of experience). 
    \CIRCLE \phantom{ } signifies statistically significant increase in AIR after AI implementation. Gray shading signifies the ideal AIR 5-12\% according to national benchmarks.}
    \label{fig:air-by-reader-all-gray}
\end{figure}

\begin{figure}
    \centering
    \includegraphics[width=0.8\linewidth]{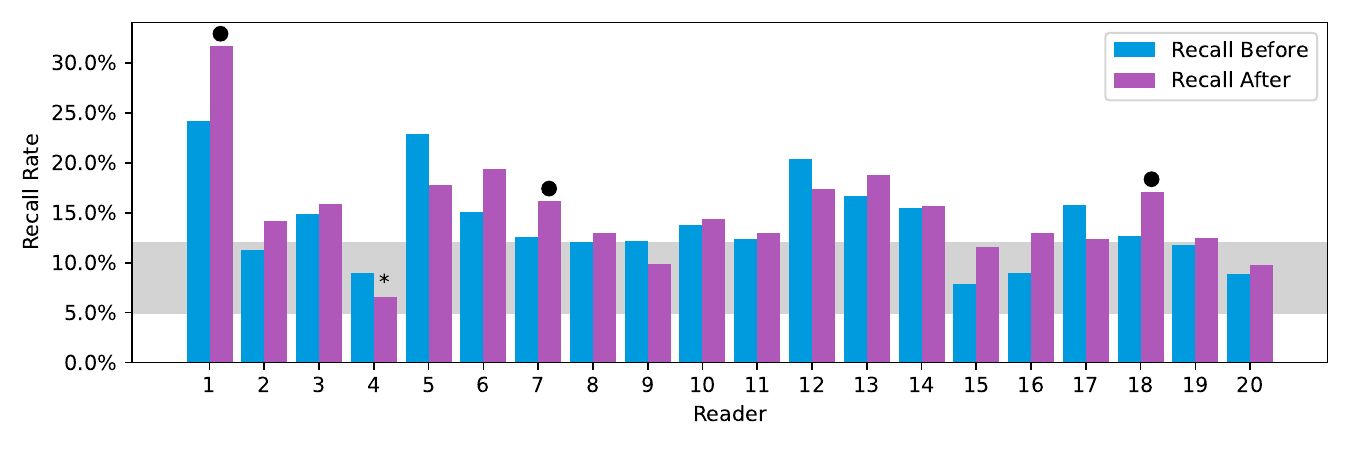}
    \caption{Comparison of AIR by Reader for Mixed Cases. 
    15 of 20 radiologists (3-35 years of experience; interpreted 564-4,719 mixed exams) demonstrated increase in recall rates after AI clinical implementation; increase in recall rates was statistically significant for 3 of 15 radiologists: 24.2\% (CI: 19.7\% - 28.7\%) to 31.6\% (CI: 27.3\% - 36.5\%) (p=0.02134, z=2.302, Cohen's h=0.167; interpreted 414 mixed exams prior to AI implementation and 620 mixed exams after AI implementation, 7 years of experience), 12.6\% (CI: 11.3\% - 14.0\%) to 16.1\% (CI: 14.7\% - 17.7\%) (p=0.00048, z=3.493, Cohen's h=0.102; interpreted 2322 mixed exams prior to AI implementation and 2397 mixed exams after AI implementation, 8 years of experience), 12.7\% (CI: 11.3\% - 14.2\%) to 17.1\% (CI: 15.1\% - 19.2\%) (p=0.00029, z=3.624, Cohen's h=0.123; interpreted 1951 mixed exams prior to AI implementation and 1540 mixed exams after AI implementation, 28 years of experience). 5 of 20 radiologists (6-38 years of experience; interpreted 318-2511 mixed exams) demonstrated decrease in recall rates after AI clinical implementation, 1 of which was statistically significant: 9.0\% (CI: 7.5\% - 10.5\%) to 6.6\% (CI: 5.3\% - 7.9\%) (p=0.02319, z=-2.270, Cohen's h =-0.091; interpreted 1265 mixed exams prior to AI implementation and 1246 mixed exams after AI implementation, 38 years of experience). 
    * signifies statistically significant decrease in AIR after AI implementation. \CIRCLE \phantom{ } signifies statistically significant increase in AIR after AI implementation. Gray shading signifies the ideal AIR 5-12\% according to national benchmarks.}
    \label{fig:air-by-reader-mixed-cases}
\end{figure}

After AI implementation, the overall recall rates for all cases increased from 11.6\% (CI: 11.3\% - 11.9\%) to 12.6\% (CI: 12.3\% - 12.9\%) (p = 0.00002, z = 4.292, Cohen's h = 0.030) (Fig.~\ref{fig:air-all-cases-by-reader}).
Additionally, we compared recall rates before and after AI implementation by case classification (Fig.~\ref{fig:air-all-cases}).
For ``all green'' cases, recall rate decreased from 7.6\% (CI: 7.2\% - 8.1\%) to 5.7\% (CI: 5.4\% - 6.2\%) (p < 0.00001, z = -6.134, Cohen's h = -0.075), with reduction observed among radiologists with 3-38 years of experience (Fig.~\ref{fig:air-by-reader-all-green}). 
For ``gray'' cases, recall rate increased from 14.3\% (CI: 13.7\% - 14.8\%) to 16.9\% (CI: 16.3\% - 17.6\%) (p < 0.00001, z = 6.223, Cohen's h = 0.074) (Fig.~\ref{fig:air-by-reader-all-gray}). 
For ``mixed'' cases, recall rate increased from 13.0\% (CI: 12.5\% - 13.6\%) to 14.5\% (CI: 13.9\% - 15.1\%) (p = 0.00067, z = 3.4, Cohen's h = 0.043) (Fig.~\ref{fig:air-by-reader-mixed-cases}).  

To assess the false-negative rate of our AI system, we reviewed the 1,090 cases that were given a ``green'' score for both breasts by AI but a final assessment of BI-RADS 0 by the interpreting radiologist during this study period. 
AI missed 4 cancers, yielding a false-negative rate of 0.37\%.

\section{Discussion}\label{sec12}

We present an AI system that is capable of automatically identifying breasts with cancer in screening mammography by integrating multiple imaging modalities (FFDM, C-View, and DBT) and retaining information from all DBT slices.
These predictions are interpretable because the model makes bounding-box predictions, highlighting suspicious regions for review.
Trained and evaluated on an extensive dataset from NYU Langone Health, our initial AI system achieved a high standalone performance, demonstrating an AUROC of 0.945 and a breast-level AUFROC\_1 of 0.848 on the V1 test set.
Moreover, an improved model trained on the V2 dataset achieved an AUROC of 0.953 and a breast-level AUFROC\_1 of 0.945 on the V2 test set.
This strong performance extended to external datasets of diverse demographics and imaging protocols, underscoring the robust generalizability of our approach.
Importantly, these systems remain effective even when only a subset of the modalities is available, supporting its applicability across varied clinical environments.

Our study improves upon previous works that explored deep learning models for breast imaging by leveraging DBT's ability to reduce tissue overlap between nearby structures, while integrating FFDM and C-View.
For example, by ensembling single-modal inferences which yield AUROC values of 0.910, 0.915, and 0.933 for FFDM, C-View, and DBT, respectively, the V1 model achieves an overall AUROC of 0.945 on the V1 test set.
External validation confirmed strong generalizability, reducing the gap to a perfect AUROC by 35.31\%-69.14\% relative to strong baselines such as GMIC and 3D-GMIC (Supplementary Table~\ref{tab:test_gmic_v1}). 
The V2 model further improved upon V1, reducing the gap to a perfect AUROC by 18.86\%-56.62\% and the gap to a perfect AUFROC\_1 by 34.17\%-37.58\% on large external datasets (Supplementary Table~\ref{tab:test_v1_v2}).
These improvements demonstrate the system’s robustness despite variations in demographics, label definitions, inclusion criteria, and imaging protocols, supporting broader real-world adoption.

This study highlights the impact of iterative development and dataset expansion in AI for medical imaging. 
The initial model laid the foundation for further progress by demonstrating promising performance in retrospective and prospective studies. 
Building on this success, we expanded the dataset, incorporated more diverse annotation labels, and refined the model design.
These efforts resulted in an enhanced system with significant gains in both AUROC and AUFROC\_1 across large external datasets.
These findings affirm that iterative development is a powerful strategy for creating AI systems for clinical implementation.

Our study suggests a tangible clinical impact. 
In the prospective clinical study, recall rates significantly decreased for ``all green'' cases among radiologists with 3-38 years of experience, suggesting AI as a clinical support tool benefits radiologists with a wide range of experience and reduces unnecessary diagnostic exams and biopsies in exams with low probability of malignancy.
With AI support, the false-negative rate for ``all green'' cases is below previously determined benchmarks for US mammographic interpretation, suggesting potential standalone use for this subset of examinations.

For ``all gray'' and ``mixed'' cases, recall rates significantly increased with AI assistance.
The patient population at our tertiary medical center includes a high-risk population with personal history of breast cancer, family history of breast cancer, and genetic mutations. 
Additionally, the radiologists were not provided with an AI suspicion score for each breast nor bounding boxes highlighting the suspicious imaging findings. 
The combined effect of the novelty of the system, the high risk population and the lack of specific information regarding the imaging findings likely contributed to the increased recall rate. 

Parallel work from our group explored the impact of AI on recall rates further in a prospective study involving 15,825 patients who underwent screening mammography (manuscript under review). 
Unlike this study, the parallel work evaluated how AI could prompt second reads for exams that were initially interpreted as negative by radiologists.
These cases were subsequently reviewed by a second radiologist, who was provided with both the breast-level predictions and bounding-box predictions.
The AI system flagged 1,647 cases (10.4\%) for a second read, leading to the detection of 17 additional cancers initially missed by the first reader, corresponding to a cancer detection rate of 10.3 per 1,000 within the AI-selected subgroup. These findings suggest that the increased recall observed in the present study for ``all gray'' and ``mixed'' cases may lead to improved cancer detection rather than unnecessary false positives.  

Additionally, a retrospective study suggested that our AI system could potentially reduce false-positive recalls by 31.7\% and decrease radiologists’ workload by 43.8\%, all without missing any malignancies. 
These findings are particularly relevant in high-volume screening environments, where even a modest reduction in recall rate can translate into considerable time savings and decreased patient anxiety.
With this time saving, radiologists could allocate more attention to genuinely suspicious cases, improving overall efficiency.

This study has a few limitations. 
First, while we observed that AI predictions influenced radiologists’ recall decisions, we did not evaluate whether this influence led to a net improvement in clinical outcomes. 
For example, we did not investigate whether additional cancers were detected or whether any cancers were missed due to the introduction of AI.
We also chose to deploy our model at the breast-level score only, reserving bounding-box capability for future iterations.
These outcome metrics are currently being tracked and will be reported in future publications.
Second, our model analyzes mammography exams without comparing to priors.
In clinical practice, radiologists reduce their false-positive rate in part by comparing current mammography exams to prior ones \cite{roelofs2007importance, hakim2015effect, hayward2016improving}.
A promising research direction could be to build an AI system that detects temporal changes.
Third, we focused exclusively on screening mammography, while clinical practice integrates same-day screening ultrasound for many patients.
Future research could aim to combine information from these other complementary modalities, as explored in our recent work~\cite{shen2023leveraging}.

In conclusion, our AI system demonstrates strong performance in identifying breasts with cancers, as well as in locating cancer lesions by effectively integrating multiple modalities and retaining high-resolution 3D information. 
Its robustness across internal and external datasets, coupled with its interpretability and potential to assist radiologists, underscores its promise for real-world clinical deployment. 
Future prospective studies are needed to determine how these promising early findings translate into tangible patient benefits, including improved cancer detection. 
Nonetheless, our findings suggest that AI can play a crucial role in refining breast cancer screening protocols, reducing the burden of unnecessary procedures, and supporting radiologists in their image interpretation to reduce diagnostic exams in specific patient populations. 
This system may lead to more accurate, less invasive, and more patient-centered breast cancer screening.

\section{Methods}\label{sec11}

\begin{figure}
    \centering
    \includegraphics[width=\linewidth]{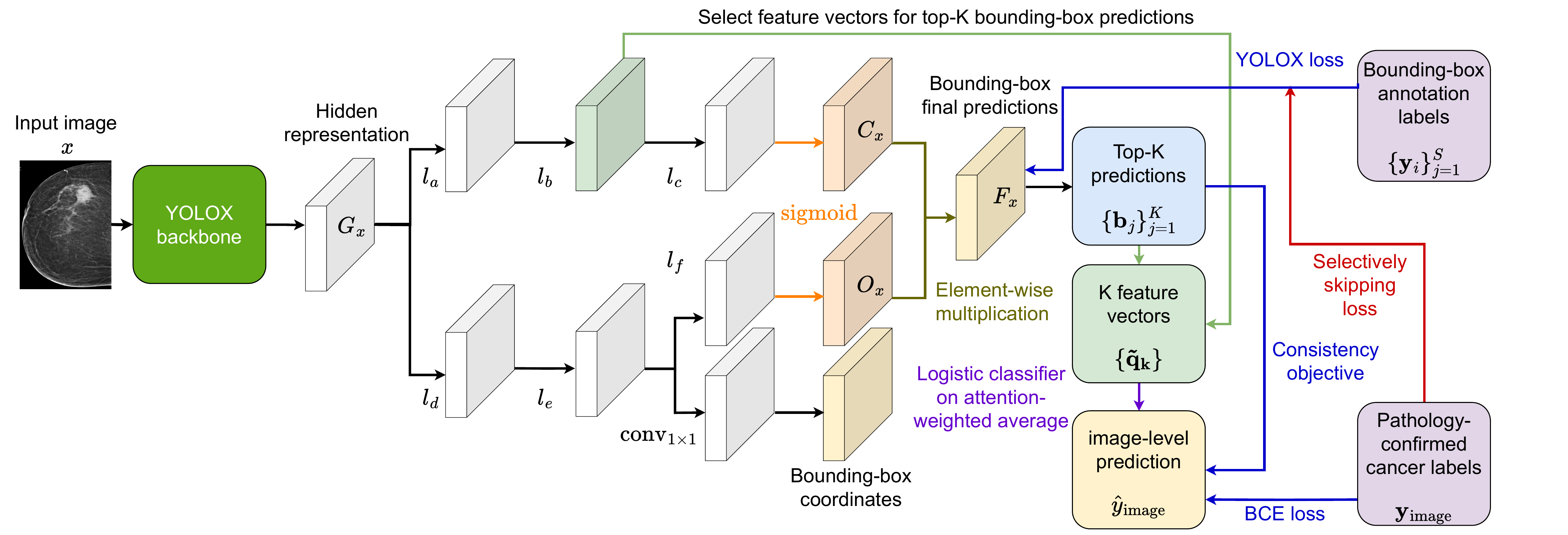}
    \caption{Overview of an individual model contained in the proposed AI system. Each model processes one mammography image $x$ at a time to create bounding-box predictions $F_x$. For the top-K bounding-box predictions $\{\mathbf{b}_j\}_{j=1}^{K}$, the model extracts the corresponding feature vectors $\{\mathbf{\tilde{q}_k}\}$. These features are combined into an attention-weighted average. The resulting representation is used to create image-level predictions $\hat{y}_{\text{image}}$ using logistic regression. The bounding-box prediction head of the model is trained with the radiologist-annotated bounding-box labels $\{\mathbf{y}_i\}_{j=1}^{S}$, and the image-level head of the model is trained with pathology-confirmed cancer labels $\mathbf{y}_{\text{image}}$. In addition, for the images that have positive cancer labels but are missing bounding-box annotation label, we skip the loss calculation for the bounding-box prediction to avoid penalizing the model for detecting mammographically-occult lesions. Furthermore, we apply consistency objective to encourage the image-level predictions to have high correlation with the top-1 bounding-box prediction of the model.}
    \label{fig:ai-system}
\end{figure}

\subsection{NYU Comprehensive Mammography Dataset}~\label{sec:dataset}

The NYU Langone Health Mammography V1 dataset consists of mammography images acquired at NYU Langone Health from 2010 to 2020. 
First, we conducted data preprocessing and cleaning to reject invalid images and exams as described in Supplementary Section~\ref{sec:preprocessing}.
We reserved 19,684 screening exams from January to March 2020 as the test set, all of which contain the FFDM, C-View, and DBT images. 
The exams from the rest of the patients are used in training and validating the model.
We refer to this dataset as training-validation set.
The V1 training-validation set consists of 376,339 screening exams and 120,493 diagnostic exams from 212,487 patients. 
Finally, we reserved 3,241 screening exams for selecting the models that form the best ensemble (ensemble-selection set). 
The details of constructing an ensemble of models is described in Section~\ref{sec:ensemble}.
There is no overlap of patients between the training-validation set, ensemble-selection set, and test set.
The statistics of the training-validation set, the test set, and the ensemble-selection set are summarized in Table~\ref{tab:stats_v1} and Table~\ref{tab:stats_v1_cont}.

Later, we constructed the NYU Langone Health Mammography V2 dataset which consists of mammography images acquired at NYU Langone Health from 2010 to 2022. 
We reserved 45,504 screening exams from January to May 2022 as the test set, all of which contain the FFDM, C-View, and DBT images. 
The V2 training-validation set consists of 562,287 screening exams and 206,206 diagnostic exams from 255,044 patients. 
The V2 ensemble-selection set consists of 20,000 screening exams. 

Race and ethnicity were self-reported by the patients at the time of their medical visits. The terms reflect those used by the healthcare system during data collection, which aligns with standard demographic reporting practices. The race and ethnicity data were included solely to describe the diversity of the study population. These variables were not used in model development, training, or during inference. Additionally, no assumptions or analyses were made using race/ethnicity as proxies for socioeconomic status or other variables.

Male patients are excluded because male breast cancers are rare \cite{fentiman2006male} and few males undergo mammography exams (less than 1\% of patients undergoing a mammography exam at NYU Langone Health are male).

\begin{table}
    \centering
\caption{Patient-level and exam-level statistics of the overall NYU Comprehensive Mammography Dataset V1 and V2. These datasets were collected at NYU Langone Health from 2010 to 2022. Exam-level BI-RADS were issued by radiologists based on patients' mammography exams. Abbreviations: N, number; SD, standard deviation.}
\label{tab:stats_v1}
    \begin{tabular}{>{\raggedleft\arraybackslash}p{5cm} | >{\raggedleft\arraybackslash}p{1.3cm} | >{\raggedleft\arraybackslash}p{1.3cm} | >{\raggedleft\arraybackslash}p{1.3cm} | >{\raggedleft\arraybackslash}p{1.3cm} | >{\raggedleft\arraybackslash}p{1.3cm} | >{\raggedleft\arraybackslash}p{1.3cm}}

         \hline
         \textbf{Characteristics, unit} & \multicolumn{2}{|>{\centering\arraybackslash}p{2.7cm}|}{\textbf{\makecell{Training-\\validation set}}} & \multicolumn{2}{|>{\centering\arraybackslash}p{2.7cm}|}{\textbf{\makecell{Ensemble-\\selection set}}} & \multicolumn{2}{|>{\centering\arraybackslash}p{2.7cm}}{\textbf{Test set}}\\
         \cline{2-7}
         & V1 & V2 & V1 & V2 & V1 & V2 \\
         \hline \hline
         Patients, N& 212,487&  255,044
&  3,117&  19,073
&  19,684& 
45,504
\\ \hline
         Race/Ethnicity& &  &  &  &  & 
\\ 
         White (N, \%)&115,392 (54.31)& 135,678 (53.20)& 2,120 (68.01)& 11,701 (61.35)&12,127 (61.61)& 26,665 (58.60)
\\ 
         African American (N, \%)&22,001 (10.35)& 27,216 (10.67)& 290 (9.30)& 1,830 (9.59)& 1,940 (9.86)& 5,132 (11.28)
\\ 
         Hispanic (N, \%)& 3,956 (1.86)& 4,761 (1.87)&32 (1.03)& 282 (1.48)& 276 (1.40)& 584 (1.28)
\\ 
         Asian (N, \%)&10,645 (5.01)&13,465 (5.28)& 103 (3.30)& 860 (4.51)& 894 (4.54)& 2,253 (4.95)
\\ 
         Other (N, \%)& 7,956 (3.74)& 9,595 (3.76)& 114 (3.66)& 736 (3.86)& 718 (3.65)& 1,794 (3.94)
\\ 
         Unknown (N, \%)& 52,537 (24.72)& 64,329 (25.22)& 458 (14.69)& 3,664 (19.21)& 3,729 (18.94)& 9,076 (19.95)\\ \hline
         Age, mean years (SD)&  57.3 (13.0)&  57.5 (13.3)&  58.6 (11.4)&  59.4 (11.7)&  58.9 (11.4)& 
59.1 (11.6)
\\ 
         $<$ 40 yrs old, N (\%)&  14,615 (6.88)&  18470 (7.24)&  61 (1.96)&  377 (1.98)&  332 (1.69)& 
901 (1.98)
\\ 
         40 - 49 yrs old, N (\%)&  54,471 (25.63)&  64,418 (25.26)&  762 (24.45)&  4,378 (22.95)&  4,625 (23.50)& 
10,583 (23.26)
\\ 
         50 - 59 yrs old, N (\%)&  55,996 (26.35)&  65,149 (25.54)&  896 (28.75)&  5,234 (27.44)&  5,735 (29.14)& 
12,939 (28.43)
\\ 
         60 - 69 yrs old, N (\%)&  48,173 (22.67)&  57,338 (22.48)&  845 (27.11)&  5,183 (27.17)&  5,313 (26.99)& 
12,225 (26.87)
\\ 
         $\geq$ 70 yrs old, N (\%)&  39,232 (18.46)&  49,669 (19.47)&  553 (17.74)&  3,901 (20.45)&  3,679 (18.69)& 
8,856 (19.46)\\ \hline
         Exams, N&  496,832&  768,493
&  3,241&  20,000
&  19,684& 
45,504
\\ 
         Screening exams, N&  376,339&  562,287
&  3,241&  20,000
&  19,684& 
45,504
\\ 
         Diagnostic exams, N&  120,493&  206,206
&  0&  0
&  0& 0
\\ \hline
         Exam-level BI-RADS&  &  
&  &  
&  & 

\\ 
         BI-RADS 0, N (\%)&  62,714 (12.62)&  96,632 (12.57)
&  1,374 (42.39)&  2,776 (13.88)&  2,470 (12.55)& 
5,990 (13.16)
\\ 
         BI-RADS 1, N (\%)&  189,838 (38.21)&  292,576 (38.07)
&  845 (26.07)&  9,017 (45.08)&  8,983 (45.64)& 
24,843 (54.60)
\\ 
         BI-RADS 2, N (\%)&  196,744 (39.60)&  294,259 (38.29)
&  848 (26.16)&  7,729 (38.65)&  8,231 (41.82)& 
14,671 (32.24)\\ 
         BI-RADS 3, N (\%)&  19,910 (4.01)&  37,532 (4.88)
&  18 (0.56)&  84 (0.42)
&  0 (0.00)& 
0 (0.00)
\\ 
         BI-RADS 4, N (\%)&  10,632 (2.14)&  20,075 (2.61)
&  58 (1.79)&  35 (0.18)
&  0 (0.00)& 
0 (0.00)
\\ 
         BI-RADS 5, N (\%)&  920 (0.19)&  1,526 (0.20)
&  0 (0.00)&  2 (0.01)
&  0 (0.00)& 
0 (0.00)
\\ 
         BI-RADS 6, N (\%)&  602 (0.12)&  1,052 (0.14)
&  0 (0.00)&  1 (0.01)
&  0 (0.00)& 
0 (0.00)
\\ 
         Unknown BI-RADS, N (\%)&  15,472 (3.11)&  24,841 (3.23)&  98 (3.02)&  356 (1.78)&  0 (0.00)& 0 (0.00)\\ \hline
    \end{tabular}
\end{table}

\begin{table}
    \centering
\caption{Breast-level and image-level statistics of the overall NYU Comprehensive Mammography Dataset V1 and V2. Malignant and benign findings at the breast level are based on pathology reports, while ground-truth bounding-box labels were manually annotated by 52 radiologists from NYU Langone Health. Abbreviations: N, number; SD, standard deviation; GT, ground-truth.}
\label{tab:stats_v1_cont}
    \begin{tabular}{>{\raggedleft\arraybackslash}p{5.5cm} | >{\raggedleft\arraybackslash}p{1.3cm} | >{\raggedleft\arraybackslash}p{1.3cm} | >{\raggedleft\arraybackslash}p{1.3cm} | >{\raggedleft\arraybackslash}p{1.3cm} | >{\raggedleft\arraybackslash}p{1.3cm} | >{\raggedleft\arraybackslash}p{1.3cm}}

         \hline
         \textbf{Characteristics, unit} & \multicolumn{2}{|>{\centering\arraybackslash}p{2.7cm}|}{\textbf{\makecell{Training-\\validation set}}} & \multicolumn{2}{|>{\centering\arraybackslash}p{2.7cm}|}{\textbf{\makecell{Ensemble-\\selection set}}} & \multicolumn{2}{|>{\centering\arraybackslash}p{2.7cm}}{\textbf{Test set}}\\
         \cline{2-7}
         & V1 & V2 & V1 & V2 & V1 & V2 \\
         \hline \hline
         Breasts, N&  938,289&  1,436,762
&  6,482&  40,000
&  39,368& 
91,008
\\ 
         Breasts with malignant findings, N&  8,193&  13,136
&  188&  147
&  128& 
256
\\  
         Breasts with benign findings, N&  27,672&  
45,333&  1,569&  
610&  524& 

1,059\\  \hline
         FFDM Images, N&  2,162,627&  3,289,332
&  13,952&  86,531
&  85,578& 
199,023
\\ 
         FFDM w/ malignant GT boxes, N&  3,795&  26,043
&  0&  236
&  196& 
530
\\ 
         FFDM w/ benign GT boxes, N&  7,555&  
7,475&  0&  
92&  6& 

0\\ 
         C-View Images, N&  799,903&  1,827,074
&  13,152&  81,290
&  79,813& 
184,736
\\ 
         C-View w/ malignant GT boxes, N&  1,728&  12,926
&  0&  224
&  186& 
484
\\ 
         C-View w/ benign GT boxes, N&  2,022&  
2,533&  0&  
92&  6& 

0\\ 
         DBT Images, N&  799,903&  1,827,074
&  13,152&  81,290
&  79,813& 
184,736
\\ 
         DBT w/ malignant GT boxes, N&  1,728&  12,838
&  0&  222
&  177& 
484
\\  
         DBT w/ benign GT boxes, N&  2,021&  
2,532&  0&  
92&  6& 

0\\  \hline
    \end{tabular}
\end{table}

\begin{figure}[ht]
    \centering
    \includegraphics[width=0.7\linewidth]{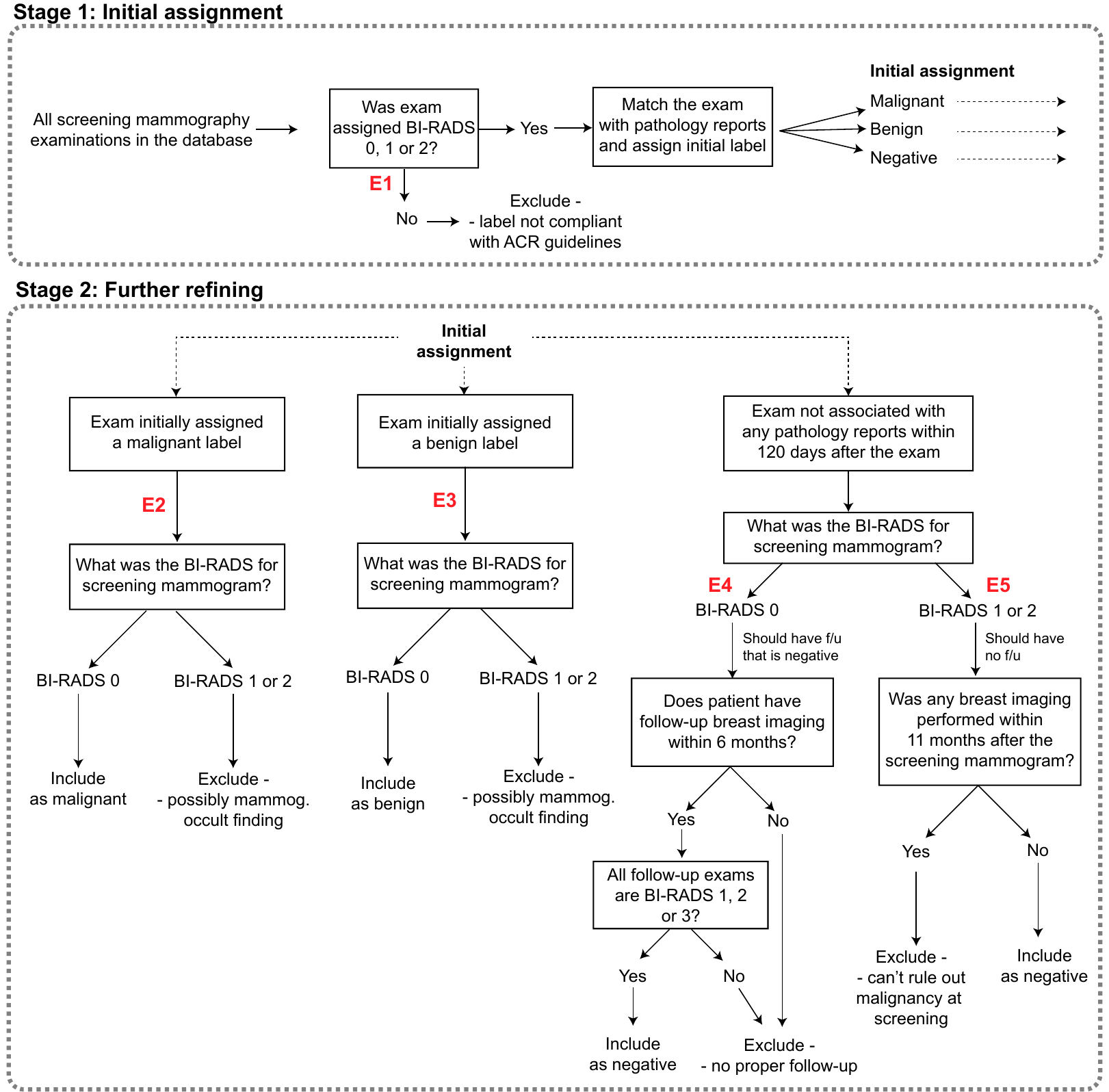}
    \caption{\textbf{Filtering of the test subset of screening mammography dataset.}
    \\Stage 1: we collect radiology reports and pathology reports for all screening mammography exams in the dataset. We use radiology reports to collect a BI-RADS category assigned to each exam, and we use pathology reports to find information about the final diagnosis associated with the exam. \textbf{E1:} According to the American College of Radiology guidelines, a screening exam should be assigned BI-RADS 0, 1 or 2. At this stage, we exclude all exams with other labels as they might not properly represent screening mammography. For all remaining screening exams, we match them with breast pathology reports, and use the diagnosis information to make an \textit{initial assignment} and classify them as malignant, benign, or negative. An exam is considered ``negative'' if it is not associated with any pathology report, and ``benign'' when there is a matching pathology report from a biopsy, however the tissue contained only benign cells. 
    \\Stage 2: we use the initial assignment and perform exam filtering based on that. \textbf{E2:} If an exam was labeled as malignant, we expect screening mammography to be given BI-RADS 0, and not BI-RADS 1 or 2. All BI-RADS 1 and 2 screening exams have been excluded, as they could have been possibly mammographically occult. \textbf{E3:} If an exam was labeled as benign, we would also expect it to be BI-RADS 0. Same as before, we exclude all BI-RADS 1 and 2 exams. If an exam was assigned a negative label, we need to confirm that it is truly negative. \textbf{E4:} If a mammography exam with negative label was given BI-RADS 0, we suspect that there were suspicious findings which were later deemed to be benign or not found at all. To confirm this, we require all such cases to have at least one follow-up exam within 6 months. Furthermore, all exams in the next 6 months must be BI-RADS 1, 2 or 3. \textbf{E5:} If a patient had a mammography exam with negative label and BI-RADS 1 or 2, she should not have any breast imaging done before the next screening exam, which we usually expect in 11-13 months. If that patient had any breast imaging performed within 11 months after the negative mammography exam, we exclude this exam. \\
    Furthermore, after all stages are completed, we reject any exams that are determined to contain only mammographically-occult cancer and no other visible cancer.}
    \label{fig:dataset_filtering}
\end{figure}

\subsection{Breast-level ground-truth cancer labels}~\label{sec:image_label}

We extracted the two binary labels from the pathology reports for each breast.
The malignant label is positive if biopsy-confirmed cancer is present.
The benign label is positive if benign findings are reported.
Since a breast can have multiple pathology findings of different types, both labels can be positive simultaneously.
Breasts that are not associated with malignant or benign findings in any pathology reports created within 120 days of the exam date are assigned negative labels.
The statistics of the breast-level cancer label are shown in Table~\ref{tab:stats_v1} and Table~\ref{tab:stats_v1_cont}.

Under the above labeling scheme, it is possible that the breast-level cancer labels might be inaccurately assigned for some exams. 
For example, for screening exams that are assigned negative labels, it is possible that the patients might have undergone additional diagnostic exams afterwards and yielded cancer sometime later than 120 days after the screening exam.
In this case, it is likely that such cancer findings are visible in the original screening exams.
Thus, a data cleaning procedure is applied to the test set to filter out exams whose labels could potentially be incorrect (Figure~\ref{fig:dataset_filtering}).

While this data cleaning procedure is applied to the test set, we do not perform similar filtering on the training data. This approach is supported by several studies which demonstrated that neural networks are robust to some degree of label noise and can learn meaningful patterns despite label inaccuracies~\cite{arpit2017closer,rolnick2017deep}. 
In addition, regularization techniques (e.g., weight decay) reduce overfitting to incorrect labels without reducing the model’s ability to learn and generalize to the test set~\cite{arpit2017closer}.

\subsection{Bounding-box ground-truth annotation labels}~\label{sec:lesion_label} 

With the help of 52 radiologists from our institution, we collected bounding-box annotations for visible biopsied lesions.
We primarily focused on collecting annotations for malignant lesions, while still collecting some benign lesions.
When annotating DBT exams, the radiologists were instructed to leave annotation on all slices where lesions are clearly visible.
For model training, we utilize all bounding-box labels from all annotated slices.
For model evaluation, we use the bounding-box labels from the central slice of each lesion as ground-truth.
The statistics of the bounding-box annotation label are shown in Table~\ref{tab:stats_v1} and Table~\ref{tab:stats_v1_cont}.

\subsection{External datasets}~\label{sec:external}

To evaluate the generalization of our AI system, we evaluated our models on external datasets whose images have not been used to train our AI system: the Chinese Mammography Database (CMMD)~\cite{Cui2021}, OPTIMAM~\cite{halling2020optimam}, and CSAW-CC~\cite{dembrower2020multi}, a subset of Emory's EMBED dataset~\cite{jeong2023emory}, the Curated Breast Imaging Subset of Digital Database for Screening Mammography (CBIS-DDSM)~\cite{lee2017curated, shen2019deep, Sawyer-Lee2016}, INbreast~\cite{moreira2012inbreast}, and BCS-DBT~\cite{buda2021data, buda2020data}.

All cancer cases in described data sets have been pathology-confirmed, i.e., the diagnosis has been made based on a pathology report, either from a biopsy or surgery.

\begin{enumerate}
\item The OPTIMAM mammography image database (OMI-DB)~\cite{halling2020optimam} provides a collection of screening and diagnostic mammograms, along with clinical data, gathered from three UK screening sites since 2011. Access to this dataset is restricted, and only partnering academic groups can gain partial access. For our study, we utilized a subset of the OPTIMAM database, which included data for 5,999 patients and 11,633 screening mammography studies. The OPTIMAM dataset contains both processing and presentation images, but we exclusively used for-presentation images in our research.

\item The Chinese Mammography Database (CMMD)~\cite{Cui2021}, published by The Cancer Imaging Archive (TCIA) in 2021, consists of 1,775 mammography studies from 1,775 patients collected between 2012 and 2016 across several Chinese institutions, including Sun Yat-sen University Cancer Center and Nanhai Hospital of Southern Medical University in Foshan. The dataset includes biopsy-confirmed breast-level labels for benign and malignant findings. It also provides patient age, finding type (e.g., calcification, mass, or both) for all cases, and immunohistochemical markers for 749 patients with invasive carcinoma.
We exclude one exam (\#D1-0951) following Stadnick et al.~\cite{stadnick2021meta} as the pre-processing algorithm failed on it.

\item The CSAW-CC dataset~\cite{dembrower2020multi} is a subset of the larger CSAW (Cohort of Screen-Aged Women) dataset, introduced in 2019. 
The published CSAW-CC dataset originally comprises repeated screening mammograms over time from 873 patients with cancer and 7,850 health control patients screened at Karolinska University Hospital from 2008 to 2016.
All mammograms in CSAW-CC are acquired on Hologic devices.
For our analyses, we included only screening-detected cancers and negative cases following Stadnick et al.~\cite{stadnick2021meta}.

\item The EMBED dataset from Emory~\cite{jeong2023emory} is a mammography collection with 3.4 million screening and diagnostic images gathered from 110,000 patients between 2013 and 2020. This dataset originally included FFDM, C-View, and DBT images, although we only used FFDM images.
We use a subset of the EMBED dataset comprising FFDM images from 9,998 exams to be consistent with another manuscript under preparation using the same subset of the data.
Concretely, we first excluded exams with breast implants and missing views and selected 35,404 patients who have at least 3 screening exams with FFDM images. We then chose the middle exams for each patient, but when the number of exams for a given patient is an even number, we broke tie by preferring the exams which have visible cancer or have additional imaging modalities available within the last 2 years. Finally, we selected a subset of 9,998 exams by (a) including all exams which have prior ultrasound exams within the last 2 years, (b) including all exams with currently-visible cancer, (c) randomly sampling from the rest of the exams, and (d) rejecting 2 exams with BI-RADS of 6.

\item The original Digital Database for Screening Mammography (DDSM)~\cite{heath2001digital}, released in 1999, includes 2,620 exams and 10,480 images of digitized film mammograms from several US institutions, such as Massachusetts General Hospital, Wake Forest University School of Medicine, Sacred Heart Hospital, and Washington University of St. Louis School of Medicine. A 2017 update, the Curated Breast Imaging Subset of DDSM (CBIS-DDSM)~\cite{lee2017curated}, features improved annotations. 
For our study, we used a modified test set of 188 exams, following Shen et al.~\cite{shen2019deep} to enable comparisons with their work.
The patient identifiers are provided in Stadnick et al.~\cite{stadnick2021meta}.

\item The INbreast dataset~\cite{moreira2012inbreast}, collected in 2010 and released in 2012, comprises 115 exams from Centro Hospitalar de S. João in Portugal. This dataset consists entirely of digital mammograms. We used a test set of 31 exams from INbreast. 
Of the 31 exams, 4 contain only malignant lesions, 16 only benign lesions, and 11 have both malignant and benign lesions. 
A list of image names for the test set is available in Stadnick et al.~\cite{stadnick2021meta}.

\item BCS-DBT~\cite{buda2021data, buda2020data}, also known as Breast-Cancer-Screening-DBT, is the dataset from Duke University Hospital consisting of patients who underwent a DBT exams at Duke Health system from January 2014 to January 2018. 
All images in this dataset are from the `Selenia Dimensions' model by `HOLOGIC, Inc.' 
The BCS-DBT training dataset contains 19,148 DBT images from 4,838 studies involving 4,362 patients, with 87 bounding-box annotations for malignant lesions and 137 for benign lesions. 
The validation dataset comprises 1,163 DBT images from 312 studies of 280 patients, with 37 bounding-box annotations for malignant lesions and 38 for benign lesions.
The test dataset comprises 1,721 DBT images from 460 studies of 418 patients, with 66 bounding-box annotations for malignant lesions and 70 for benign lesions.
The labels of the test dataset were not released until January 2024 as it was used in SPIE-AAPM-NCI DAIR Digital Breast Tomosynthesis Lesion Detection (DBTex) Challenge~\cite{konz2023competition,park2021lessons}.
Because of this, some relevant works including Park et al.~\cite{park2023efficient} used the training set for reporting the performance.
We report model performance on the combined training+validation+test dataset of BCS-DBT.
The reported performances on BCS-DBT dataset are evaluated at image-level for classification and lesion-level for bounding-box predictions, rather than breast-level.
This is because this dataset does not guarantee the presence of both CC and MLO views for some breasts.
\end{enumerate}

\subsection{AI system architecture}
\label{sec:yolox}

Our AI system (Fig.~\ref{fig:ai-system}) consists of multiple deep convolutional neural networks that take two-dimensional images as input. 
Specifically, we adapted the YOLOX~\cite{ge2021yolox} architecture to identify lesion locations in mammography images in the form of bounding-box predictions. These bounding box predictions are made for each 2D image and each DBT image slice.
In addition, we use hidden representations from the highest-rated bounding-box predictions from each image to generate an overall image-level probability of breast cancer presence.
Breast-level predictions for each imaging modality are formed by averaging image-level probabilities across all images for each breast. 
Ultimately, modality-specific predictions are combined to create a multi-modal ensemble, leveraging the distinct advantages of each imaging technique.

\paragraph{Inputs and outputs}

Let $\mathbf{x} \in \mathbb{R}^{3 \times H \times W}$ be an input image such as FFDM, C-View, or a two-dimensional slice of a DBT image.\footnote{Even though the mammography images originally have one channel, we repeat it three times to build three-channel images. This is a necessary step in utilizing deep neural networks that are originally designed for natural images, since natural images have three channels representing red, green, and blue colors.}
$H$ and $W$ denote the height and width of the input image.
The AI system generates two image-level predictions $\hat{\mathbf{y}}_{\text{image}}=(\hat{y}_x^m, \hat{y}_x^b) \in [0, 1]$, indicating the probability of the presence of at least 
 one malignant and at least one benign finding in the input image, respectively.
In addition to the overall image-level prediction, the model also outputs a set of bounding-box predictions $\{\mathbf{b}_i\}_{i=1}^{M}$, where each bounding box $\mathbf{b}_i = (u_{i}, v_{i}, \xi_{i}, \kappa_{i}, s_i^m, s_i^b)$ represents a rectangular region in the image that is predicted to contain a region of interest. Here, $M$ is the number of bounding-box predictions, and $(u_{i}, v_{i})$ represents the center coordinate of the $i$-th bounding box, $(\xi_{i}, \kappa_{i})$ represents width and height of the $i$-th bounding box, and $(s_i^m, s_i^b)$ represents the probability of the $i$-th bounding box representing malignant and benign lesions, respectively.

\paragraph{Ensembling multiple imaging modalities} \label{sec:ensemble}

The three modalities capture different properties of the tissue. 
DBT is a 3D modality consisting of multiple slices which show structures appearing at corresponding depths.
This reduces tissue overlap between nearby structures and thus increases lesion conspicuity, which is particularly beneficial in dense breast tissue.
However, DBT and C-View image quality could be compromised by artifacts created around calcifications~\cite{horvat2019calcifications}.
In addition, 2D images such as FFDM and C-View are more useful than DBT for quickly scanning for clusters of calcifications.
Since no single input modality is strictly superior, our modeling utilizes all available input modalities to leverage their complementary information.
Concretely, we first compute the predictions for the three input modalities separately from models trained on each modality and average the predictions for each breast.
The technical details on how to select the models for the ensemble and how to combine the model predictions are described in Supplementary Section~\ref{sec:appendix-ensemble}

\paragraph{Creating bounding-box predictions}

The YOLOX architecture decouples the computation of the probability prediction of each bounding-box prediction into two components: the predicted probability of each target (benign and malignant lesions) and the objectness prediction.
The objectness prediction represents the probability that a predicted bounding box corresponds to a lesion, regardless of whether it is benign or malignant.
YOLOX multiplies the target-specific probability of each bounding-box prediction with the corresponding objectness prediction to compute the final probability for each bounding-box. 

Concretely, for an input image $\mathbf{x}$, the backbone network outputs the hidden representation $\mathbf{E_x} \in \mathbb{R}^{h \times w \times \phi}$ where $h, w, \phi$ are dimensions of the representation.
YOLOX further processes the hidden representation $\mathbf{E_x}$ to output the probability of malignancy and the objectness prediction.
First, a 1$\times$1 convolutional layer with SiLU nonlinearity is applied to the hidden representation $\mathbf{E_x}$ to output another hidden representations $\mathbf{G_x} \in \mathbb{R}^{h \times w \times \psi}$ as 
\begin{equation*}
    \mathbf{G_x} =  \text{conv}_{1\times1}(\mathbf{E_x}).
\end{equation*}
where $\psi$=256 for YOLOX-L architecture and $\psi$=320 for YOLOX-X architecture. 
The purpose of creating this second hidden representation $\mathbf{G_x}$ is to adjust the size of the channel dimension for the subsequent computation.
Secondly, a series of convolutional layers is applied to the hidden representation $\mathbf{G_x}$ to make predictions. 
The prediction of target-specific probability for malignant and benign targets $\mathbf{C_x} \in \mathbb{R}^{h \times w \times 2}$ is computed as
\begin{equation*}
    \mathbf{C_x} =  \text{sigmoid}(l_c(l_b(l_a(\mathbf{G_x})))),
\end{equation*}
where $l_a$ and $l_b$ are parameterized as $3\times3$ convolutional layers with SiLU nonlinearity and $l_c$ is parameterized as a $1\times1$ convolutional layer. 
The objectness predictions $\mathbf{O_x} \in \mathbb{R}^{h \times w \times 1}$ are computed as
\begin{equation*}
    \mathbf{O_x} =  \text{sigmoid}(l_f(l_e(l_d(\mathbf{G_x})))),
\end{equation*}
where $l_d$ and $l_e$ are parameterized as $3\times3$ convolutional layers with SiLU nonlinearity and $l_f$ is parameteraized as a $1\times1$ convolutional layer.
Lastly, the final probabilities for the bounding boxes $\mathbf{F_x} \in \mathbb{R}^{h \times w \times 2}$ are computed by element-wise multiplying the target-specific probability $\mathbf{C_x}$ and the objectness prediction $\mathbf{O_x}$ by broadcasting the singleton dimension of $\mathbf{O_x}$ to match the corresopnding dimension of $\mathbf{C_x}$ as
\begin{equation*}
    \mathbf{F_x} =  \mathbf{C_x} \odot \mathbf{O_x}.
\end{equation*}

The aforementioned convolutional layers are divided into two distinct groups, with independent calculations.
The first group, comprising the $l_a$, $l_b$, and $l_c$ layers, is dedicated to distinguishing between different targets: malignant and benign.
The second group, consisting of the $l_d$, $l_e$, and $l_f$ layers, is responsible for determining the objectness score.
Additionally, the outputs of the layers $l_d$ and $l_e$ are utilized to predict the bounding-box coordinates, as illustrated in Figure~\ref{fig:ai-system}.
In other words, the second group of layers specializes in precise localization tasks, such as refining bounding boxes, without being influenced by classification-related features. 

\paragraph{Creating image-level predictions}
\label{sec:partial}

When utilizing breast-level labels, we have to create image-level predictions $\hat{\mathbf{y}}_{\text{image}} \in \mathbb{R}$ and then calculate binary cross-entropy loss with the breast-level label.
If the image-level prediction $\hat{\mathbf{y}}_{\text{image}}$ is calculated by aggregating the bounding-box predictions $\mathbf{F_x}$ directly, then backpropagation from the image-level classification will update not only the convolutional layers $l_a$, $l_b$, $l_c$ but also $l_d$, $l_e$, $l_f$.
This is suboptimal because breast-level label lacks the necessary information to train the convolutional layers $l_d$, $l_e$, $l_f$, the group of layers that focus on precise localization task.
Since the convolutional layers $l_d$, $l_e$, $l_f$ must be trained with high accuracy utilizing the bounding-box labels to output the objectness predictions, training signal from $\hat{\mathbf{y}}_{\text{image}}$ created solely from the breast-level label would likely corrupt the behaviors of these layers.
This could lead to creating many false-positive bounding-box predictions, decreasing the overall performance.

Instead, we compute the image-level probability prediction $\hat{\mathbf{y}}_{\text{image}}$ in a way which does not involve the convolutional layers $l_d$, $l_e$, $l_f$.
First, we identify the top-$\mathbf{K}$ bounding-box predictions $\{\mathbf{b}_j\}_{j=1}^{K}$ for the input image $\mathbf{x}$ after applying non-maximum suppression to the model predictions $\mathbf{F_x}$.\footnote{
For models trained on the V2 dataset, we additionally include the global feature vectors acquired by applying max pooling and average pooling in each feature map.
This ensures that the image-level prediction can be made using the necessary information from the input image, even if the top-K bounding-box predictions did not capture the cancer lesion.}
Second, from the feature map $l_b(l_a(\mathbf{G_x}))$, an intermediate tensor in the process of computing $\mathbf{C_x}$, we select a set of $\mathbf{K}$ vectors $\{\mathbf{\tilde{q}_k}\}$ which correspond to the top-$\mathbf{K}$ bounding-box predictions.
Third, we calculate the attention weights $\alpha_k$ for the $\mathbf{K}$ feature vectors $\{\mathbf{\tilde{q}_k}\}$
using a gated attention mechanism~\cite{ilse2018attention} as
\begin{equation*}
    \alpha_k = \frac{\text{exp}\{\mathbf{w}^\intercal (\text{tanh}(\mathbf{V}\mathbf{\tilde{q}}_k^{\intercal}) \odot \text{sigmoid}(\mathbf{U}\mathbf{\tilde{q}}_k^{\intercal}) )\}}{\sum^K_{j=1}\text{exp}\{\mathbf{w}^\intercal (\text{tanh}(\mathbf{V}\mathbf{\tilde{q}}_j^{\intercal}) \odot \text{sigmoid}(\mathbf{U}\mathbf{\tilde{q}}_j^{\intercal}) )\}},
\end{equation*}
where $\mathbf{w} \in \mathbb{R}^{L \times 1}$, $\mathbf{V} \in \mathbb{R}^{L \times S}$ and $\mathbf{U} \in \mathbb{R}^{L \times S}$ are learnable parameters. 
For models with YOLOX-L architecture, we set $L = 64$ and $S = 256$.
For models with YOLOX-X architecture, we set $L = 80$ and $S = 320$.
Fourth, we create a representation $\mathbf{z}_{\text{attn}}$ which is an attention-weighted average of the feature vectors $\{\mathbf{\tilde{q}_k}\}$ as
\begin{equation*}
    \mathbf{z}_{\text{attn}} = \sum_{k=1}^{K} \alpha_k \tilde{\mathbf{q}}_k.
\end{equation*}
Fifth, we apply a fully connected layer with sigmoid nonlinearity to $\mathbf{z}_{\text{attn}}$ to generate the image-level prediction 
    $\hat{\mathbf{y}}_{\text{image}} = \text{sigmoid}(\mathbf{w_{\text{image}}}^T \mathbf{z}_{\text{attn}})$,
where $ \mathbf{w}_{\text{image}} \in \mathbb{R}^{S \times 2}$ are learnable parameters.

As a result, the image-level prediction $\hat{\mathbf{y}}_{\text{image}}$ is created by utilizing the accurate bounding-box predictions $\mathbf{F_x}$ indirectly.
The training signal from the breast-level label does not update the convolutional layers $l_d$, $l_e$, $l_f$, which prevents deteriorating the performance of the bounding-box predictions $\mathbf{F_x}$.

For DBT images during model training, we only load one slice at a time and choose top-$\mathbf{K}$ boxes for the loaded slice. 
For model inference with DBT images, we load all slices and choose top-$\mathbf{K}$ boxes from the entire DBT image. 
This way, we do not require loading the entire DBT image during training, yet our model generalizes well to inference on entire DBT image by aggregating information from all slices.
More information about slice-level training as well as inference with three-dimensional image can be found in Supplementary Section~\ref{sec:dbt_training} and Supplementary Section~\ref{sec:mss}.

\paragraph{Training objective}

We use two types of ground-truth labels, breast-level and bounding-box-level, which have distinct advantages.
Breast-level labels based on pathology reports (section~\ref{sec:image_label}) are easier to collect and available for all patients which have undergone biopsies, but do not provide precise location, size or shape of lesions.
Bounding-box labels (section~\ref{sec:lesion_label}) can teach models exactly where each lesion is located and what it looks like and thus provide richer and more precise training signal to the model. However, the bounding-box annotations are laborious to collect. 
In addition, radiologists cannot locate mammographically-occult cancer lesions on the images even if they retrospectively knew that they were present based on reports from other imaging procedures. 

To leverage the strengths of both types of labels while mitigating their individual limitations, we utilize them both in training our models. 
Concretely, our models are simultaneously trained with two objective functions for two different tasks: object detection and image classification.
Object detection is the task of predicting bounding-boxes for regions of interests with ground-truth bounding-box labels.
We refer to the object-detection loss function used in the YOLOX architecture, defined in the original paper~\cite{ge2021yolox}, as YOLOXLoss.
Image classification is the task of predicting image-level probabilities for benign and malignant targets with pathology-confirmed ground-truth breast-level labels, whose loss function is binary cross-entropy for each target. 

However, it can be problematic to utilize these images with a positive breast-level label and missing bounding-box label in training the AI models for the object detection task since we might end up incorrectly teaching our AI models that there are no lesions in these images.
We chose a simple way of handling these images: we used these images when calculating the image classification objective (calculated with image-level predictions and labels), but ignored them when calculating object detection objective (calculated with bounding-box predictions and labels).
This prevents incorrectly teaching the model that there are no lesions in breasts with pathology-confirmed findings.

In addition, for models trained on the V2 dataset, we introduce a consistency objective, an additional term to the training objective to make the image-level prediction similar to the top-1 bounding-box prediction for each image.
Concretely, we calculate the L1 distance between the two predictions.
This calibrates the bounding-box predictions and the image-level predictions, which could help radiologists interpret and utilize the model predictions.

The final training objective $\mathcal{L}$ for the proposed AI system for one training example is as follows: 
\begin{align*}
    \mathcal{L} = &\mathbf{1}(S \neq 0  \text{ or }  \text{sum}(\mathbf{y}_{\text{image}}) < 1) \cdot \text{YOLOXLoss}(\{\mathbf{b}_j\}_{j=1}^{M}, \{\mathbf{y}_i\}_{j=1}^{S})\\
    + &\text{BCE}(\hat{\mathbf{y}}_{\text{image}}, \mathbf{y}_{\text{image}})\\
    + &10 \cdot |\text{detach}(\mathbf{b}_1) - \hat{\mathbf{y}}_{\text{image}}|,
\end{align*}
where $\mathbf{1}$ is an indicator function, $\mathbf{y}_{\text{image}}$ is the corresponding pathology-confirmed breast-level label (for both benign and malignant categories) for the given image, $M$ is the number of bounding-box predictions created for an image, $S$ is the number of ground-truth bounding-box labels available for an image, $\{\mathbf{y}_i\}_{j=1}^{S}$ represents the ground-truth bounding-box labels, $\mathbf{b}_1$ represents top-1 bounding-box prediction for the image, and $\text{detach}$ represents a procedure to detach a variable from the computation graph, preventing gradient flow through this term during backpropagation. 
This ensures that the consistency objective affects only the optimization of $\hat{\mathbf{y}}_{\text{image}}$ without influencing the bounding-box predictions.

\subsection{Statistical analysis}~\label{sec:stats}

To measure the model performance in breast-level classification, we computed the area under the ROC and PR curves (AUROC and AUPRC) using a nonparametric (trapezoidal) method.
To determine p-values for comparing AUCs between models, we conducted a one-tailed permutation test~\cite{chihara2018mathematical}. 
The detailed description on the permutation test can be found in the Supplementary Section~\ref{sec:permutation_test}.

The difference in AUROC between models was used as an effect size to quantify the magnitude of performance differences. Confidence intervals for AUROC differences were calculated using 1,000 bootstrap resamples, providing a measure of uncertainty for these effect sizes.
The bootstrap analysis assumes that the sample distribution is representative of the population.

To measure model performance of the bounding-box predictions, we employ the Free-Response Operating Characteristic (FROC) analysis. 
FROC evaluates the trade-off between the sensitivity of detecting the ground-truth labels (y-axis) and the number of false-positive predictions per image (x-axis), making it a suitable metric for bounding-box prediction task.
Specifically, we consider a prediction true-positive if the distance between the center point of the prediction and the center point of the ground-truth label is less than (a) half the diagonal of the ground-truth label or (b) 100 pixels, as per Buda et al.~\cite{buda2021data}. 
In addition, for DBT images, there is an additional criterion for the depth dimension: for a prediction to be considered true-positive, the distance between the center slice of the ground-truth lesion and the center slice of the prediction must be within 25\% of the number of total slices in the corresponding image.
We report the area under the free-response ROC curve (AUFROC) for the interval on the x-axis between 0 and 1 false-positive predictions per image (AUFROC\_1), as well as the sensitivity values at the thresholds that lead to 0.5, 1, and 2 false-positive predictions per image.
These sensitivity values are assessed at one of the two levels: lesion-level (the percentage of ground-truth malignant lesions that are detected) and breast-level (the percentage of breasts with at least one ground-truth malignant lesion that are detected with at least one true-positive bounding-box prediction across any lesion on any view).

We compare the AUCs between the models and display the differences in AUCs as well as ``reduction of errors'' as defined below.
The concept of ``reduction of errors'' provides a clearer perspective on performance improvements. 
Instead of focusing solely on the absolute increase in performance metrics, this approach emphasizes the proportionate reduction in errors made by the model. 
This is particularly meaningful in screening mammography because the task is inherently constrained by factors such as image quality, the subtle nature of early-stage abnormalities, and the variability in breast tissue density. 
These limitations mean that the theoretical maximum performance is below 1 (perfect accuracy), even for the best possible models. 
Reducing the error rate narrows the gap between current performance and the practical upper limit of what can be achieved, highlighting the substantial impact of even incremental absolute improvements in AUC.
We calculate the proportional reduction of error achieved by the baseline model, $m$, relative to the improved model, $m^*$, using the formula:
\[
\text{Reduction of Error} = 1 - \frac{\text{Error}_{m}}{\text{Error}_{m^*}}
\]
where \(\text{Error} = 1 - \text{AUC}\) and the subscripts refer to the respective models, $m$ and $m^*$.

No covariates were included in the analysis as the model exclusively used imaging data as input. Patient demographic and clinical variables such as age or race were not used as model input.
Each train-validation split was made randomly, and the test sets were defined as all exams within specific time ranges.
Bayesian analysis was not conducted in this study, as the statistical methods employed were based on bootstrapping for confidence intervals and permutation testing for comparisons.
The study design does not involve hierarchical or complex experimental structures.

The statistical significance of the difference in recall rates between the non-AI and AI groups was assessed using a two-tailed two-proportion z-test~\cite{chihara2018mathematical}. Recall rates were defined as the proportion of recalled patients in each group, denoted as \( \hat{p}_1 = \frac{r_1}{n_1} \) for the non-AI group and \( \hat{p}_2 = \frac{r_2}{n_2} \) for the AI group, where \( r_1 \) and \( r_2 \) represent the number of recalled patients, and \( n_1 \) and \( n_2 \) represent the total number of patients in each group, respectively. The test statistic (\( z \)) was calculated as:
\[
z = \frac{(\hat{p}_1 - \hat{p}_2)}{\sqrt{\hat{p}(1 - \hat{p}) \left( \frac{1}{n_1} + \frac{1}{n_2} \right)}}
\quad \text{where} \quad
\hat{p} = \frac{r_1 + r_2}{n_1 + n_2} .
\] 
The p-value was derived from the standard normal distribution based on the absolute value of \( z \). 
A p-value less than 0.05 was considered statistically significant.
In addition, we also report Cohen's h as the effect size of the two-proportion z-test, calculated as follows:
\[
h = 2 \arcsin\left(\sqrt{p_1}\right) - 2 \arcsin\left(\sqrt{p_2}\right).
\]
The required sample size \( n \) per group for a two-proportion z-test is calculated using the following equation~\cite{Wang2007Sample}:
\[
n = \frac{(Z_{\alpha/2} + Z_{\beta})^2 \cdot \left[ p_1(1-p_1) + p_2(1-p_2) \right]}{(p_1 - p_2)^2}
\]
where \( Z_{\alpha/2} \) is the critical value of the standard normal distribution for a two-tailed test at significance level \(\alpha\), \( Z_{\beta} \) is the critical value of the standard normal distribution corresponding to the desired power \((1 - \beta)\), \( p_1 \) is the estimated recall rate in the group before AI implementation, \( p_2 \) is the estimated recall rate in the group after AI implementation.
We set \(Z_{\alpha/2} = 1.960\) for \(\alpha = 0.05\), and \(Z_{\beta} = 0.842\) for 80\% power.
Since gathering exams before AI implementation is trivial, we initially prepared the 40,415 exams before AI implementation without sample size estimation.
This contained 13,776 gray cases, 12,755 mixed cases, and 13,884 green cases.
Afterwards, we conducted 5,000 pilot studies after AI implementation to estimate the required sample sizes for this study.
For all cases where the recall rate before AI implementation is 11.6\% and the recall rate after AI implementation was 12.5\% according to the pilot studies. The required sample size for each group was 20,535.
The number of exams before and AI implementation, 40,415 and 40,603 respectively, exceeds the estimated sample size requirement.
For gray cases where the recall rate before AI implementation is 14.3\% and the recall rate after AI implementation was 17.5\% according to the pilot studies, the required sample size for each group was 2,046.
The number of gray cases before and AI implementation, 13,776 and 14,767 respectively, exceeds the estimated sample size requirement.
For mixed cases where the recall rate before AI implementation is 13.0\% and the recall rate after AI implementation was 14.6\% according to the pilot studies, the required sample size for each group was 7,291.
The number of mixed cases before and AI implementation, 12,755 and 12,789 respectively, exceeds the estimated sample size requirement.
For green cases where the recall rate before AI implementation is 7.6\% and the recall rate after AI implementation was 5.3\% according to the pilot studies, the required sample size for each group was 1,787.
The number of green cases before and AI implementation, 13,884 and 13,047 respectively, exceeds the estimated sample size requirement.

\section*{Acknowledgements}  \label{sec14}

The authors would like to thank Mario Videna and Abdul Khaja for supporting our computing environment. They also like to thank the support of Nvidia Corporation with the donation of some of the GPUs used in this research.

\section{Ethics Declarations} \label{sec15}

\subsection{Funding}

This research was supported by the National Institutes of Health (NIH), through grants TL1TR001447 (the National Center for Advancing Translational Sciences) and P41EB017183 (the National Institute of Biomedical Imaging and Bioengineering), the Gordon and Betty Moore Foundation (9683), and the National Science Foundation (1922658). The content is solely the responsibility of the authors and does not necessarily represent the official views of any of the bodies funding this work. 

\subsection{Conflict of interest/Competing interests}

The authors declare that they have no conflict of interest.

\subsection{Ethics approval and consent to participate}

The collection of NYU Breast Cancer Screening Datasets was approved by NYU Langone Health IRB protocol ID\#i18-00712\_CR3 and the prospective study was approved by NYU Langone Health IRB protocol ID\#s21-01154. 
They were compliant with the Health Insurance Portability and Accountability Act. 
Informed consent was waived.

\subsection{Consent for publication}

All authors agreed to the submission of this manuscript.

\subsection{Data Availability}

CBIS-DDSM, INbreast, CMMD, and BCS-DBT contain de-identified images and are publicly available for download.
Although the INbreast dataset is no longer maintained by the original research group, its authors encourage users to reach out for access to the images. 
CBIS-DDSM can be found on the Cancer Imaging Archive (\url{https://doi.org/10.7937/K9/TCIA.2016.7O02S9CY}). 
Similarly, CMMD is available through the Cancer Imaging Archive  (\url{https://doi.org/10.7937/tcia.eqde-4b16}).
Likewise, BCS-DBT is available through the Cancer Imaging Archive (\url{https://doi.org/10.7937/E4WT-CD02}).
The OPTIMAM and CSAW-CC datasets are restricted-access resources, requiring interested users to submit an access request to their respective investigators. 
A small subset of EMBED dataset is hosted on the AWS Open Data Program (\url{https://registry.opendata.aws/emory-breast-imaging-dataset-embed}).
However, the full EMBED dataset is restricted-access resource, requiring interested users to submit an access request to its investigators. 
The EMBED used in this paper is a part of the full EMBED dataset.

The NYU Breast Cancer Screening Dataset was obtained under NYU Langone Health IRB protocol ID\#i18-00712\_CR3 and is from the NYU Langone Health private database. 
Therefore, it cannot be made publicly accessible. 
We previously published the following report explaining how the dataset was created for reproducibility~\cite{wu2019nyu}. 
While this report only explains algorithms for importing FFDM images, importing C-View and DBT images follow the same general principles as explained in the supplementary materials.
Although we cannot make the dataset public, we can evaluate models from other research institutions on the test portion of the dataset upon request.

For further information regarding data availability, please contact the corresponding author at k.j.geras@nyu.edu. Requests will be responded to within one week.

\subsection{Materials availability}

Not applicable.

\subsection{Code Availability}

The prior work cited in this work are open-sourced and publicy available: GMIC - \url{https://github.com/nyukat/GMIC} and 3D-GMIC - \url{https://github.com/nyukat/3D_GMIC}.
We used open-source libraries to conduct our experiments, such as PyTorch (\url{https://pytorch.org}) machine
learning framework. 
The code for the YOLOX deep neural network which our models are based upon is open-source and is available at \url{https://github.com/Megvii-BaseDetection/YOLOX}.
The code for FROC analysis is available at the following open-source repository: \url{https://github.com/mazurowski-lab/duke-dbt-data}.
At this point, we are not publicly sharing the code for the modified neural networks or the trained model weights to protect the potential commercial viability of our system.

\subsection{Author contribution}

JP and KJG designed experiments using neural networks. JP built the data preprocessing pipeline, preprocessed the internal datasets and the external test sets (EMBED and BCS-DBT), and carried out the experiments using neural networks. JW developed the test-set filtering algorithm and preprocessed the external test sets (CMMD, OPTIMAM, CSAW-CC, InBreast, CBIS-DDSM). AL, LH, and LM analyzed the results from a clinical viewpoint and conducted the prospective study. MW, AL and JP analyzed the data from the prospective study. HT, JG and BB developed the external test set (EMBED) and provided guidance on using it in the evaluation. KJG supervised the project. JP, JW, YX, HT, JG, BB, MW, LM, LH, AL, and KJG contributed to writing and reviewing the manuscript.

\subsection{Inclusion \& Ethics Statement}

Local researchers and collaborators were actively involved in data collection and analysis to ensure the research aligns with institutional priorities. Their contributions are acknowledged through co-authorship.
This research was conducted using a diverse dataset of screening and diagnostic mammography exams at NYU Langone Health and seven external datasets collected across three continents, ensuring broad representation across populations and imaging settings. 
Roles and responsibilities were clearly outlined. 
No capacity-building initiatives were discussed.
No restrictions or prohibitions were encountered. 
Ethical approval for the study was obtained from the NYU Langone institutional review board.
This study does not involve animal welfare or biorisk considerations.
We expect that the research does not result in stigmatization, incrimination, discrimination, or otherwise personal risk to participants.
No specific health, safety, or security risks to researchers were identified during this study.
No biological materials or cultural artifacts were transferred out of the country.
The study includes citations of relevant local and regional research to ensure alignment with existing knowledge.

\clearpage
\printbibliography[]
\clearpage
\appendix

\setcounter{figure}{0}
\setcounter{table}{0}

\renewcommand{\thefigure}{A\arabic{figure}}
\renewcommand{\thetable}{A\arabic{table}}

\section{Appendix}

This document includes the supplementary material for the article ``A Multi-Modal AI System for Screening Mammography: Integrating 2D and 3D Imaging to Improve Breast Cancer Detection in a Prospective Clinical Study'', constituting an extended version of the methods section of the main paper. Section~\ref{sec:appendix_dataset_construction} provides more details on the constructions of datasets, Section~\ref{sec:appendix_data_augmentation} describes the data augmentation method, Section~\ref{sec:appendix_hyperparameters} provides the ranges of model hyperparameters used in hyperparameter tuning, Section~\ref{sec:appendix_2d_3d} explains how to utilize neural networks for 2D images with 3D images, Section~\ref{sec:appendix-ensemble} describes more details about ensembling model predictions, Section~\ref{sec:permutation_test} describes the details about the permutation test used in this study, Section~\ref{sec:retrospective_tests} provides more detailed results from the statistical tests including the differences in AUCs and the percentage of errors reduced, Section~\ref{sec:appendix_external} provides an additional external validation result, and Section~\ref{sec:appendix_clinical} provides more details for the clinical implementation.

\subsection{Construction of datasets}~\label{sec:appendix_dataset_construction}

\subsubsection{Preprocessing of the dataset}~\label{sec:preprocessing}

We developed and executed a data preprocessing pipeline for the mammography exams including FFDM, C-View, and/or DBT images, based on our data report for FFDM-only mammography exams~\cite{wu2019nyu}.
This process involved the steps already described in the aforementioned data report: an algorithm to filter out invalid images and exams, an algorithm to remove background pixels and retain the foreground, and an algorithm to calculate the best window placement.

Invalid images and exams included the following: exams from non-female patients, exams with breast implants, duplicate images based on SOPInstanceUID, and invalid images based on ImageType, PerformedProcedureStepDescription, SeriesDescription, StudyDescription, ProtocolName metadata.
Specifically, images with ImageType of ORIGINAL were rejected because they lack necessary post-processing.
For the description metadata, we only accepted images if their expected study type (screening or diagnostic) matches with the DICOM metadata.
In addition, we only accept certain views depending on the exam type: CC and MLO for screening and CC, MLO, LM, ML, XCCL, XCC, TAN, XCCM, AT, RL, RM for diagnostic.

To improve image loading time, we removed the black background from DICOM images and saved only relevant regions.
This was achieved by capturing the largest connected component after morphological operations to select the foreground, and saving the rectangle region that contains the foreground with a safety margin.
The details on this process are in our earlier data report~\cite{wu2019nyu}.
For DBT images, we leveraged the corresponding C-View images to determine the cropping coordinates. 
Specifically, when selecting the coordinates for foreground extraction in C-View images, we applied the same cropping parameters to the corresponding DBT images across all slices.

Even after removing the black background, the images often remain very large, necessitating a strategy to focus on the most important regions.
In addition, fixing the image tensor size is essential for efficient batching in deep learning.
To achieve this, we determined a fixed-size window so that we can obtain fixed-sized tensors from all images: $2866 \times 1814$ for FFDM and $2166 \times 1339$ for DBT and C-View images.
The window sizes were determined to capture comparable amounts of information between the imaging modalities.
For each image, we calculated the optimal location to place this window so that we can focus on the most important region, containing the outmost part of the breasts. 
A more detailed description of the algorithm can be found in our earlier data report~\cite{wu2019nyu}.

\subsubsection{Detailed image-level statistics on mammographic views of the images in the dataset}

Supplementary Table~\ref{tab:stats_views} shows the number of images with different views for each subset of the NYU Comprehensive Mammography Dataset V1 and V2.

\begin{table}
\renewcommand{\tablename}{Supplementary Table}
    \centering
\caption{Image-level statistics for the views of the overall NYU Comprehensive Mammography Dataset V1 and V2. Abbreviations: N, number.}
\label{tab:stats_views}
    \begin{tabular}{>{\raggedleft\arraybackslash}p{5.5cm} | >{\raggedleft\arraybackslash}p{1.3cm} | >{\raggedleft\arraybackslash}p{1.3cm} | >{\raggedleft\arraybackslash}p{1.3cm} | >{\raggedleft\arraybackslash}p{1.3cm} | >{\raggedleft\arraybackslash}p{1.3cm} | >{\raggedleft\arraybackslash}p{1.3cm}}

         \hline
         \textbf{Characteristics, unit} & \multicolumn{2}{|>{\centering\arraybackslash}p{2.7cm}|}{\textbf{\makecell{Training-\\validation set}}} & \multicolumn{2}{|>{\centering\arraybackslash}p{2.7cm}|}{\textbf{\makecell{Ensemble-\\selection set}}} & \multicolumn{2}{|>{\centering\arraybackslash}p{2.7cm}}{\textbf{Test set}}\\
         \cline{2-7}
         & V1 & V2 & V1 & V2 & V1 & V2 \\
         \hline \hline
         FFDM Images, N&  2,162,627&  3,289,332
&  13,952&  86,531
&  85,578& 
199,023
\\  
         CC, N & 1,030,650& 1,562,207& 6,884& 42,495& 42,197& 97,593\\
         MLO, N & 1,033,577& 1,559,949& 7,068& 44,036& 43,381& 101,430\\
         LM, N & 29,728& 55,546& 0
& 0
& 0& 0\\
         ML, N & 47,407& 77,385& 0
& 0
& 0& 0\\
         XCCL, N & 15,422& 24,426& 0
& 0
& 0& 0\\
         XCC, N & 2,106& 3,894& 0
& 0
& 0& 0\\
         TAN, N & 307& 374& 0
& 0
& 0& 0\\
         XCCM, N & 1,193& 1,973& 0
& 0
& 0& 0\\
         AT, N & 1,368& 2,044& 0
& 0
& 0& 0\\
         RL, N & 442& 780& 0
& 0
& 0& 0\\
         RM, N & 427& 754& 0& 0& 0& 0\\ \hline
         C-View Images, N&  799,903&  1,827,074
&  13,152&  81,290
&  79,813& 
184,736
\\ 
         CC, N & 398676& 892,438& 6,556& 40,507& 39,833& 91,964\\
         MLO, N & 401227& 895,625& 6,596& 40,783& 39,980& 92,772\\
         LM, N & 0& 153& 0
& 0
& 0& 0\\
         ML, N & 0& 28,600& 0
& 0
& 0& 0\\
         XCCL, N & 0& 8,430& 0
& 0
& 0& 0\\
         XCC, N & 0& 0& 0
& 0
& 0& 0\\
         TAN, N & 0& 20& 0
& 0
& 0& 0\\
         XCCM, N & 0& 963& 0
& 0
& 0& 0\\
         AT, N & 0& 845& 0
& 0
& 0& 0\\
         RL, N & 0& 0& 0
& 0
& 0& 0\\
         RM, N & 0& 0& 0& 0& 0& 0\\ \hline
         DBT Images, N&  799,903&  1,827,074
&  13,152&  81,290
&  79,813& 
184,736
\\ 
         CC, N & 398,676& 892,438& 6,556& 40,507& 39,833& 91,964\\
         MLO, N & 401,227& 895,625& 6,596& 40,783& 39,980& 92,772\\
         LM, N & 0& 153
& 0
& 0
& 0& 0\\
         ML, N & 0& 28,600& 0
& 0
& 0& 0\\
         XCCL, N & 0& 8,430& 0
& 0
& 0& 0\\
         XCC, N & 0& 0
& 0
& 0
& 0& 0\\
         TAN, N & 0& 20
& 0
& 0
& 0& 0\\
         XCCM, N & 0& 963
& 0
& 0
& 0& 0\\
         AT, N & 0& 845
& 0
& 0
& 0& 0\\
         RL, N & 0& 0
& 0
& 0
& 0& 0\\
         RM, N & 0& 0& 0& 0& 0& 0\\ \hline
    \end{tabular}
\end{table}

\subsubsection{Pseudo bounding-box label generation using breast-level labels}

Naively training models on both screening and diagnostic FFDM exams from the V1 dataset resultd in lower performance compared to models trained solely on screening FFDM exams.
This was unexpected, as increasing the size of training data usually leads to performance improvements. 
We hypothesized that this decline is due to the higher proportion of positive images without bounding-box labels in the combined dataset, as most diagnostic exams lacked annotations at the time. 
The missing bounding-box labels amplified an imbalance: the model is constantly penalized for negative images but rarely encouraged to detect lesions in positive images.

To address this, we created pseudo bounding-box labels for the positive images without bounding-box labels in the combined screening+diagnostic dataset in the V1 training-validation set.
Specifically, we performed model inference on the positive images without bounding-box labels using a model previously trained on screening dataset, and use resulting predictions as pseudo bounding-box labels.
To improve the quality of pseudo bounding-box labels, we incorporated breast-level pathology-confirmed labels: the top-1 malignant bounding-box prediction was assigned as the pseudo-label for the images with malignant pathology-confirmed label, and the top-1 benign bounding-box prediction was assigned as the pseudo-label for the images with benign pathology-confirmed label.
With these additional pseudo-labels, we no longer observed the unexpected drop of performance.
However, we did not create the pseudolabels for the V2 dataset, as the V2 dataset had a much higher proportion of annotated exams.

\subsubsection{Identifying matching images between modalities}~\label{sec:matching_modalities}

To ensemble bounding-box predictions, images from different modalities need to have pixel-level field-of-view alignment.
If they have different resolutions, they must show the said alignment after resizing to match the resolutions.  
It is straightforward to identify such matches between C-View and DBT images: C-View images are created from the source DBT image, and they share the same IrradiationEventUID or FrameOfReferenceUID metadata in the DICOM files.
In contrast, FFDM images, acquired through separate radiation exposures, do not always align perfectly with DBT images.
Despite these limitations, for the DBT images in the V1 and V2 test sets, we identified the matching FFDM images with pixel-level alignment in over 99.7\% and 99.4\% of cases respectively, enabling multi-modal bounding-box ensembling in most cases. 
We describe the specific algorithm of identifying these matching triplets below.

Since DBT images and the corresponding C-View images already have the same field-of-view, we only need to compare the FFDM and C-View images to identify the matching FFDM-C-View-DBT triplets.
To facilitate this comparison, we first resize the FFDM images to match the resolution of the C-View images.
FFDM images in FFDM-DBT combo exams acquired by Hologic Selenia Dimensions model are acquired in two sizes: $4096 \times 3328$ and  $3328 \times 2560$.
C-View images from the same system are acquired in two sizes: $2457 \times 1996$ and  $2457 \times 1890$.
Sicne the FFDM image with size $4096 \times 3328$ has the same aspect ratio as the C-View image with size $2457 \times 1996$, the former is resized to the latter's dimensions.
Likewise, the same principle applies to the FFDM image with size $3328 \times 2560$, which is resized to match the C-View image with size $2457 \times 1890$ due to their identical aspect ratios.

Second, we compare the intersection over union of the breast shapes from the FFDM and C-View images.
However, directly comparing the nonzero pixels in the images is problematic due to text overlays and artifacts such as zero-valued rows at the image edges in some of the image modalities. 
To address this, we extracted the largest connected components of the nonzero mask of each image and excluded the top and bottom 50 rows from the analysis to avoid artifacts.
The intersection over union calculated on these selected pixels provided a more reliable measure of field-of-view alignment.

Third, we generate all possible FFDM-C-View pairs for each view in each exam.
For example, if there are 3 FFDM images and 2 C-View images for a view in an exam, this leads to 6 candidate FFDM-C-View pairs.
We initially filtered out pairs with lower than 0.96589 intersection-over-union as invalid pairs.
This threshold was empirically chosen because it successfully separated incorrect pairs and correct pairs in manual inspection.
Lastly, for each C-View image, we chose the pair with the highest intersection-over-union among all remaining pairs.
By adding the corresponding DBT images to each chosen FFDM-C-View pair, this process yields the final selection of FFDM-C-View-DBT triplets.

\subsection{Data augmentation}~\label{sec:appendix_data_augmentation}

\subsubsection{Data augmentation during training}

Mammography images are initially loaded as image tensors of pre-determined sizes as described in Section~\ref{sec:preprocessing}: $2866 \times 1814$ for FFDM and $2166 \times 1339$ for DBT and C-View images.
These cropped images are then resized to smaller-sized images for model processing, where this final image size is a hyperparameter.

We in turn apply random horizontal flips as well as affine transformation to the input image.
The flip probability is 0.5, and the affine transformation parameters are as follows: random rotation ($-15^{\circ}$ to $15^{\circ}$), random translation ($\pm$ up to 10\% of image size), scaling by a random factor between 0.8 and 1.6, random shearing ($-25^{\circ}$ to $25^{\circ}$).
If the pixel-level segmentation label is available, we apply the same transformations that were applied to the corresponding input image to the pixel-level segmentation label, and extract the bounding boxes again from the transformed segmentation to ensure the tightest fit to the lesion.

If the pixel-level segmentation label is not available for this image and only the bounding-box label is available, affine transformation could distort the bounding-box labels to include an area much larger than the lesions, degrading the quality of the annotation labels. 
To mitigate this, we employ a simpler set of transformations for the images that have bounding-box labels but not the segmentation labels. 
Concretely, after cropping a window of predetermined size but before the final resizing, we take a slightly smaller crop with randomness in the location of the crop (up to 100 pixels) and the size of the crop (100-200 pixels smaller than the original crop).
Since this does not involve rotation and shearing, the quality of bounding-box labels does not change.
In addition, this is consistent with the test-time data augmentation in which we do not want to distort the input too much.

\subsubsection{Data augmentation during validation}

The cropping window is placed in the location we determined to be optimal as described in Section~\ref{sec:preprocessing}.
We do not perform any random augmentation of input over multiple window placements during validation because we want to minimize the computation required for this phase by only processing each image once per epoch.

\subsubsection{Test time data augmentation}~\label{sec:test_time_augmentation}

At test time, we generate multiple predictions for the same image, and then aggregate them into a final prediction. 
These predictions are derived from slightly modified versions of the original image, created by placing the cropping window in multiple different locations to cover the entire breast.
In addition, we apply the aforementioned simple set of transformations that only involves minor translation and resizing, rather than the more complex affine transformation.
This approach still introduces sufficient diversity to enhance robustness of model predictions while ensuring the input remains close to its original representation, avoiding excessive distortion.

\subsection{Training Details}~\label{sec:appendix_hyperparameters}

The network is initialized with pretrained weights released by the authors of the original YOLOX paper~\cite{ge2021yolox}. 
These weights are from the YOLOX models trained with COCO train2017 dataset~\cite{lin2014microsoft} without any previous pretraining.
The training objective is optimized using stochastic gradient descent (SGD) with Nesterov momentum~\cite{sutskever2013importance} and weight decay.

We performed hyperparameter optimization using random search~\cite{bergstra2012random}.
For the V1 models, randomly sample the learning rate $\eta$ from log-uniform distribution $\log_{10}(\eta) \sim \mathcal{U}(-6.046, -5.456)$ and the weight decay hyperparameter $\lambda$ from log-uniform distribution $\log_{10}(\eta) \sim \mathcal{U}(-3.523, -3.155)$ and the momentum $\mu$ from uniform distribution $\mu \sim \mathcal{U}(0.80, 0.92)$, the number of bounding-box predictions $K \in \{4, 5, 6, 7, 8, 9, 10\}$ with equal probabilities, the complexity of model architecture $A \in \{s, m, l, x\}$ with equal probabilities, and the image height $H \in \{1536, 1664, 1792, 1920, 2048\}$ with equal probabilities.
For the V2 models, we chose a slightly different range of some hyperparameters compared to the V1 models. 
We randomly sample the learning rate $\eta$ from log-uniform distribution $\log_{10}(\eta) \sim \mathcal{U}(-6.046, -5.398)$ and the weight decay hyperparameter $\lambda$ from log-uniform distribution $\log_{10}(\eta) \sim \mathcal{U}(-3.523, -3.260)$ and the momentum $\mu$ from uniform distribution 
$\mu \sim \mathcal{U}(0.80, 0.92)$, the number of bounding-box predictions $K \in \{5, 6, 7, 8, 9\}$ with equal probabilities, the complexity of model architecture $A \in \{l, x\}$ with equal probabilities.
For the V2 models for DBT and C-View modalities, we sample the image height $H \in \{1536, 1664, 1792, 1920, 2048\}$ with equal probabilities.
For the V2 models for FFDM modality, we sample the image height $H \in \{2048, 2176, 2304, 2432, 2560, 2688, 2816\}$ with equal probabilities.

For each modality, we optimized the model hyperparameters with 20 random search trials.
All models were trained for up to 40 epochs.
In each epoch, training was performed on a subset of images composed of all positive images (i.e., those associated with either benign or malignant biopsy-confirmed findings) and an equal number of randomly-sampled negative images.
While each modality-specific model was trained primarily using images from the corresponding modality, the C-View and DBT models additionally incorporated FFDM images from exams that did not include C-View or DBT.
We saved and used the model weights from the training epoch with the highest AUROC in identifying images with malignant findings on the validation set.

\subsection{Utilizing neural networks for 2D images with 3D images}~\label{sec:appendix_2d_3d}

\subsubsection{Deciding which DBT slice to load during training and validation}\label{sec:dbt_training}

Our AI system is based on convolutional neural networks for 2D images, processing each slice of DBT image separately.
While inference is performed on all slices separately and aggregated, doing so during the training and validation phases is infeasible due to about 70-fold increase in processing time and GPU memory usage compared to processing a single slice.
Therefore, we only utilize one slice of each DBT image at a time during training and validation phases.

For the positive images, loading any random slice is suboptimal because not all slices contain visible cancer lesions.
For example, training the model to predict high malignancy prediction for a slice without visible lesion could be detrimental to the model performance.
Instead, we utilize the radiologists' annotation labels and C-View images during training as follows.
For the images with annotation labels, we load one of the annotated slices randomly during training which guarantees that there is a visible lesion on the loaded slice.
For the images without any annotation labels, if they have positive breast-level classification labels, then we load the corresponding C-View image which should contain all information from the entire breast.
For the images without any annotation labels, if they are negative images, we choose to load a random slice.
This approach allows us to perform efficient training by loading the most informative slice and avoiding loading the entire 3D image.

During validation, we use a slightly different strategy for consistent evaluation between epochs and models. 
For positive images with annotations, we choose to load the center slice of one of the annotated lesions.
For positive images without annotations, we continue to load the corresponding C-View image.
For negative images, we choose to load the central slice of the DBT images.

\subsubsection{Performing test-time inference on 3D images: aggregating 2D slice-level predictions for 3D images}~\label{sec:mss}

\begin{figure}
\renewcommand\figurename{Supplementary Figure}
    \centering
    \includegraphics[width=\linewidth, trim={0 5mm 0 0}, clip]{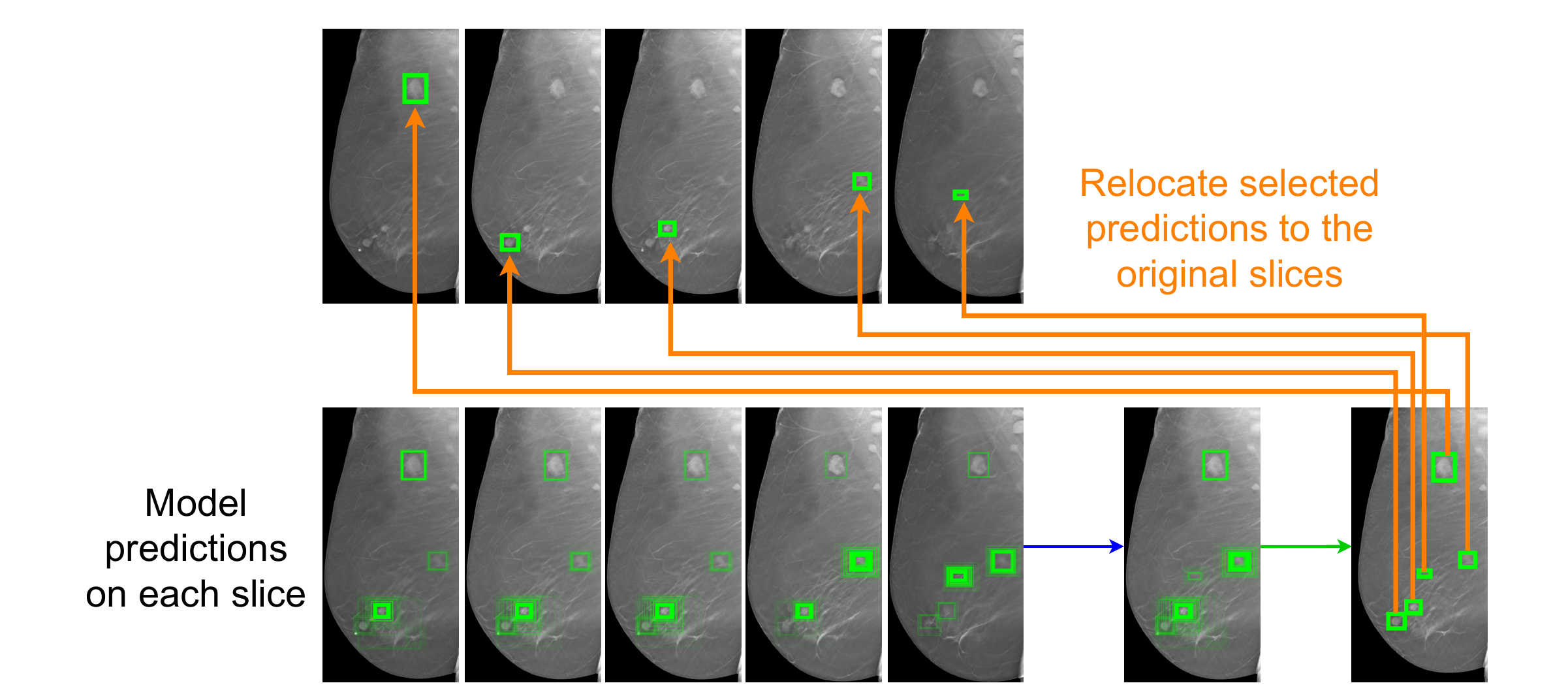}
    \includegraphics[width=\linewidth, trim={0 0 0 5mm}, clip]{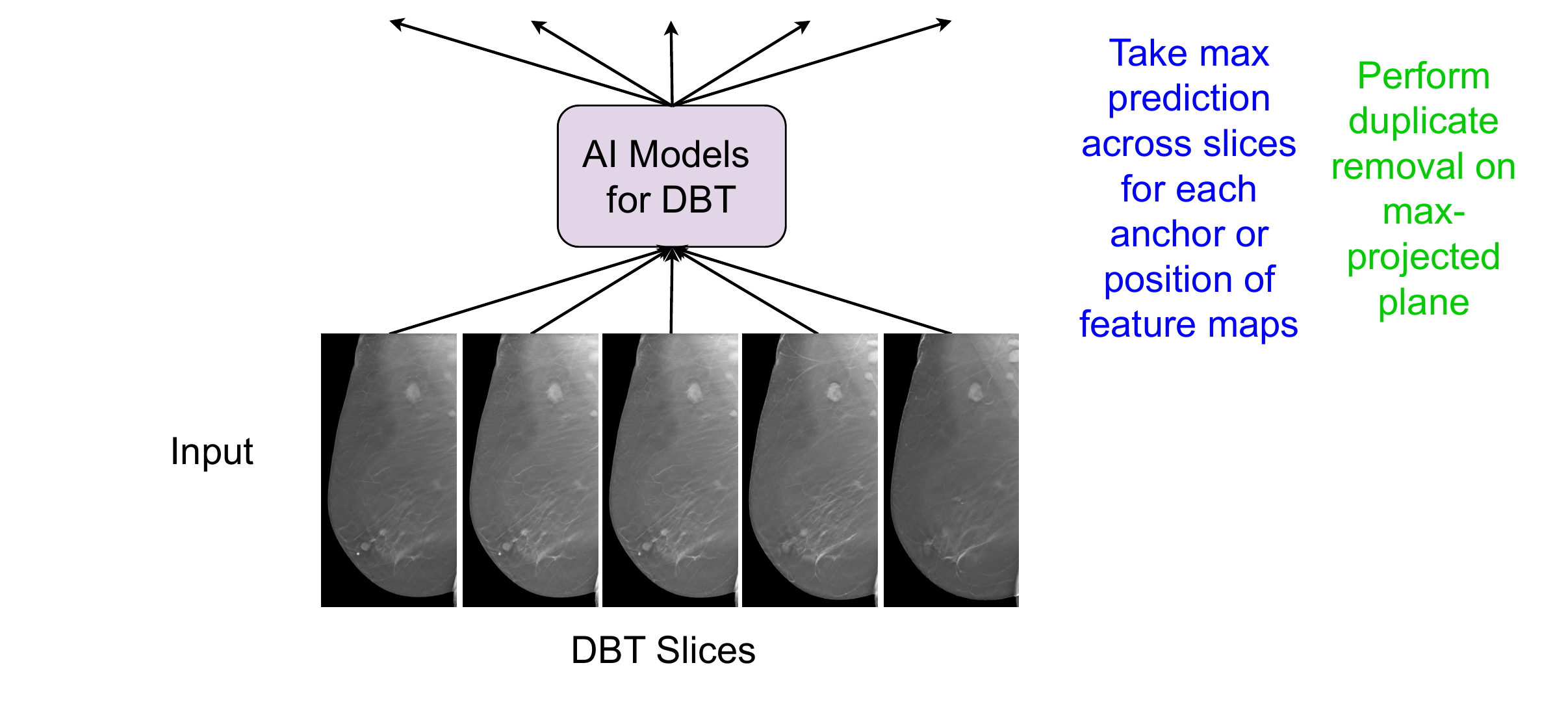}
    \caption{An overview of the Max-Slice-Selection (MSS) algorithm. This algorithm takes advantage of the characteristics of the DBT images and saves the required computation compared to naively running non-maximum selection algorithm on the concatenated collection of predictions from all slices.}
    \label{fig:mss}
\end{figure}

DBT images are three-dimensional and involve around 70 slices on average.
To utilize the slices with the most lesion conspicuity, it is necessary to create bounding-box predictions for every single slice during test-time inference on DBT images. 
However, this process creates numerous duplicate predictions for the same lesion on nearby slices.
To efficiently remove these duplicates before applying non-maximum suppression (NMS), we introduce a novel Max-Slice-Selection (MSS) algorithm as shown in Figure~\ref{fig:mss}.

MSS efficiently aggregates 2D slice-level predictions into a final set of bounding-box predictions for an entire 3D image during test-time inference. 
This algorithm was briefly mentioned in our prior work~\cite{park2021lessons} but had not been explained in full detail before.
In short, MSS first removes duplicates across the depth dimension before the standard NMS algorithm is applied to remove the remaining duplicates across width and height dimensions.
Specifically, let $\mathbf{A_{x,i}} \in \mathbb{R}^{h \times w}$ be the collection of probability scores of bounding-box predictions made on $\mathbf{i}$th slice of image $\mathbf{x}$.
These bounding-box predictions are generated from all locations of the feature maps with height $h$ and width $w$, and thus can be arranged as 2D tensors corresponding to the feature map.
First, we aggregate the box predictions $\mathbf{A_{x,i}}$ from all $\mathbf{n}$ slices and concatenate in a new depth dimension to create $\mathbf{B_x} \in \mathbb{R}^{n \times h \times w}$, a tensor for all bounding-box probabilities for the 3D image $\mathbf{x}$. 
Then, we choose the max prediction along the depth dimension, meaning we choose the slice with the highest prediction at each location of the feature map (or each anchor in anchor-based networks).
This creates a new tensor $\mathbf{D_{x}} \in \mathbb{R}^{1 \times h \times w}$ which contains the bounding-boxes with the highest probabilities at each xy-location of the tensor $\mathbf{B_x}$.
The assumption behind this procedure is that there is a single slice where the lesion is the most clearly visible (likely the center of the lesion) and that no two distinct lesions will have highly overlapping xy-coordinates.
At this point, the total number of box predictions is the same as the original number of box predictions which would have been generated form a single slice of image.
We then treat these predictions as if they all come from the same slice and then apply NMS algorithm.
Finally, we relocate the surviving bounding-box predictions to the corresponding slices which they originated from. 

MSS has a similar effect as running NMS with the concatenated collection of bounding-box predictions from all slices.
However, NMS has an algorithmic complexity of $O(m^2)$ where $m$ is the number of total bounding-box predictions before the duplicate removal procedure.
Therefore, NMS can be prohibitively slow when using all predictions from all slices as input.
A common approach to address this is to select the top K box predictions before applying NMS to reduce the runtime.
However, in DBT images where there are numerous duplicate predictions from each xy-location across multiple slices, this approach becomes less effective.
MSS efficiently reduces the number of bounding-box predictions before NMS is called by removing duplicates along the depth dimension, effectively reducing the required runtime while minimizing the risk of missing important predictions.

In addition, when creating image-level predictions with DBT images, we use the bounding-box predictions after this MSS algorithm and aggregate their feature vectors.
In this process, we only take bounding-box predictions for the malignant category since there already are a lot of predictions in the malignant category when we combine predictions across all slices.

\subsection{Ensembling model predictions} \label{sec:appendix-ensemble}

\subsubsection{Ensembling predictions between different images from the same exam}

\paragraph{Creating breast-level predictions with multi-modal mammography images}

Each individual model in our AI system is trained using data from one imaging modality (FFDM, C-view, or DBT).
For a given breast, the system processes both CC and MLO view images across all three imaging modalities using the corresponding models. 
These predictions are ensembled to produce a comprehensive breast-level prediction.
This final prediction indicates whether the breast is deemed suspicious (above a predefined threshold, shown as gray) or non-suspicious (below the threshold, shown as green).
This process is illustrated in
Supplementary Figure~\ref{fig:ai-system-inference}.a.

\paragraph{Multi-modal bounding-box ensemble}~\label{sec:multimodal_bbox_ensemble}

For each breast view (CC or MLO), bounding-box predictions are generated independently by AI models trained on each imaging modality.
When FFDM image has the matching C-View and DBT images with the same field-of-view as described in Supplementary Section~\ref{sec:matching_modalities}, the bounding-box predictions can be ensembled between different modalities.
First, we ensemble model predictions within each imaging modality to prepare one set of predictions per modality.
Second, we transform the bounding-box coordinates into a shared coordinate space.
Specifically, we resize the coordinates from the bounding-box predictions from FFDM images to match the coordinate space of the matching C-View images.
Also, we hide the depth information from the coordinates of bounding-box predictions from DBT images to match the 2-dimensional coordinate space of the matching C-View images.
Third, we perform multi-modal bounding-box ensemble.
Out of different methods we have tried, we find that simply performing NMS algorithm with a intersection-over-union threshold of 0.05 leads to the best performance.
The final bounding-box predictions displayed on the FFDM and C-view images originate from this ensemble process. 
In the case of DBT images, however, only the predictions that are either generated directly from the DBT modality or have been merged with such predictions are displayed. 
This is because the some of the predictions created from this ensemble process only involved predictions from 2D images and thus do not contain any depth information required to display them on DBT images.
For multi-modal bounding-box ensemble, we report the performance of the ensembled bounding-box predictions in the C-View coordinate space using C-View bounding-box labels as ground-truth.
This ensembling approach leverages complementary information from all three imaging modalities, enhancing the robustness and accuracy of the bounding-box predictions for radiologists.
This process is illustrated in Supplementary Figure~\ref{fig:ai-system-inference}.b.

\begin{figure}
\renewcommand\figurename{Supplementary Figure}
    \centering
    \begin{picture}(0,0) 
    \put(-3,290){\textbf{a}} 
    \end{picture}
    \includegraphics[width=0.45\linewidth]{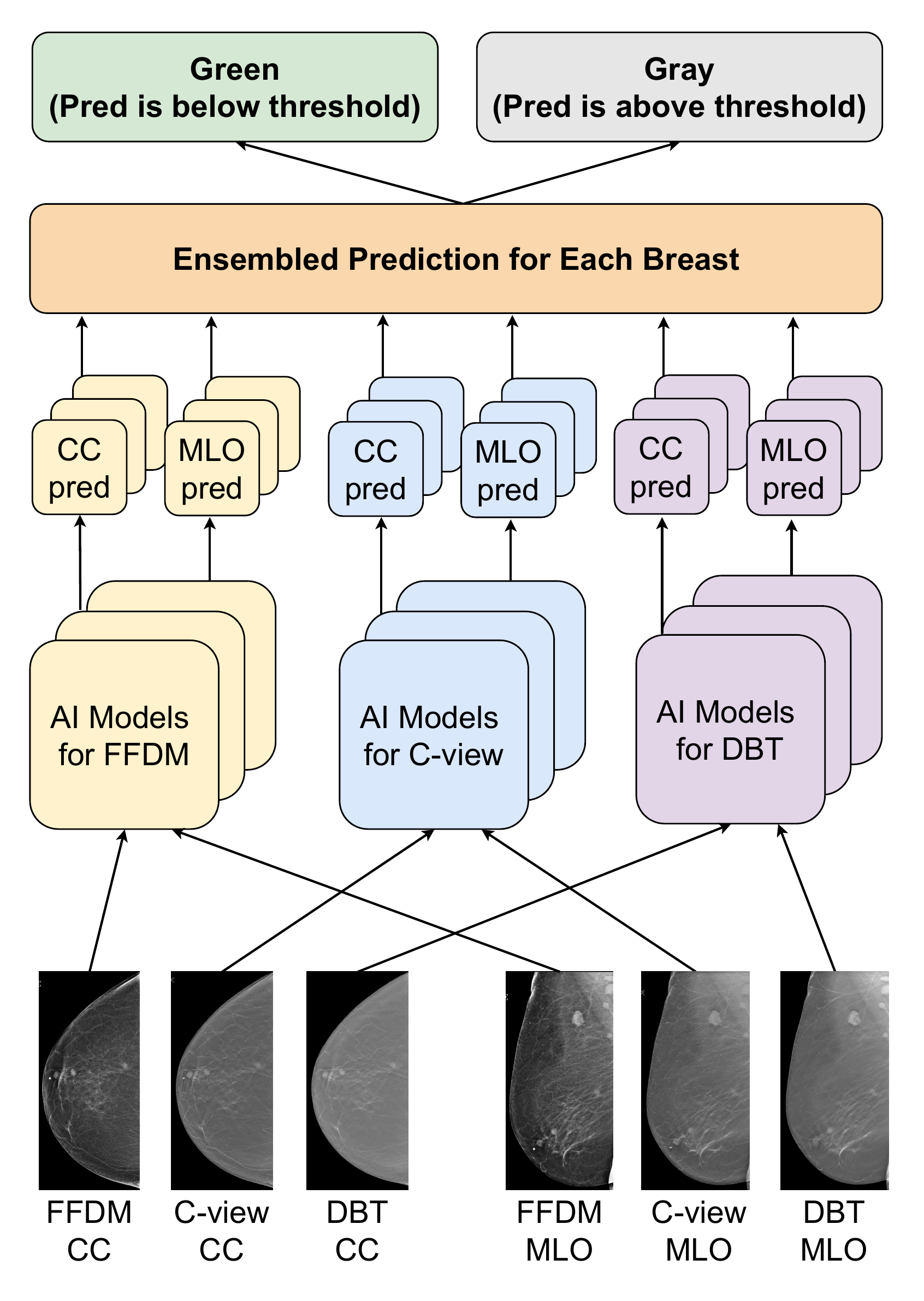}
    \begin{picture}(0,0) 
    \put(-3,360){\textbf{b}} 
    \end{picture}
    \includegraphics[width=0.45\linewidth]{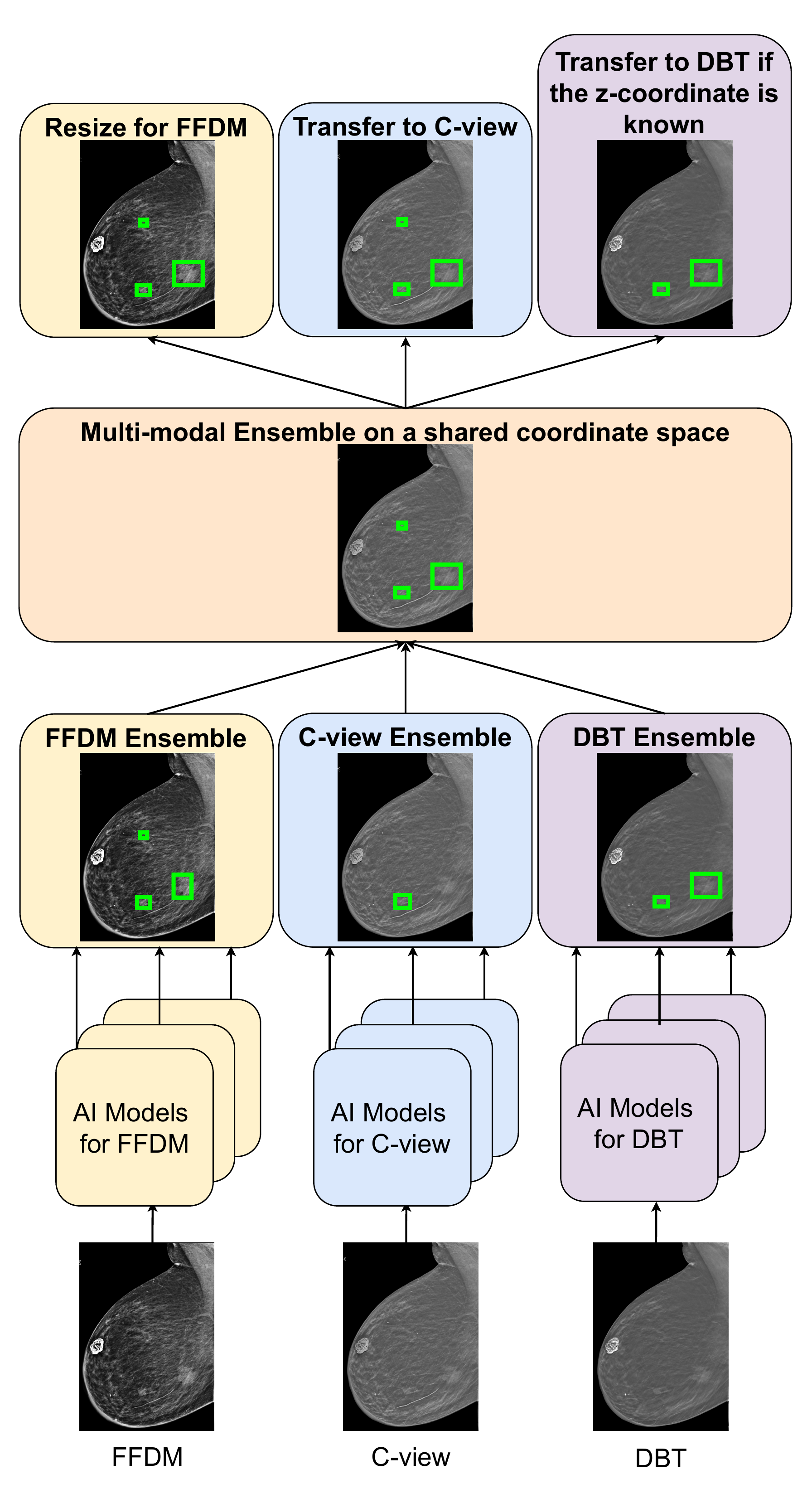}
    \caption{Overview of the AI system’s inference pipeline. a: For each breast, we prepare both CC and MLO view images in all of FFDM, C-View, and DBT imaging modalities. These images are fed into the respective models included in the AI system, each of which create image-level predictions. These predictions for each breast are aggregated to create the final ensemble prediction, which informs radiologists whether they are deemed suspicious or not. b: For each view, the bounding-box predictions from FFDM, C-View, and DBT modalities are combined to create a unified set of predictions. }
    \label{fig:ai-system-inference}
\end{figure}
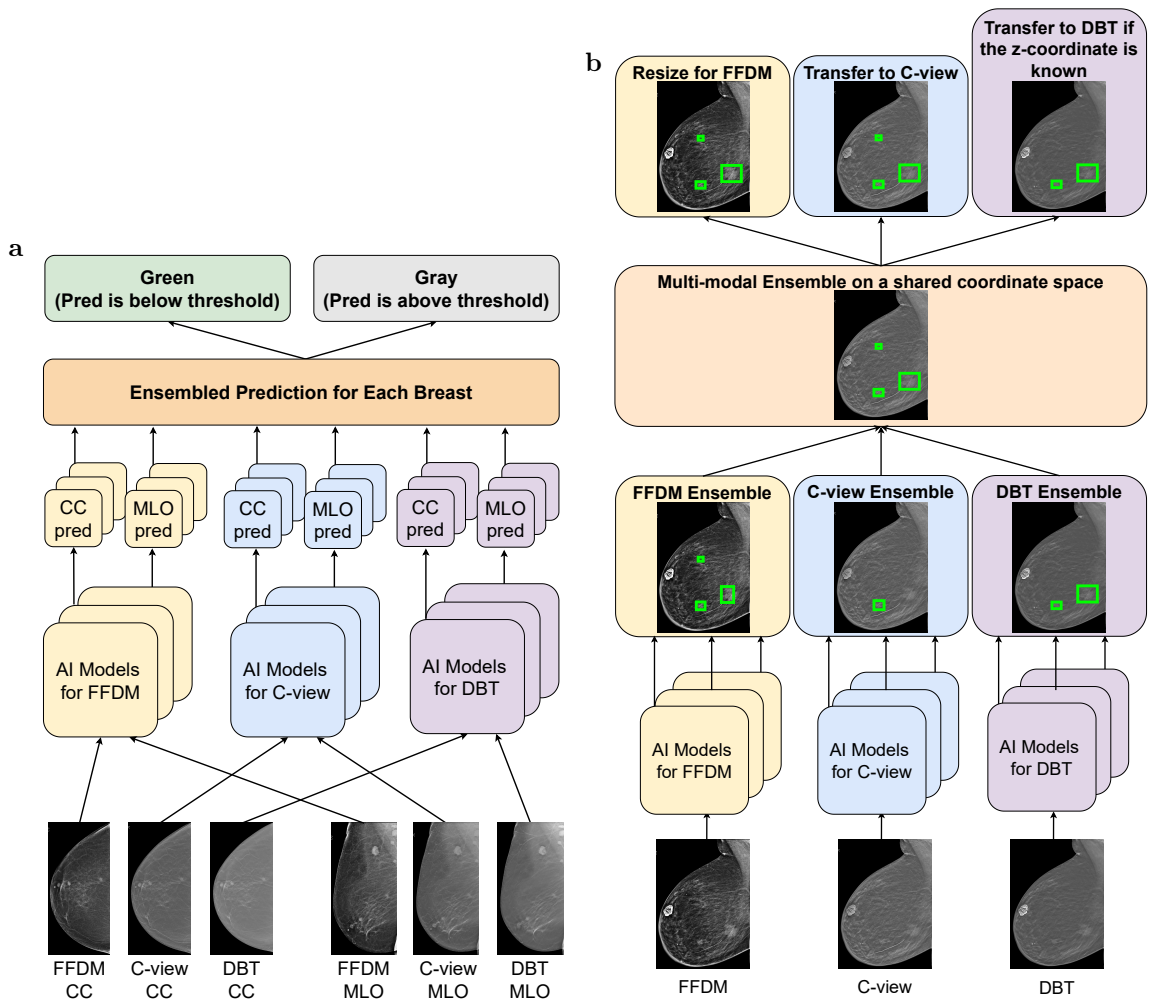

\subsubsection{Greedy ensemble selection}

To maximize the diversity between models in an ensemble, we employed a 10-fold cross-validation strategy with the training-validation sets of the V1 and V2 datasets.
First, we divided the training-validation sets into 10 subsets.
Second, we created 10 training-validation splits where one of the subsets is used as a validation set and the remaining subsets were used as a training set for each split.
Third, we performed hyperparameter-tuning on one of these splits to identify the top-3 hyperparameter configurations.
Fourth, in each training-validation split, we trained three models using the top-3 hyperparameters from the step 3. The training was conducted with different random seeds for the three imaging modalities: FFDM, C-View, and DBT.
Fifth, with each model, we performed five iterations of inference on each image in the ensemble-selection sets.
Sixth, all image-level classification predictions are averaged for each breast to create breast-level predictions for each model.
Seventh, using breast-level classification labels, we apply the greedy ensemble selection algorithm~\cite{caruana2004ensemble} to identify a small subset of models that lead to maximal breast-level classification performance, with an additional constraint that the order of models to be selected alternate between models trained with different imaging modalities.
Eighth, for the 12 models selected with repetition by the greedy ensemble selection algorithm, we performed five iterations of inference on each image from the test dataset using test-time input augmentation as described in section~\ref{sec:test_time_augmentation}.
Ninth, all image-level classification predictions in the test-time inference outputs are averaged for each breast to create breast-level test predictions for each model.
Tenth, the breast-level test predictions from different models are ensembled according to the 12 models selected in the eighth step.

\subsection{Details on the permutation tests performed in this study}~\label{sec:permutation_test}

Permutation testing is a nonparametric statistical approach used to assess whether the observed difference in model performance could have arisen by chance. 
In this study, we performed permutation tests with 10,000 iterations for each comparison.
In each of 10,000 iterations, we randomly swapped predictions between the two models for each case, generating a distribution of differences under the null hypothesis. 
The null hypothesis assumed no difference in AUC between the models.
A one-sided p-value was computed as the proportion of times the null distribution produced a difference as large as or larger than the observed difference, reflecting how extreme the observed statistic is under the null hypothesis.
The significance level is 0.05.
The predictions were transformed into percentile ranks with respect to each model predictions before they were swapped between models, ensuring that the test is not biased by absolute score magnitudes but rather reflects relative ranking across cases. 
For bounding-box predictions, depending on whether the AUFROC was calculated lesion-level or breast-level, predictions for each image or each breast are randomly swapped, respectively.
As for the percentile ranking for bounding-box predictions for lesion-level and breast-level AUFROC calculation, the collection of top-1 bounding-box predictions per image or per breast for each model was used as a basis for calculating the percentile ranking for all bounding-box predictions for each model.

\subsection{Detailed statistical test results for retrospective evaluation}~\label{sec:retrospective_tests}

The p-values of permutation tests, differences in AUCs, and the reduction of errors between GMIC and the V1 model are shown in Supplementary Table~\ref{tab:test_gmic_v1}. 
The p-values of permutation tests, differences in AUCs, and the reduction of errors between the V1 model and the V2 model are shown in Supplementary Table~\ref{tab:test_v1_v2}. 
The differences between the classification performance of the GMIC and the V1 model are statistically significant in CMMD, OPTIMAM, CSAW-CC, EMBED, InBreast, CBIS-DDSM, and BCS-DBT for AUROC and AUPRC.
The V1 model reduced the errors in AUROC of GMIC or 3D-GMIC by 35.31\% to 69.14\% across CMMD, OPTIMAM, CSAW-CC, EMBED, CBIS-DDSM, and BCS-DBT.
The differences between the classification performance of the V1 model and the V2 model are statistically significant in the OPTIMAM, CSAW-CC, EMBED, and BCS-DBT for AUROC and AUPRC.
The V2 model further reduced the errors in AUROC of the V1 model by 18.86\% to 56.62\% across large external datasets: OPTIMAM, CSAW-CC, EMBED, and BCS-DBT.
In addition, the differences between the AUFROC\_1 of the V1 model and the V2 model are statistically significant in the EMBED and BCS-DBT.
The V2 model further reduced the errors in AUFROC\_1 of the V1 model by 37.58\% for EMBED and 34.17\% for BCS-DBT.

\begin{table}
\renewcommand{\tablename}{Supplementary Table}
    \centering
\caption{Comparison of model performance metrics and statistical significance for pairwise model evaluations, highlighting the improvement of the V1 model over GMIC or 3D-GMIC. The table presents p-values of permutation tests, differences in AUROC and AUPRC ($\Delta \text{AUC}$) with 95\% confidence interval, and the reduction of errors with 95\% confidence interval. The significance level is 0.05.}
\label{tab:test_gmic_v1}
    \begin{tabular}{cc|ccc}
    \hline
         Data set&   AUC
&  p-value
& $\Delta \text{AUC}$ &Reduction of errors\\ \hline
         CMMD
&  ROC&  P < 0.0001
& 0.060 (0.048 - 0.072)&35.31\% (29.19\% - 41.28\%)
\\
         CMMD
&  PR&  P < 0.0001
& 0.049 (0.039 - 0.060)&34.88\% (28.94\% - 40.47\%)
\\
         OPTIMAM
&  ROC
&  P < 0.0001
& 0.096 (0.090 - 0.103)&57.48\% (55.23\% - 59.57\%)
\\
         OPTIMAM
&  PR&  P < 0.0001
& 0.166 (0.155 - 0.176)&45.20\% (43.18\% - 47.20\%)
\\
         CSAW-CC
&  ROC
&  P < 0.0001
& 0.039 (0.030 - 0.049)&69.14\% (59.31\% - 77.93\%)
\\
         CSAW-CC
&  PR&  P < 0.0001
& 0.268 (0.229 - 0.306)&53.11\% (47.37\% - 58.97\%)
\\
         EMBED
&  ROC
&  P < 0.0001
& 0.107 (0.071 - 0.147)&49.39\% (36.72\% - 60.18\%)
\\
 EMBED
& PR& P < 0.0001
&0.149 (0.091 - 0.208)&15.89\% (9.70\% - 22.57\%)
\\
 InBreast& ROC
& P = 0.1475
&0.016 (-0.009 - 0.057)&81.82\% (-150.00\% - 100.00\%)
\\
 InBreast& 
PR& P = 0.1382
&0.033 (-0.024 - 0.129)&79.19\% (-151.43\% - 100.00\%)
\\
 CBIS-DDSM& ROC& P < 0.0001
&0.217 (0.124 - 0.309)&55.56\% (36.55\% - 70.05\%)
\\
 CBIS-DDSM& PR& P < 0.0001
&0.269 (0.170 - 0.374)&62.93\% (47.66\% - 74.03\%)
\\ 
 BCS-DBT& ROC& P < 0.0001
&0.087 (0.061 - 0.113)&60.74\% (48.58\% - 71.41\%)
\\
 BCS-DBT& PR& P < 0.0001&0.240 (0.172 - 0.313)&28.73\% (20.74\% - 37.42\%)\\ \hline
    \end{tabular}

\end{table}

\begin{table}
\renewcommand{\tablename}{Supplementary Table}
    \centering
\caption{Comparison of model performance metrics and statistical significance for pairwise model evaluations, highlighting the improvement of the V2 model over the V1 model. The table presents p-values of permutation tests, differences in AUROC, AUPRC, and AUFROC\_1 ($\Delta \text{AUC}$) with 95\% confidence interval, and the reduction of errors with 95\% confidence interval. The significance level is 0.05. The V2 model reduced the errors in AUROC of the V1 model by 18.86\% to 56.62\% across large external datasets (OPTIMAM, CSAW-CC, EMBED, and BCS-DBT) and  in AUFROC\_1 by 37.58\% for EMBED and 34.17\% for BCS-DBT.}
\label{tab:test_v1_v2}
    \begin{tabular}{cc|ccc}
    \hline
         Data set&   AUC
&  p-value
& $\Delta \text{AUC}$ &Reduction of errors\\ \hline
         CMMD
&  ROC&  P = 0.3440
& 0.002 (-0.007 - 0.009)&1.50\% (-6.21\% - 8.26\%)
\\
         CMMD
&  PR&  P = 0.3744
& 0.001 (-0.005 - 0.006)&1.04\% (-5.25\% - 6.87\%)
\\
         OPTIMAM
&  ROC
&  P < 0.0001
& 0.013 (0.011 - 0.016)&18.86\% (15.82\% - 21.83\%)
\\
         OPTIMAM
&  PR&  P < 0.0001
& 0.029 (0.024 - 0.034)&14.44\% (12.06\% - 16.73\%)
\\
         CSAW-CC
&  ROC
&  P = 0.0005
& 0.006 (0.002 - 0.010)&33.97\% (15.04\% - 52.10\%)
\\
         CSAW-CC
&  PR&  P < 0.0001
& 0.034 (0.018 - 0.049)&14.22\% (7.64\% - 20.53\%)
\\
         EMBED
&  ROC
&  P = 0.0001
& 0.032 (0.018 - 0.046)&29.15\% (18.37\% - 39.67\%)
\\
 EMBED
& PR& P = 0.0001
&0.076 (0.038 - 0.120)&9.61\% (4.83\% - 15.42\%)
\\
 InBreast& ROC
& P = 0.2502
&0.004 (0.000 - 0.020)&100.00\% (100.00\% - 100.00\%)
\\
 InBreast& 
PR& P = 0.2734
&0.009 (0.000 - 0.044)&100.00\% (0.00\% - 100.00\%)
\\
 CBIS-DDSM& ROC& P = 0.9556
&-0.036 (-0.078 - 0.007)&-20.92\% (-50.29\% - 3.69\%)
\\
 CBIS-DDSM& PR& P = 0.9195
&-0.028 (-0.067 - 0.011)&-17.83\% (-49.57\% - 5.83\%)
\\ 
 BCS-DBT& ROC& P < 0.0001
&0.032 (0.018 - 0.046)&56.62\% (38.68\% - 70.75\%)
\\
 BCS-DBT& PR& P = 0.0027&0.085 (0.027 - 0.148)&14.30\% (4.92\% - 24.00\%)\\
 EMBED& FROC\_1& P = 0.0146
&0.035 (-0.011, 0.078)&
37.58\% (-23.36\%, 63.01\%)\\
 BCS-DBT& FROC\_1& P = 0.0003&0.060 (0.023, 0.099)&34.17\% (14.60\%, 51.61\%)\\ \hline
    \end{tabular}

\end{table}

\subsection{Additional external validation}~\label{sec:appendix_external}

VinDr-mammo dataset~\cite{nguyen2023vindr} consists of 5,000 mammography exams sampled from the exams taken between 2018 and 2020 at Hanoi Medical University Hospital and Hospital 108.
Unlike other datasets, VinDr-mammo dataset does not contain information about biopsy-confirmed breast cancer. 
This dataset, however, contains the BI-RADS labels of each breast. 
Breast-level statistics is as follows: 6,703 breasts have BI-RADS of 1, 2,338 breasts have BI-RADS of 2, 465 breasts have BI-RADS of 3, 381 breasts have BI-RADS of 4, and 113 breasts have BI-RADS of 5. 

Since VinDr-mammo dataset did not contain biopsy results but only contained radiologists' interpretation of mammography images in terms of BI-RADS, we defined a task of identifying suspicious findings which are defined as BI-RADS of 4 and 5.
The negative examples in this task are defined as exams with BI-RADS of 1, 2 and 3.
We measure the efficacy of our AI systems in separating the exams with and without suspicious findings using the model's probability predictions for the presence of breast cancer.
Despite not being trained for this particular task, our AI systems show decent performance for the internal test sets and the VinDr-mammo dataset.
On VinDr-mammo, the V1 model achieved 0.801 AUROC [CI: 0.776 to 0.824] and 0.457 AUPRC [CI: 0.412 to 0.501]. 
In comparison, GMIC achieved 0.793 AUROC [CI: 0.770 to 0.815] and 0.422 AUPRC [CI: 0.378 to 0.465]. 
The difference between the AUROC of the GMIC and the V1 model on VinDr-mammo was not statistically significant (permutation test, AUC difference: 0.008 [CI: -0.017 to 0.035], reduction of error: 3.83\% [CI: -8.65\% to 16.00\%], P = 0.2761).
The difference between the AUPRC of the GMIC and the V1 model on VinDr-mammo was statistically significant (permutation test, AUC difference: 0.036 [CI: 0.008 to 0.063], reduction of error: 6.16\% [CI: 1.32\% to 10.64\%], P = 0.0061).
In addition, the V2 model achieved 0.811 AUROC [CI: 0.787 to 0.834] and 0.492 AUPRC [CI: 0.447 to 0.535]. 
The difference between the AUROC of the V1 and V2 models on VinDr-mammo was not statistically significant (permutation test, AUC difference: 0.010 [CI: -0.006 to 0.027], reduction of error: 5.09\% [CI: -3.56\% to 12.88\%], P = 0.1150).
The difference between the AUPRC of the V1 and V2 models on VinDr-mammo was statistically significant (permutation test, AUC difference: 0.034 [CI: 0.011 to 0.059], reduction of error: 6.31\% [CI: 2.12\% to 10.69\%], P = 0.0022).

\subsection{Detailed clinical implementation results}~\label{sec:appendix_clinical}

Supplementary Table~\ref{tab:my_table} presents a detailed comparison of the abnormal interpretation rate (AIR) for each radiologist before and after the implementation of the AI model. 
The table includes the AIR values prior to and following AI implementation, the statistical significance of any changes (p-values), the total number of mammography exams interpreted during the study period, and the years of experience for each radiologist (ranging from 3 to 28 years). 
While 2 radiologists demonstrated a significant reduction in their recall rates, 3 radiologists exhibited a significant increase. 
For the remaining 15 radiologists, no statistically significant changes in AIR were observed. 
This table provides insight into the variability of radiologist responses to the clinical implementation of AI in screening mammography.

\begin{table}
\renewcommand{\tablename}{Supplementary Table}
\centering
\caption{Comparison of abnormal interpretation rate (AIR) by radiologist for all cases before and after AI implementation. Each AIR is accompanied by 95\% confidence interval. The p-value was calculated using a two-proportion z-test. The significance level is 0.05.}
\label{tab:my_table}

\resizebox{\textwidth}{!}{%
\begin{tabular}{| l  |l  |l  |r|r|p{3cm}|p{2cm} |}
\hline
AIR before AI & AIR after AI & p-value &    z&Cohen's h&number of mammogrphy exams interpreted & years of experience \\
\hline
20.9\% (18.5\% - 23.3\%)& 24.2\% (21.9\% - 26.6\%)& P = 0.05454&    1.923&0.079
&2,369&7
\\
\hline
10.4\% (8.8\% - 12.2\%)& 11.0\% (9.3\% - 12.8\%)& P = 0.6006&    0.524&0.020
&2,766&14
\\
\hline
11.9\% (10.2\% - 13.6\%)& 12.0\% (10.4\% - 13.8\%)& P = 0.94957&    0.063&0.002
&2,785&15
\\
\hline
9.4\% (8.5\% - 10.3\%)& 8.6\% (7.8\% - 9.5\%)& P = 0.2411&    -1.172&-0.025
&8,513&38
\\
\hline
20.5\% (17.0\% - 24.5\%)& 15.2\% (12.4\% - 17.8\%)& P = 0.02406&    -2.256&-0.139
&1,074&6
\\
\hline
13.4\% (11.8\% - 15.2\%)& 16.0\% (13.8\% - 17.9\%)& P = 0.07786&    1.763&0.072
&2,425&10
\\
\hline
10.6\% (9.9\% - 11.4\%)& 13.3\% (12.5\% - 14.2\%)& P < 0.00001&    5.016&0.083
&14,602&8
\\
\hline
9.5\% (6.9\% - 12.0\%)& 10.7\% (9.6\% - 11.8\%)& P = 0.39043&    0.859&0.041
&3,302&7
\\
\hline
9.6\% (8.3\% - 10.9\%)& 9.0\% (7.5\% - 10.8\%)& P = 0.57764&    -0.557&-0.020
&3,253&17
\\
\hline
11.9\% (10.5\% - 13.4\%)& 10.9\% (9.4\% - 12.6\%)& P = 0.39522&    -0.850&-0.032
&2,918&13
\\
\hline
12.5\% (10.5\% - 14.3\%)& 9.9\% (8.1\% - 11.9\%)& P = 0.05632&    -1.909&-0.084
&2,109&3
\\
\hline
20.1\% (17.8\% - 22.1\%)& 16.3\% (14.2\% - 18.5\%)& P = 0.01115&    -2.538&-0.099
&2,703&21
\\
\hline
15.1\% (12.9\% - 17.4\%)& 14.6\% (12.5\% - 16.9\%)& P = 0.74082&    -0.331&-0.015
&1,852&28
\\
\hline
15.2\% (13.9\% - 16.7\%)& 16.5\% (15.1\% - 18.1\%)& P = 0.20215&    1.275&0.036
&4,998&8
\\
\hline
6.0\% (4.9\% - 7.2\%)& 8.2\% (6.8\% - 9.8\%)& P = 0.01750&    2.376&0.088
&2,937&3
\\
\hline
8.3\% (6.8\% - 9.9\%)& 9.1\% (7.5\% - 10.9\%)& P = 0.47356&    0.717&0.029
&2,437&13
\\
\hline
14.2\% (12.5\% - 16.1\%)& 14.5\% (13.0\% - 16.1\%)& P = 0.82752&    0.218&0.008
&3,322&12
\\
\hline
10.6\% (9.9\% - 11.4\%)& 15.6\% (14.7\% - 16.7\%)& P < 0.00001&    7.631&0.148
&10,631&28
\\
\hline
11.9\% (10.0\% - 13.9\%)& 9.9\% (8.5\% - 11.5\%)& P = 0.11554&    -1.574&-0.063
&2,529&4
\\
\hline
7.6\% (6.2\% - 9.1\%)& 7.1\% (6.1\% - 8.2\%)& P = 0.57250&    -0.564&-0.019&3493&35\\
\hline
\end{tabular}
}
\end{table}

Supplementary Table~\ref{tab:stats_prospective} shows the distribution of covariates that could influence recall rates in the two groups in the prospective study: patient age, breast density, and family history of relevant cancer.
We categorize family history into three groups: ``has family history of relevant cancer,'' ``no family history of relevant cancer,'' and ``no mention of family history,'' defining relevant cancer broadly to include any type of cancer mentioned in the radiology report of each exam.
We use an NLP algorithm to extract this information from the radiology reports of each exam.
While a narrower definition focusing on family history of breast cancer might be more clinically meaningful, it is infeasible due to vague or incomplete description in some reports (e.g., "positive family history" with no further explanation).
Supplementary Table~\ref{tab:subgroup_test_prospective}, Supplementary Table~\ref{tab:subgroup_test_prospective_gray}, Supplementary Table~\ref{tab:subgroup_test_prospective_mixed}, Supplementary Table~\ref{tab:subgroup_test_prospective_green} shows the comparison of AIR before and after the AI implementation as well as p-value across subgroups defined by age, breast density, and family history of relevant cancer, respectively.
Specifically:
\begin{enumerate}
    \item Supplementary Table~\ref{tab:subgroup_test_prospective} is for all cases.
    \item Supplementary Table~\ref{tab:subgroup_test_prospective_gray} is for gray cases only.
    \item Supplementary Table~\ref{tab:subgroup_test_prospective_mixed} is for mixed cases only.
    \item Supplementary Table~\ref{tab:subgroup_test_prospective_green} is for green cases only.
\end{enumerate}
The results for the all, gray, mixed, and green cases show that the changes between AIR before AI and AIR after AI for most subgroups align with the overall trends observed in the analysis without subgroups.
This subgroup analysis suggests that the observed difference in AIR is primarily attributable to the introduction of AI, rather than other covariates.

\begin{table}
\renewcommand{\tablename}{Supplementary Table}
    \centering
\caption{Statistics of the exams in the prospective study. Abbreviations: N, number; SD, standard deviation.}
\label{tab:stats_prospective}
    \begin{tabular}{>{\raggedleft\arraybackslash}p{8cm} | >{\raggedleft\arraybackslash}p{3.7cm} | >{\raggedleft\arraybackslash}p{3.3cm}}

         \hline
         \textbf{Characteristics, unit} & Interpreted without AI & Interpreted with AI  \\
         \hline \hline
         Exams, N& 40,415&  
40,603\\ \hline
         Age, mean years (SD)&  59.1 (12.2)&  
59.9 (12.3)\\ 
         $<$ 40 yrs old, N (\%)&  900 (2.23\%)&  
870 (2.14\%)
\\ 
         40 - 49 yrs old, N (\%)&  10,267 (25.40\%)&  
9,674 (23.83\%)
\\ 
         50 - 59 yrs old, N (\%)&  10,458 (25.88\%)&  
10,072 (24.81\%)
\\ 
         60 - 69 yrs old, N (\%)&  10,054 (24.88\%)&  
10,279 (25.32\%)
\\ 
         $\geq$ 70 yrs old, N (\%)&  8,736 (21.62\%)&  
9,708 (23.91\%)\\ \hline
         Breast density& &  
\\ 
         A (breasts are almost entirely fatty), N (\%)&3,475 (8.60\%)&  2,635 (6.49\%)
\\ 
         B (scattered areas of fibroglandular density), N (\%)&16,651 (41.20\%)& 16,144 (39.76\%)
\\ 
         C  (breasts are heterogeneously dense), N (\%)& 17,647 (43.66\%)& 19,238 (47.38\%)
\\ 
         D (the breasts are extremely dense), N (\%)&2,630 (6.51\%)&2,579 (6.35\%)
\\ 
         Unknown density, N (\%)& 12 (0.03\%)&  7 (0.02\%)\\ \hline
         Family history& &   
\\ 
         Has family history of relevant cancer, N (\%)&11,562 (28.61\%)&  11,292 (27.81\%)\\ 
         No family history of relevant cancer, N (\%)&16,561 (40.98\%)&  15,949 (39.28\%)\\ 
         No mention of family history, N (\%)& 12,292 (30.41\%)&  13,362 (32.91\%)\\ \hline
    \end{tabular}
\end{table}

\begin{table}
\renewcommand{\tablename}{Supplementary Table}
    \centering
\caption{Subgroup analysis of abnormal interpretation rate (AIR) for all cases before and after AI implementation by age, breast density, and family history of relevant cancer. Each AIR is accompanied by 95\% confidence interval. The p-value was calculated using a two-proportion z-test. The significance level is 0.05.}
\label{tab:subgroup_test_prospective}

\resizebox{\textwidth}{!}{%
    \begin{tabular}{l|rrrrr}
    \hline
         Subgroup&  AIR before AI &  AIR after AI & p-value  & z&Cohen's h\\ \hline
         Age & & &  & &\\
          < 40 yrs old&  19.2\% (16.4\% - 21.8\%)& 20.6\% (17.9\% - 23.4\%)& P = 0.47605& 0.713&0.034
\\
          40-49 yrs old&  15.7\% (15.0\% - 16.4\%)& 17.2\% (16.4\% - 18.0\%)& P = 0.00482& 2.819&0.040
\\
          50-59 yrs old&  10.3\% (9.7\% - 10.9\%)& 11.6\% (11.0\% - 12.2\%)& P = 0.00214& 3.070&0.043
\\
          60-69 yrs old&  10.2\% (9.6\% - 10.8\%)& 10.9\% (10.4\% - 11.5\%)& P = 0.09055& 1.693&0.024
\\ 
 >=70 yrs old& 9.2\% (8.6\% - 9.8\%)& 10.1\% (9.5\% - 10.7\%)&P = 0.05087& 1.953&0.029\\ \hline
         Breast density & & &  & &\\
          Nondense (A+B)&  9.9\% (9.5\% - 10.3\%)& 10.7\% (10.3\% - 11.2\%)& P = 0.00727& 2.684&0.027
\\
          Dense (C+D)&  13.3\% (12.8\% - 13.7\%)& 14.2\% (13.7\% - 14.6\%)& P = 0.00728& 2.684&0.026\\ \hline
         Family history of relevant cancer& & &  & &\\
          Has family history&  10.7\% (10.2\% - 11.3\%)& 12.2\% (11.7\% - 12.8\%)& P = 0.00045& 3.509&0.046
\\
          No family history&  11.6\% (11.1\% - 12.0\%)& 13.0\% (12.5\% - 13.5\%)& P = 0.00011& 3.873&0.043
\\
          No mention of family history&  12.4\% (11.9\% - 13.1\%)& 12.4\% (11.8\% - 13.0\%)& P = 0.94071& -0.074&-0.001\\ \hline
    
    \end{tabular}
}
\end{table}

\begin{table}
\renewcommand{\tablename}{Supplementary Table}
    \centering
\caption{Subgroup analysis of abnormal interpretation rate (AIR) for gray cases before and after AI implementation by age, breast density, and family history of relevant cancer. Each AIR is accompanied by 95\% confidence interval. The p-value was calculated using a two-proportion z-test. The significance level is 0.05.}
\label{tab:subgroup_test_prospective_gray}

\resizebox{\textwidth}{!}{%
    \begin{tabular}{l|rrrrr}
    \hline
         Subgroup&  AIR before AI &  AIR after AI & p-value  & z&Cohen's h\\ \hline
         Age & & &  & &\\
          < 40 yrs old&  25.2\% (18.0\% - 32.4\%)& 29.9\% (22.2\% - 37.5\%)& P = 0.37827& 0.881&0.105
\\
          40-49 yrs old&  20.6\% (19.2\% - 22.3\%)& 25.5\% (23.9\% - 27.4\%)& P = 0.00005& 4.046&0.117
\\
          50-59 yrs old&  13.4\% (12.2\% - 14.7\%)& 16.4\% (15.3\% - 17.7\%)& P = 0.00041& 3.534&0.086
\\
          60-69 yrs old&  14.1\% (13.0\% - 15.2\%)& 16.4\% (15.2\% - 17.5\%)& P = 0.00444& 2.845&0.065
\\ 
 >=70 yrs old& 11.1\% (10.1\% - 12.1\%)& 13.2\% (12.3\% - 14.2\%)&P = 0.00208& 3.079&0.065\\ \hline
         Breast density & & &  & &\\
          Nondense (A+B)&  12.9\% (12.1\% - 13.8\%)& 15.6\% (14.8\% - 16.5\%)& P = 0.00001& 4.343&0.077
\\
          Dense (C+D)&  15.5\% (14.6\% - 16.3\%)& 18.0\% (17.2\% - 18.9\%)& P = 0.00003& 4.208&0.068\\ \hline
         Family history of relevant cancer& & &  & &\\
          Has family history&  13.0\% (12.0\% - 14.1\%)& 16.1\% (15.0\% - 17.2\%)& P = 0.00005& 4.036&0.087
\\
          No family history&  14.3\% (13.4\% - 15.2\%)& 17.9\% (17.0\% - 18.8\%)& P < 0.00001& 5.237&0.098
\\
          No mention of family history&  15.6\% (14.5\% - 16.7\%)& 16.6\% (15.5\% - 17.7\%)& P = 0.22411& 1.216&0.027\\ \hline
    
    \end{tabular}
}
\end{table}

\begin{table}
\renewcommand{\tablename}{Supplementary Table}
    \centering
\caption{Subgroup analysis of abnormal interpretation rate (AIR) for mixed cases before and after AI implementation by age, breast density, and family history of relevant cancer. Each AIR is accompanied by 95\% confidence interval. The p-value was calculated using a two-proportion z-test. The significance level is 0.05.}
\label{tab:subgroup_test_prospective_mixed}

\resizebox{\textwidth}{!}{%
    \begin{tabular}{l|rrrrr}
    \hline
         Subgroup&  AIR before AI &  AIR after AI & p-value  & z&Cohen's h\\ \hline
         Age & & &  & &\\
          < 40 yrs old&  25.1\% (19.8\% - 30.9\%)& 25.8\% (20.3\% - 31.6\%)& P = 0.86195& 0.174&0.016
\\
          40-49 yrs old&  18.8\% (17.4\% - 20.3\%)& 23.6\% (22.1\% - 25.2\%)& P < 0.00001& 4.604&0.117
\\
          50-59 yrs old&  12.3\% (11.3\% - 13.4\%)& 12.9\% (11.7\% - 14.1\%)& P = 0.51235& 0.655&0.016
\\
          60-69 yrs old&  10.4\% (9.4\% - 11.5\%)& 11.1\% (10.0\% - 12.2\%)& P = 0.36993& 0.897&0.022
\\ 
 >=70 yrs old& 9.1\% (8.0\% - 10.3\%)& 9.8\% (8.7\% - 10.9\%)&P = 0.36773& 0.901&0.025\\ \hline
         Breast density & & &  & &\\
          Nondense (A+B)&  11.3\% (10.5\% - 12.0\%)& 12.4\% (11.7\% - 13.2\%)& P = 0.04137& 2.040&0.036
\\
          Dense (C+D)&  15.0\% (14.1\% - 15.9\%)& 16.5\% (15.7\% - 17.5\%)& P = 0.01707& 2.385&0.043\\ \hline
         Family history of relevant cancer& & &  & &\\
          Has family history&  12.1\% (11.0\% - 13.2\%)& 12.9\% (11.8\% - 14.0\%)& P = 0.31966& 0.995&0.024
\\
          No family history&  13.3\% (12.4\% - 14.3\%)& 15.0\% (14.1\% - 16.0\%)& P = 0.01294& 2.485&0.049
\\
          No mention of family history&  13.6\% (12.5\% - 14.7\%)& 15.3\% (14.2\% - 16.5\%)& P = 0.02825& 2.194&0.049\\ \hline
    
    \end{tabular}
}
\end{table}

\begin{table}
\renewcommand{\tablename}{Supplementary Table}
    \centering
\caption{Subgroup analysis of abnormal interpretation rate (AIR) for green cases before and after AI implementation by age, breast density, and family history of relevant cancer. Each AIR is accompanied by 95\% confidence interval. The p-value was calculated using a two-proportion z-test. The significance level is 0.05.}
\label{tab:subgroup_test_prospective_green}

\resizebox{\textwidth}{!}{%
    \begin{tabular}{l|rrrrr}
    \hline
         Subgroup&  AIR before AI & AIR after AI & p-value  & z&Cohen's h\\ \hline
         Age & & &  & &\\
          < 40 yrs old&  14.9\% (11.8\% - 18.0\%)& 14.9\% (11.7\% - 18.1\%)& P = 0.98988& 0.013&0.001
\\
          40-49 yrs old&  11.0\% (10.1\% - 11.9\%)& 8.3\% (7.5\% - 9.1\%)& P = 0.00001& -4.347&-0.092
\\
          50-59 yrs old&  5.5\% (4.8\% - 6.3\%)& 5.1\% (4.3\% - 5.9\%)& P = 0.40936& -0.825&-0.020
\\
          60-69 yrs old&  5.3\% (4.6\% - 6.2\%)& 3.2\% (2.6\% - 3.9\%)& P = 0.00006& -4.006&-0.105
\\ 
 >=70 yrs old& 5.3\% (4.3\% - 6.2\%)& 3.0\% (2.3\% - 3.7\%)&P = 0.00023& -3.684&-0.117\\ \hline
         Breast density & & &  & &\\
          Nondense (A+B)&  5.8\% (5.3\% - 6.4\%)& 3.9\% (3.4\% - 4.4\%)& P < 0.00001& -5.129&-0.091
\\
          Dense (C+D)&  9.4\% (8.7\% - 10.1\%)& 7.4\% (6.7\% - 8.0\%)& P = 0.00001& -4.339&-0.074\\ \hline
         Family history of relevant cancer& & &  & &\\
          Has family history&  6.8\% (6.0\% - 7.6\%)& 6.3\% (5.5\% - 7.2\%)& P = 0.45264& -0.751&-0.018
\\
          No family history&  7.3\% (6.7\% - 8.0\%)& 5.4\% (4.8\% - 6.0\%)& P = 0.00005& -4.046&-0.079
\\
          No mention of family history&  8.7\% (7.9\% - 9.6\%)& 5.7\% (5.1\% - 6.4\%)& P < 0.00001& -5.528&-0.116\\ \hline
    
    \end{tabular}
}
\end{table}

\end{document}